\documentclass[%
 reprint,
 superscriptaddress,
 amsmath,amssymb,
 aps,
 pra, 
]{revtex4-2}

\usepackage{graphicx}
\usepackage{dcolumn}
\usepackage{multirow} 
\usepackage{xcolor}
\usepackage[normalem]{ulem}
\usepackage{bm}
\usepackage{hyperref}
\usepackage{physics}
\usepackage{braket}
\usepackage{amsmath}
\usepackage{amsfonts}
\usepackage{amssymb}
\usepackage{amsthm}
\usepackage{mathtools}
\usepackage{color} 
\usepackage{gensymb}
\usepackage{here}

\begin{document}

\title{Hamiltonian simulation-based quantum-selected configuration interaction for large-scale electronic structure calculations with a quantum computer}

\author{Kenji Sugisaki}
\email{ksugisaki@keio.jp}
\affiliation{Quantum Computing Center, Keio University, 3-14-1 Hiyoshi, Kohoku-ku, Yokohama, Kanagawa 223-8522, Japan}
\affiliation{Graduate School of Science and Technology, Keio University, 7-1 Shinkawasaki, Saiwai-ku, Kawasaki, Kanagawa 212-0032, Japan}
\affiliation{Keio University Sustainable Quantum Artificial Intelligence Center (KSQAIC), Keio University, 2-15-45 Mita, Minato-ku, Tokyo 108-8345, Japan}
\affiliation{Centre for Quantum Engineering, Research and Education (CQuERE), TCG Centres for Research and Education in Science and Technology (TCG CREST), Sector V, Salt Lake, Kolkata 700091, India}

\author{Shu Kanno}
\affiliation{Mitsubishi Chemical Corporation, Science \& Innovation Center, Yokohama, Kanagawa 227-8502, Japan}
\affiliation{Quantum Computing Center, Keio University, 3-14-1 Hiyoshi, Kohoku-ku, Yokohama, Kanagawa 223-8522, Japan}

\author{Toshinari Itoko}
\affiliation{IBM Quantum, IBM Research--Tokyo, 19-21 Nihonbashi Hakozaki-cho, Chuo-ku, Tokyo 103-8510, Japan}
\affiliation{Quantum Computing Center, Keio University, 3-14-1 Hiyoshi, Kohoku-ku, Yokohama, Kanagawa 223-8522, Japan}

\author{Rei Sakuma}
\affiliation{Materials Informatics Initiative, RD Technology \& Digital Transformation Center, JSR Corporation, 3-103-9 Tonomachi, Kawasaki-ku, Kawasaki, Kanagawa 210-0821, Japan}
\affiliation{Quantum Computing Center, Keio University, 3-14-1 Hiyoshi, Kohoku-ku, Yokohama, Kanagawa 223-8522, Japan}

\author{Naoki Yamamoto}
\affiliation{Department of Applied Physics and Physico-Informatics, Keio University, 3-14-1 Hiyoshi, Kohoku-ku, Yokohama, Kanagawa 223-8522, Japan}
\affiliation{Quantum Computing Center, Keio University, 3-14-1 Hiyoshi, Kohoku-ku, Yokohama, Kanagawa 223-8522, Japan}
\affiliation{Keio University Sustainable Quantum Artificial Intelligence Center (KSQAIC), Keio University, 2-15-45 Mita, Minato-ku, Tokyo 108-8345, Japan}

\date{\today}

\begin{abstract}
Quantum-selected configuration interaction (QSCI) is an approach for quantum chemical calculations using current quantum computers. 
In conventional QSCI, Slater determinants used for the wave function expansion are sampled by iteratively performing approximate wave function preparation and subsequent measurement in the computational basis, and then the subspace Hamiltonian matrix is diagonalized on a classical computer. 
In this approach, preparation of a high-quality approximate wave function is necessary to accurately compute total energies. 
Here we propose a Hamiltonian simulation-based QSCI (HSB-QSCI) to avoid this difficulty, by sampling the Slater determinants from quantum states generated by the real-time evolution of approximate wave functions. 
We provide numerical simulations for the lowest spin-singlet and triplet states of oligoacenes (benzene, naphthalene, and anthracene), phenylene-1,4-dinitrene, and hexa-1,2,3,4,5-pentaene. We found that the HSB-QSCI is applicable not only to molecules where the Hartree--Fock provides a good approximation of the ground state, but also to strongly correlated systems where preparing a high-quality approximate wave function is hard. 
Hardware demonstrations of the HSB-QSCI are also reported for carbyne molecules expressed by up to 36 qubits, using an IBM Quantum processor.
The HSB-QSCI captures more than 99.18\% of the correlation energies in the active space by considering about 1\% of all the Slater determinants in 36 qubit systems, illustrating the ability of the proposed method to efficiently consider important electronic configurations. 
\end{abstract}
\maketitle

\section{Introduction}
Among the diverse topics in the field of quantum computing, quantum chemical calculations of atoms and molecules have attracted attention as a promising application of quantum computers. A method to perform the full-configuration interaction (full-CI) calculation using the quantum phase estimation (QPE) algorithm was proposed in 2005~\cite{Alan-2005}, and proof-of-principle experiments for the full-CI/STO-3G of the H$_2$ molecule were reported in 2010~\cite{Lanyon-2010, Du-2010}. The quantum circuit for QPE-based full-CI is usually very deep, so highly accurate QPE demonstrations were limited to small models with a few qubits~\cite{Wang-2015, OMalley-2016, Santagati-2018, Blunt-2023, Yamamoto-2024}.
Recently a QPE-type algorithm has been demonstrated in models with up to 33 qubits~\cite{Kanno-2024}, however, it is still challenging to compute total energies of larger molecules with chemical precision on quantum computers available today, where chemical precision is defined as an error from the full-CI energy [or complete active space configuration interaction (CAS-CI) energy when the active space approximation is employed] less than 1.0 kcal mol$^{-1}$.

To reduce the computational load of quantum computers, several quantum--classical hybrid algorithms have been proposed. In 2014, the variational quantum eigensolver (VQE) was proposed for quantum chemical calculations using noisy intermediate-scale quantum (NISQ) devices~\cite{Peruzzo-2014}. In VQE, the quantum state corresponding to an approximate wave function is generated using a parameterized quantum circuit, and the energy expectation value is computed by repeatedly performing state preparation and measurements. The parameters are then optimized to minimize the energy expectation value on a classical computer. These steps are iterated until convergence is reached. VQE has been extensively studied both theoretically and experimentally~\cite{Tilly-2022}, but it often suffers from challenges related to the high sampling cost required to achieve chemical precision~\cite{Gonthier-2022} and issues with variational optimization in the presence of barren plateaus~\cite{McClean-2018}.

As an alternative approach to quantum chemical calculations using current quantum devices, the quantum-selected configuration interaction (QSCI) method has been proposed~\cite{Kanno-2023}. In the QSCI, quantum computers are used to sample Slater determinants that are important contributors to the ground state wave function. This can be done by running a quantum circuit to prepare an approximate wave function and measuring the quantum state in the computational basis. The subspace Hamiltonian matrix is then constructed from the Slater determinants selected via quantum computation and diagonalized on a classical computer. Note that the concept of the selected CI has been well investigated in classical computation, and various selected CI methods have been developed, such as the CI using a perturbative selection made iteratively (CIPSI)~\cite{Huron-1973}, heat-bath CI (HCI)~\cite{Holmes-2016}, and adaptive sampling CI (ASCI)~\cite{Tubman-2020}. These methods are often combined with multi-reference perturbation theory~\cite{Garniron-2018, Sharma-2017}. In the quantum domain, QSCI deals with the variational optimization of the wave function in the selected subspace, and the effects of the remaining dynamical correlations can be taken into account by combining it with, for example, auxiliary-field quantum Monte Carlo (AFQMC)~\cite{Yoshida-2025, Danilov-2025} or general multi-configurational quasi-degenerate perturbation theory (GMC-QDPT)~\cite{Shirai-2025}, although the perturbative energy correction from the QSCI is outside the scope of this study.

In the QSCI approach, the quantum computer is anticipated to efficiently sample a polynomial number of important Slater determinants from an exponentially large Hilbert space, and the nature of the quantum states used for sampling determines the accuracy of the calculation. 
At the stage of preparing an approximate wave function in QSCI, we may employ VQE or the adiabatic state preparation method \cite{Kanno-2023}. To advance the former technique, combining an adaptive strategy for ansatz construction (ADAPT~\cite{Grimsley-2019}) to enhance the efficiency of VQE~\cite{Nakagawa-2023}, and using a simpler cost function for VQE parameter optimization~\cite{Nutzel-2024} have also been proposed. However, these approaches may still face challenges in variational optimizations when implemented on quantum hardware. It should also be noted that in the latter approach \cite{Nutzel-2024}, the VQE wave function with optimized parameters does not necessarily correspond to the minimum with respect to the given Hamiltonian.
Another strategy to prepare the quantum state for QSCI is to perform the coupled cluster singles and doubles (CCSD) calculation on a classical computer and embed the CCSD wave function using the local unitary cluster Jastrow (LUCJ) ansatz~\cite{Motta-2023}. By using the LUCJ ansatz and introducing an error mitigation technique called as a self-consistent configuration recovery (SCCR), which is inspired by the structure of chemistry problems, Robledo-Moreno and coworkers reported the QSCI calculations on the IBM superconducting quantum processor and supercomputer ``Fugaku'' for the nitrogen molecule (N$_2$) and the iron--sulfur clusters [2Fe--2S] and [4Fe--4S] with up to 77 qubits~\cite{Robledo-2024}. This approach has also been applied to interaction energy~\cite{Kaliakin-2024} and excited state calculations~\cite{Barison-2024, Liepuoniute-2024}. The combination of the QSCI with density matrix embedding theory has also been reported~\cite{Shajan-2024}. While this approach appears to be successful, it is not clear whether it is superior to the method of sampling important Slater determinants from the CCSD wave function on a classical computer.

In this work, we report a Hamiltonian simulation-based quantum state preparation for QSCI (HSB-QSCI), in which important Slater determinants are sampled from the quantum states after the real-time evolutions of initial wave functions \cite{footnote}. 
Our proposed approach has several advantages over existing methods: (1) A simple approximate wave function, such as the HF wave function, can be used as the initial wave function, and no variational optimization in VQE is required. (2) Different samples can be obtained by changing the duration of the evolution time, and higher-order excitations can be considered even for short-time evolution. (3) The accuracy of the Hamiltonian simulation does not need to be very high. As we will show later, the first-order Trotter decomposition with the time length of a single Trotter step $\Delta t = 1$ in atomic unit is sufficient to collect important Slater determinants for the organic molecules under study. In addition, Hamiltonian term truncation based on operator locality in the localized molecular orbital basis can help to reduce the computational cost. (4) Feedback of the QSCI results into quantum computations is possible; that is, Slater determinants found to be important in the QSCI wave function can be added to the initial wave functions. (5) Although short-time evolution can be accurately simulated on a classical computer, long-time evolution poses significant challenges for classical simulation. 

Note that methods of sampling basis states for Hamiltonian matrix diagonalization based on Hamiltonian simulation have already been investigated in the framework of quantum Krylov subspace (QKS) algorithms~\cite{Cortes-2022, Zhang-2024}. 
In the QKS method, the Hamiltonian matrix $H_{kl} = \braket{\Phi_k|H|\Phi_l}$ and the overlap matrix $S_{kl} = \braket{\Phi_k|\Phi_l}$ are computed using a quantum computer via a Hadamard test, and the generalized eigenvalue problem $\mathbf{Hc}=\mathbf{Sc}E$ is solved on a classical computer. Here, $\ket{\Phi_k} = U^k\ket{\Phi_0}$, where $U = e^{-iH\Delta t}$ and $\ket{\Phi_0}$ is the initial wave function. Since the Taylor expansion of the time evolution operator yields polynomials of $H$ as follows, time evolution is expected to help finding electron configurations that are important for the wave function expansion.
\begin{eqnarray}
    e^{-iHt} = 1 -iHt + \frac{(-iHt)^2}{2} + \frac{(-iHt)^3}{3!} + \cdots
\end{eqnarray}
Unlike HSB-QSCI, QKS is applicable to systems in which the correlated wave function is supported by an exponentially large number of Slater determinants. However, the overlap matrix $\mathbf{S}$ becomes ill-conditioned when the subspace basis is nearly linearly dependent~\cite{Zhang-2024, Lee-2025}, and the sampling cost in QKS becomes large in order to avoid numerical instability. In contrast, HSB-QSCI requires only measurements in the computational (Pauli-Z) basis and therefore the computational cost on a quantum computer is significantly reduced. HSB-QSCI is also expected to be numerically more stable, because Slater determinants are orthonormal, while subspace Hamiltonian diagonalization on a classical computer can be a bottleneck.
In the context of the Monte Carlo simulation, a quantum dynamics Hamiltonian Monte Carlo (qdHMC) method is also proposed~\cite{Lockwood-2024}, in which the proposal step is performed on a quantum computer by sampling after time evolution. 

As a proof-of-concept demonstration of the HSB-QSCI method, we first carried out numerical simulations for the spin-singlet ground state of the H$_2$O molecule, and compared the results obtained using different sampling strategies (randomly sampling Slater determinants preserving spatial symmetry, $M_S$, and $n_{elec}$, where $M_S$ and $n_{elec}$ represent the spin magnetic quantum number and the number of electrons, respectively, VQE-based QSCI~\cite{Kanno-2023}, and HCI). Then, we performed numerical simulations for the spin-singlet ground (S$_0$) state and the first excited spin-triplet (T$_1$) state of oligoacenes [benzene ({\bf 1}; 12 qubits), naphthalene ({\bf 2}; 20 qubits), and anthracene ({\bf 3}; 28 qubits)], phenylene-1,4-dinitrene ({\bf 4}; 20 qubits), and hexa-1,2,3,4,5-pentaene ({\bf 5}; 20 qubits) in both planar and twisted geometries. Our numerical simulations showed that the HSB-QSCI can calculate the total energy with chemical precision by sampling the Slater determinants with about 10--20 steps of Hamiltonian simulations.
We also report hardware demonstrations of the HSB-QSCI for the S$_0$ and T$_1$ states of three carbyne (one-dimensional sp-hybridized carbon atom chain) molecules including {\bf 5}, octa-1,2,3,4,5,6,7-heptaene ({\bf 6}; 28 qubits), and deca-1,2,3,4,5,6,7,8,9-nonaene ({\bf 7}; 36 qubits) using an IBM Quantum processor \texttt{ibm\_kawasaki}, with the aid of matrix product operator (MPO)-based classical compression of the quantum circuit for time evolution. The HSB-QSCI energies calculated on \texttt{ibm\_kawasaki} agree with the CAS-CI values within 0.6 kcal mol$^{-1}$ of error in {\bf 5}. In {\bf 6}, chemical precision for the total energy was not achieved with the real quantum device, but the HSB-QSCI predicted the singlet--triplet energy gap and the energy difference between planar and twisted geometries in chemical precision. The energy difference between planar and twisted geometries in the S$_0$ state of {\bf 7} was also calculated with an error of 0.29 kcal mol$^{-1}$. In {\bf 7}, the HSB-QSCI captures more than 99.18\% of correlation energies by considering only about 1\% of all the Slater determinants.

\section{Theory}

\begin{figure*}[ht]
\includegraphics[width=\textwidth]{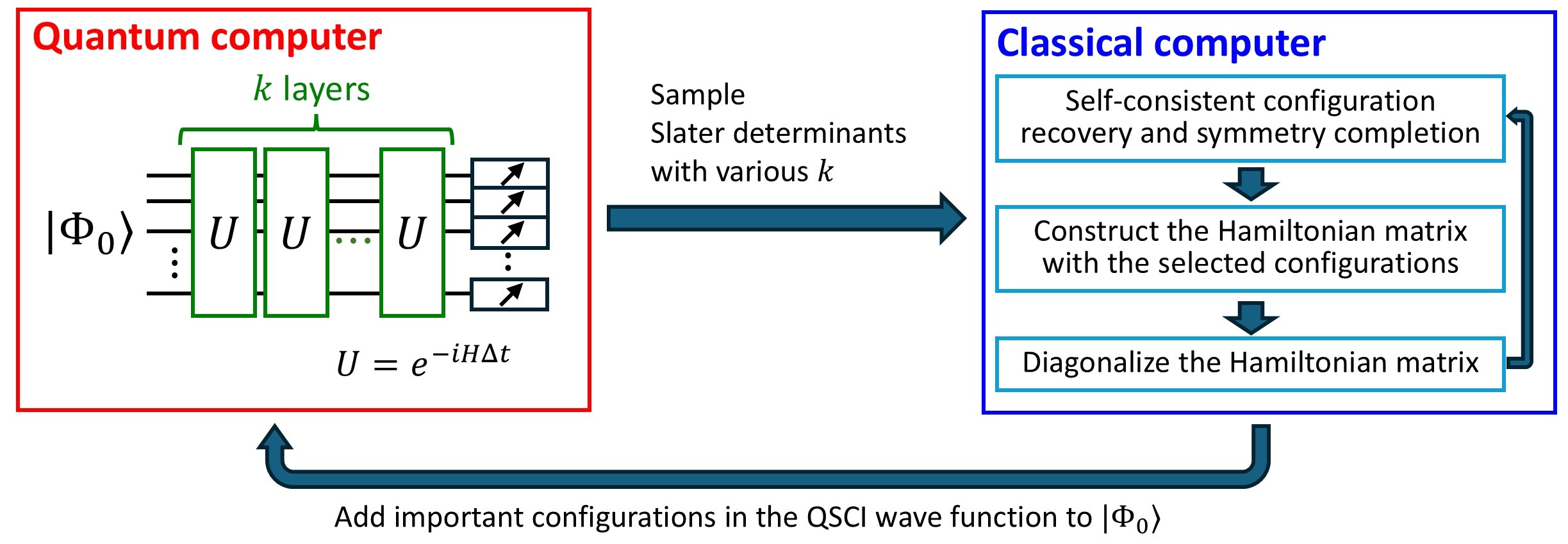}
\caption{\label{fig:fig1}Schematic view of the HSB-QSCI algorithm. }
\end{figure*}

In the {\it ab initio} molecular orbital theory, the wave function taking into account the electron correlation is expressed by a linear combination of the Slater determinants as follows:
\begin{eqnarray}
    \ket{\Phi} = c_{\mathrm{HF}}\ket{\psi_{\mathrm{HF}}} + \sum_n c_n \ket{\psi_n}.
    \label{eq1}
\end{eqnarray}

Here, $\ket{\psi_{\mathrm{HF}}}$ is the HF determinant and $\ket{\psi_n}$ are electronically excited determinants from the HF, and $c_n$ are the corresponding expansion coefficients. In the CI method, the number of Slater determinants included in the wave function expansion increases rapidly with the number of molecular orbitals and electrons, and the excitation order. To calculate the total energy with high accuracy and a low computational cost, methods to perform the CI expansion with selected Slater determinants have been investigated~\cite{Huron-1973, Holmes-2016, Schriber-2016, Zimmerman-2017, Tubman-2020}. The QSCI method uses a quantum computer to select important Slater determinants through the approximate wave function preparation and the subsequent computational-basis measurement. 
In the HSB-QSCI method, important Slater determinants are sampled from the measurement of the quantum state obtained from the time evolution of an approximate wave function. 

A schematic view of the HSB-QSCI method is shown in Figure~\ref{fig:fig1}. In the HSB-QSCI, the electronic configurations (Slater determinants) used as the basis for the Hamiltonian matrix are sampled from the time-evolved wave functions $\{\ket{\Phi_k}\}$:
\begin{eqnarray}
    \ket{\Phi_k} = e^{-iHk\Delta t}\ket{\Phi_0}.
    \label{eq2}
\end{eqnarray}

To simulate the real time evolution on a quantum computer, second quantized Hamiltonian given as
\begin{eqnarray}
    H = \sum_{pq} h_{pq} a_p^\dagger a_q + \frac{1}{2} \sum_{pqrs} g_{pqrs} a_p^\dagger a_q^\dagger a_s a_r
    \label{eq3}
\end{eqnarray}
is transformed to a qubit Hamiltonian 
\begin{eqnarray}
    H_q = \sum_j^J w_j P_j, 
    \label{eq4}
\end{eqnarray}
using a fermion--qubit mapping method. Here, $h_{pq}$ and $g_{pqrs}$ in Eq.~\eqref{eq3} are one- and two-electron integrals, respectively, and $a_p^\dagger$ and $a_p$ are the creation and annihilation operators, respectively, acting on the $p$-th spin orbital. $P_j$ is a tensor product of Pauli operators referred to as a Pauli string, and $w_j$ is its corresponding coefficient computed from $h_{pq}$ and $g_{pqrs}$. In this work, we used the Jordan--Wigner transformation (JWT)~\cite{Jordan-1928} as the fermion--qubit mapping, where each qubit stores the occupation number of the corresponding spin orbital: $\ket{1}$ if the spin orbital is occupied, and $\ket{0}$ otherwise. In the JWT, the number of qubits required for wave function mapping is equal to the number of spin orbitals in the active space. The quantum circuit for the time evolution operator $U = e^{-iH\Delta t}$ can be constructed using conventional approximate approaches such as the Trotter--Suzuki decomposition~\cite{Trotter-1959, Suzuki-1976} and qDRIFT~\cite{Ouyand-2020}, or prepared through classical optimization~\cite{Kanno-2024} or via variational quantum algorithms~\cite{Kurogi-2024}.  

The initial wave function $\ket{\Phi_0}$ should have an overlap with the target electronic state. The HF wave function $\ket{\psi_{\mathrm{HF}}}$ is a reasonable choice for $\ket{\Phi_0}$ in the electronic ground state calculations of typical closed-shell singlet molecules in their equilibrium geometries, but $\ket{\Phi_0}$ need not be a single Slater determinant. By defining the initial wave function as 
\begin{eqnarray}
    \ket{\Phi_0} = \sum_l c_l \ket{\Psi_l}, 
    \label{eq5}
\end{eqnarray}
where $\ket{\Psi_l}$ is the $l$-th eigenfunction of the Hamiltonian, the quantum state after the time evolution is written as follows:
\begin{eqnarray}
    \ket{\Phi_k} = \sum_l c_l e^{-iE_lk\Delta t}\ket{\Psi_l}. 
    \label{eq6}
\end{eqnarray}

It is clear that the total evolution time length $k\Delta t$ controls the magnitude of the interferences between the eigenstates, and the measurements of the quantum states in the computational basis with different $k$ can yield different Slater determinants. The information of the sampled Slater determinants is transferred to the classical computer, and the Hamiltonian matrix with the selected configurations is constructed and then diagonalized it to obtain the QSCI wave functions and energies. 

The effectiveness of sampling in the HSB-QSCI is mainly controlled by the evolution time length $\Delta t$, the number of time steps $k$, and the initial wave function $\ket{\Phi_0}$. Note that from Eq. (\ref{eq6}), we expect that $k\Delta t$ must be set longer for systems with smaller energy gaps with the excited states. 
In the HSB-QSCI, we can reconstruct $\ket{\Phi_0}$ using the information of the QSCI wave function by constructing a multiconfigurational wave function consisting of several Slater determinants that have large contributions in the QSCI wave function in the previous step. Performing Hamiltonian simulations with different initial wave functions $\ket{\Phi_0}, \ket{\Phi_0'}, \ket{\Phi_0''}, \dots$ and merging the measurement results to perform the QSCI is another option. The availability of such feedback from the subspace Hamiltonian diagonalization part on a classical computer to the state preparation part on a quantum computer is one of the important features of the HSB-QSCI method. 

Note that the wave functions are simultaneous eigenfunctions of the Hamiltonian and electron spin $\bf{S}^2$ operator in the non-relativistic regime, but Slater determinants are not always eigenfunctions of the $\bf{S}^2$ operator (spin eigenfunctions). In fact, open-shell Slater determinants with spin-$\beta$ unpaired electron(s) are not spin eigenfunctions. Here we have assumed that $n_\alpha \geq n_\beta$, where $n_\alpha$ and $n_\beta$ are the numbers of spin-$\alpha$ and spin-$\beta$ electrons, respectively. In order to make the QSCI wave function being spin eigenfunctions, we introduced a symmetry completion step before constructing the subspace Hamiltonian matrix. In the symmetry completion step, open-shell Slater determinants that are missing to span the spin eigenfunction are added. For example, in the spin-singlet state calculations, if the bit string corresponding to the Slater determinant $\ket{2\alpha\alpha\beta\beta0}$ is sampled (2, $\alpha$, $\beta$, and 0 mean that the corresponding molecular orbital is doubly occupied, singly occupied by a spin-$\alpha$ electron, singly occupied by a spin-$\beta$ electron, and unoccupied, respectively), then we add $\ket{2\alpha\beta\alpha\beta0}$, $\ket{2\alpha\beta\beta\alpha0}$, $\ket{2\beta\beta\alpha\alpha0}$, $\ket{2\beta\alpha\beta\alpha0}$, and $\ket{2\beta\alpha\alpha\beta0}$, if they are not sampled. Similarly for the spin-triplet state, if $\ket{2\alpha\alpha\alpha\beta0}$ is measured, then we add $\ket{2\alpha\alpha\beta\alpha0}$, $\ket{2\alpha\beta\alpha\alpha0}$, and $\ket{2\beta\alpha\alpha\alpha0}$. 

In a case when the Hamiltonian simulations are carried out on a current noisy quantum hardware, the measurements of the quantum state give bit strings that correspond to the Slater determinants with different numbers of electrons. The SCCR proposed recently~\cite{Robledo-2024} can be adopted to recover the Slater determinants with a correct number of electrons. In the SCCR, if the bit string obtained from the measurement has the wrong number of electrons, bits are flipped to recover the desired number of electrons, and the probability of bit flipping is determined from the occupation number of the molecular orbital calculated from the QSCI wave function of the previous step. This process is iterated until convergence or the iteration step reaches the predefined maximum steps. The SCCR is performed before symmetry completion. 

\section{Computational conditions}

\begin{figure}
    \centering
    \includegraphics[width=\linewidth]{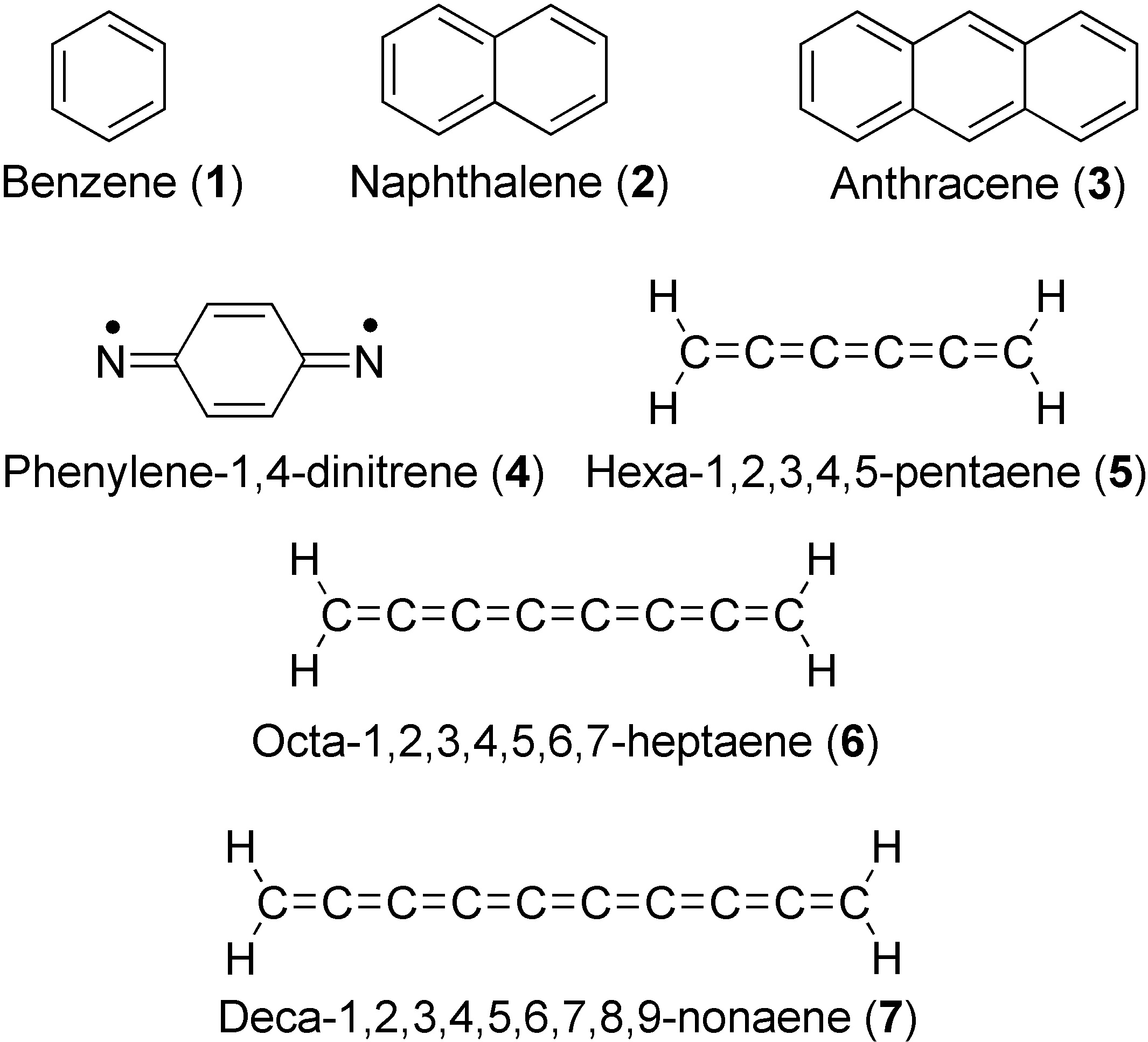}
    \caption{Target molecules being studied.}
    \label{fig:fig2}
\end{figure}

To demonstrate the HSB-QSCI, we performed numerical simulations on a classical computer. The target systems are the S$_0$ state of H$_2$O molecule, the S$_0$ and T$_1$ states of oligoacenes ({\bf 1}, {\bf 2}, and {\bf 3}), {\bf 4} as a representative system of the open-shell singlet ground state, and {\bf 5}, as shown in Figure~\ref{fig:fig2}. We also performed hardware demonstrations using an IBM Quantum processor for carbyne molecules {\bf 5}, {\bf 6}, and {\bf 7}. For the H$_2$O molecule, we used the same geometry as in the original QSCI paper~\cite{Kanno-2023}. Geometry optimizations of oligoacenes and carbynes in planar geometry were performed at the B3LYP\cite{Becke-1993, Stephens-1994}/6-31G*\cite{Ditchfield-1971, Hehre-1972} level of theory. In all DFT calculations, we used default settings in Gaussian 16~\cite{g16}. The SCF convergence thresholds were set to $10^{-8}$ and $10^{-6}$ for the RMS and maximum density changes, respectively, and $10^{-6}$ Hartree for the energy changes. For geometry optimizations, the convergence criteria were 0.000450 and 0.000300 for the maximum and RMS forces, and 0.001800 and 0.001200 for the maximum and RMS displacements, respectively. We used UltraFine grid with a pruned (99,590 grid), having 99 radial shells and 590 angular points per shell. The twisted geometry of carbynes is generated by rotating one of the CH$_2$ moieties 90 degrees from the equilibrium geometry. The geometry optimization of {\bf 4} was carried out at the CASSCF(10e,10o)/6-31G* level. Here, ($N$e,$M$o) represents the active space consisting of $N$ electrons and $M$ molecular orbitals. For the CASSCF geometry optimization, we used a gradient convergence threshold of $10^{-4}$ Hartree/Bohr, which is the default setting in the GAMESS-US~\cite{GAMESS} software. Convergence is achieved when the largest gradient component is below the specified threshold and the root mean square gradient is less than 1/3 of the threshold. Cartesian coordinates of all molecules are given in Supporting Information. 

Since the computational cost of numerical simulations of quantum circuits scales exponentially with the number of qubits, we adopted the active space approximation. For the H$_2$O molecule, we used the (6e,5o) active space, consistent with the QSCI study by Kanno and coworker~\cite{Kanno-2023}. We used the (6e,6o), (10e,10o), and (14e,14o) active spaces for {\bf 1}, {\bf 2}, and {\bf 3}, respectively, which consist of the valence $\pi$ and $\pi^*$ orbitals. The one- and two-electron integrals of H$_2$O and oligoacenes are calculated at the RHF/STO-3G~\cite{Hehre-1969} level. The active space of {\bf 4} contains the valence $\pi$ and $\pi^*$ orbitals and the in-plane 2p orbital of the nitrogen atoms, and the size is (10e,10o). The CASSCF(10e,10o)/6-31G* molecular orbitals are used to construct the second quantized Hamiltonian of {\bf 4}. In the case of carbyne molecules {\bf 5}, {\bf 6}, and {\bf 7} with (10e,10o), (14e,14o), and (18e,18o) active spaces, respectively, the RHF/6-31G* calculation is performed and then the occupied molecular orbitals are localized using the Pipek--Mezey method~\cite{Pipek-1989}. For the virtual orbitals, we first formed the singular value decomposition (SVD) quasi-atomic external orbitals using an SVD with respect to the accurate atomic minimal basis functions, and then formed the ordered external orbitals using exchange integrals (using a keyword EXTLOC=ATMNOS in GAMESS-US software~\cite{GAMESS}). 
The active orbitals of all molecules are provided in Figures S1--S10 in Supporting Information. One- and two-electron atomic orbital integrals are obtained from GAMESS-US, and the AO--MO integral transformations were performed using our own Python3 program. The qubit Hamiltonian is then generated using the \texttt{OpenFermion} library~\cite{OpenFermion}. All the DFT calculations were done with the Gaussian 16 package~\cite{g16}, and the RHF and CASSCF calculations were carried out with the GAMESS-US software~\cite{GAMESS}. 

\begin{figure*}[ht]
    \centering
    \includegraphics[width=\textwidth]{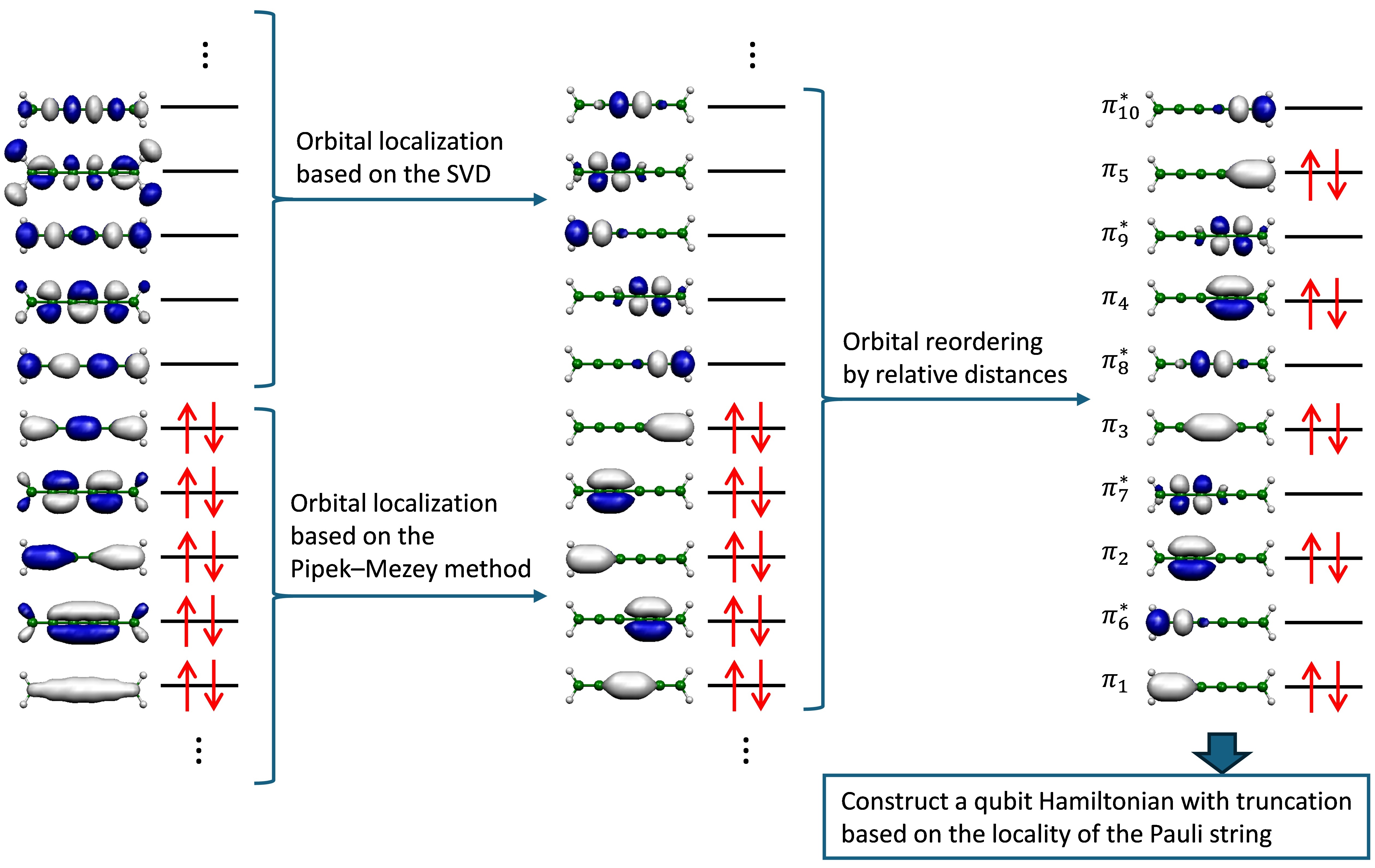}
    \caption{Procedure of orbital transformations for the Hamiltonian term truncations based on the locality of the Pauli strings, in the case of {\bf 5} in its planar geometry. Red arrows indicate the electron occupancies in the RHF wave function.}
    \label{fig:fig3}
\end{figure*}

As we mentioned in the previous section, the time evolution operation in the HSB-QSCI does not have to be exact, and we can introduce various approximations. In this work, we investigated the effect of Hamiltonian truncation on the sampled Slater determinants in the Hamiltonian simulation and on the QSCI energies. We investigated Hamiltonian truncation based on the locality of the qubit Hamiltonian terms. In this approach, we first generate the localized molecular orbitals (LMOs), and then rearrange the LMOs according to their relative spatial distances. During the qubit Hamiltonian construction using the reordered LMOs, we calculate the locality of the Pauli string (number of Pauli-X, Y, and Z operators in the Pauli string). If the locality of the Pauli string exceeds the threshold, then we exclude it from the qubit Hamiltonian. We investigated this method in {\bf 5}, keeping in mind to perform the Hamiltonian simulations on a superconducting quantum device with MPO-based classical compression of the time evolution quantum circuit. The procedure of this step is illustrated in Figure \ref{fig:fig3}. 

To sample the Slater determinants based on the real-time evolution of the initial wave function, we set the evolution time for $U$ to be $\Delta t = 1$ in atomic unit and $k = 1, 2, \dots, 10$, unless otherwise noted. To construct the quantum circuit for $U$, we adopt the first-order Trotter decomposition with a single Trotter slice. Magnitude ordering~\cite{Tranter-2018} is used for the Trotterized term ordering. The quantum circuit simulations were done using the \texttt{qsim} library~\cite{qsim}, which allows us to use GPGPU. The subspace Hamiltonian matrix construction and diagonalization was done using PyCI library~\cite{PyCI}, and the SCCR for hardware demonstrations was carried out using \texttt{qiskit-addon-sqd} library~\cite{qiskit-addon-sqd}. It should be noted that the subspace Hamiltonian diagonalization can also be done using \texttt{qiskit-addon-sqd}, but it uses a selected CI \texttt{kernel\_fixed\_space} subroutine in PySCF~\cite{PySCF} where the spin $\alpha$ and $\beta$ occupancies are stored separately, and the Slater determinants constructed from all possible combinations of spin $\alpha$ and $\beta$ strings are used as the basis for the wave function expansion. In this implementation, Slater determinants not sampled by the quantum computer can be included in the subspace Hamiltonian diagonalization, resulting in a quadratic increase in the dimension of the Hamiltonian matrix. This can be seen as a kind of subspace enlargement at the expense of compactness of the QSCI wave function. We also performed a subspace Hamiltonian diagonalization using PySCF, and the results are given in Supporting Information section D. 
Quantum circuit simulations and the subspace Hamiltonian diagonalization in the QSCI part were performed on the NVIDIA DGX H100 system.  

\section{Results and Discussion}

Subsections A, B, C, and D are devoted to show numerical simulations for the S$_0$ state of H$_2$O, S$_0$ and T$_1$ states of oligoacenes, phenylene-1,4-dinitrene, and hexa-1,2,3,4,5-pentaene, respectively. 
Subsection E gives a hardware demonstrations for the S$_0$ and T$_1$ states of carbyne molecules. 

\subsection{The S$_0$ state of the H$_2$O molecule}

\begin{figure}[ht]
    \centering
    \includegraphics[width=\linewidth]{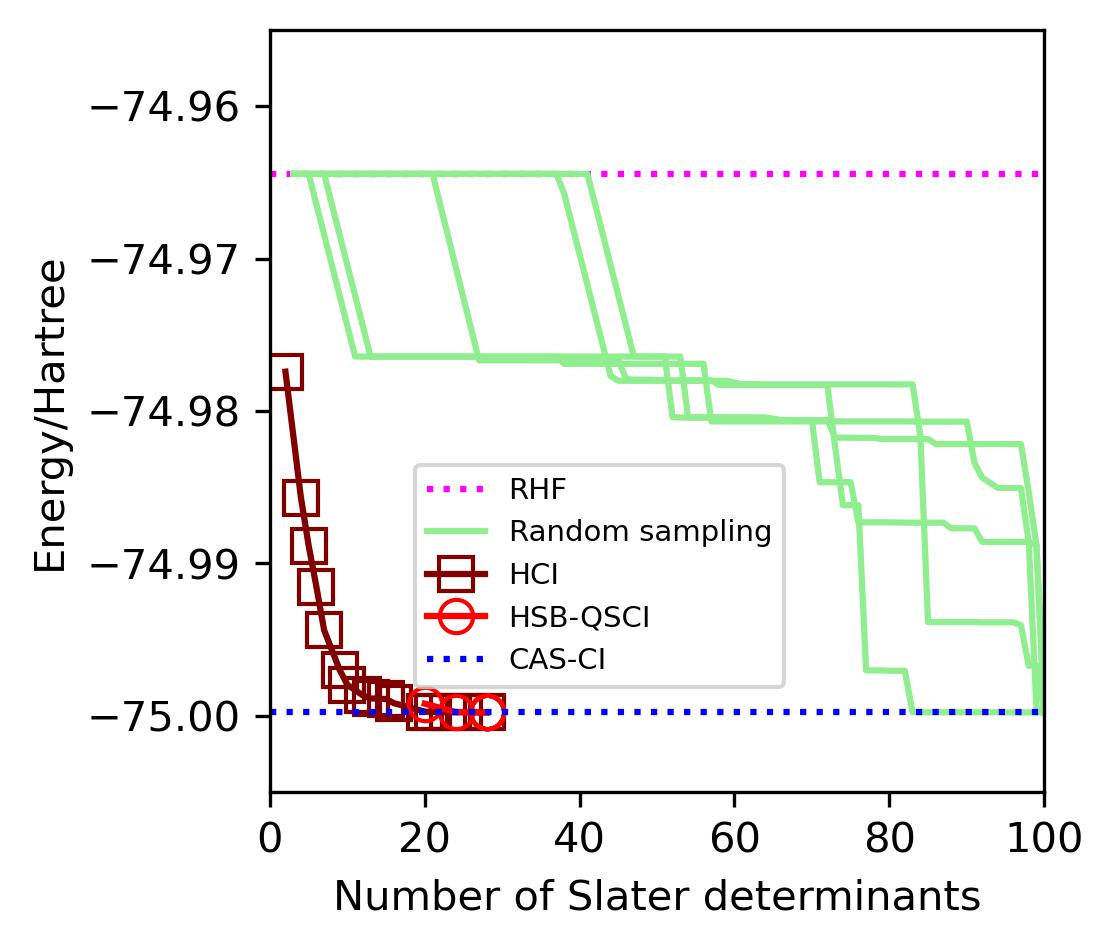}
    \caption{HSB-QSCI and random sampling-based QSCI results of the H$_2$O molecule.}
    \label{fig:newfig1}
\end{figure}

In this subsection, we discuss the performance of the sampling strategies for Slater determinants in the QSCI, using the H$_2$O molecule. We compare the HSB-QSCI, QSCI with randomly sampled Slater determinants that preserve spatial symmetry, $M_S$, and $n_{elec}$, VQE-based QSCI~\cite{Kanno-2023}, and HCI. HSB-QSCI calculations were performed with $k = 10$, and the number of shots per time step was set to $1 \times 10^4$. For the random sampling-based QSCI, we always included the HF configuration in the subspace Hamiltonian, and excited determinants belonging to the same irreducible representation of the S$_0$ state and having the correct $M_S$ and $n_{elec}$. The number of excited determinants satisfying these conditions is 99. We performed five independent simulations. The HCI is performed using the PyCI package~\cite{PyCI}. The results are summarized in Figure~\ref{fig:newfig1}. In HSB-QSCI with $k = 1$, the energy error is already less than $6 \times 10^{-4}$ Hartree (0.36 kcal mol$^{-1}$). Chemical precision is achieved with only one time evolution step, considering only 20 Slater determinants. This is in contrast to the VQE-based QSCI~\cite{Kanno-2023}, which requires about 50 VQE iterations and 32 Slater determinants to achieve an energy error below $1 \times 10^{-3}$ Hartree. Note that they also reported that 16 Slater determinants are enough to achieve chemical precision, after VQE convergence. 
It is worth noting that the probability of sampling the HF configuration in the first time step of HSB-QSCI is approximately 0.942. In contrast, in the perfect state preparation limit of VQE-based sampling, this probability is calculated to be 0.978. This indicates that HSB-QSCI has a higher chance of sampling excited determinants than the VQE-based approach. Of course, the quantum circuit is deeper in HSB-QSCI than in the VQE-based one. However, improvements in the overall number of required shots and the compactness of the QSCI wave function are significant. HCI yields slightly lower energy than HSB-QSCI with 20 Slater determinants, $\Delta E_{\text{HCI}-\text{CAS-CI}} = 5 \times 10^{-5}$ Hartree = 0.032 kcal mol$^{-1}$. However, as the number of Slater determinants increases, the difference in energy errors between HSB-QSCI and HCI diminishes, and the two methods
give identical energies when 28 Slater determinants are used. It is important to note that HSB-QSCI yields energies comparable to HCI, although starting from an approximate wave function in HSB-QSCI may lead to sampling Slater
determinants that are more relevant to excited states than to the ground state. The comparison with the random sampling-based implementation is even more illuminating. In this case, the energy error remains large even when 80\% of the Slater determinants are included in the subspace Hamiltonian diagonalization.  

\subsection{The S$_0$ and T$_1$ states of oligoacenes}

\begin{figure*}[ht]
    \centering
    \includegraphics[width=\textwidth]{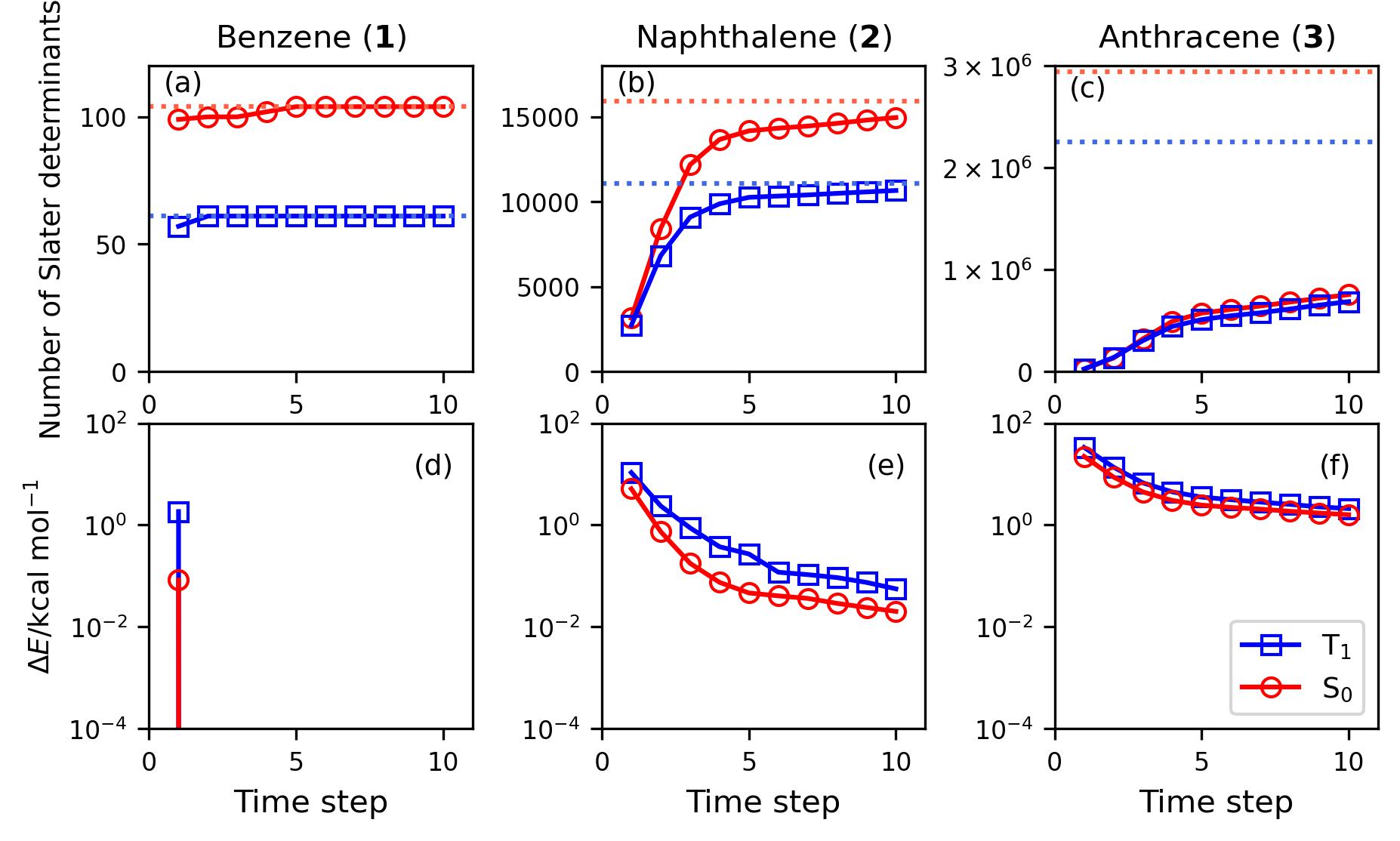}
    \caption{HSB-QSCI results of oligoacenes. The number of Slater determinants included in the Hamiltonian diagonalization is given in (a), (b), and (c) for {\bf 1}, {\bf 2}, and {\bf 3}, respectively. Red and blue indicate the S$_0$ and T$_1$ states, respectively. Dotted lines represent the number of Slater determinants in the CAS-CI wave function. The difference of the HSB-QSCI energy from the CAS-CI values in units of kcal mol$^{-1}$ are given in (d), (e), and (f) on the logarithmic scale.}
    \label{fig:fig4}
\end{figure*}

There is no doubt that aromatic rings are one of the most important molecular skeletons in chemistry, and the study of the electronic structures of oligoacenes is very important. It is known that the zigzag edge of graphene fragments exhibits strong open-shell characters~\cite{Zeng-2021}, and the contribution of the HF electronic configuration to the full-CI wave function decreases for larger oligoacenes. Electronic excited states of oligoacenes are also of interest because of their potential for various applications such as singlet fission photovoltaics~\cite{Baldacchino-2022}. 

The results of the HSB-QSCI simulations of oligoacenes are shown in Figure~\ref{fig:fig4}. The RHF and ROHF-like single configurational wave functions are used as the starting wave functions for the Hamiltonian simulations of the S$_0$ and T$_1$ states, respectively, and the number of shots for each time step was set as $1 \times 10^5$. The number of Slater determinants in the CAS-CI wave function is 104 (S$_0$) and 61 (T$_1$) for {\bf 1}, 15912 (S$_0$) and 11076 (T$_1$) for {\bf 2}, and 2945056 (S$_0$) and 2255121 (T$_1$) for {\bf 3}, in the $D_{2h}$ point group. 
Our numerical simulations revealed that the Slater determinants that are important to describe the target electronic states are efficiently sampled from the Hamiltonian simulations. In {\bf 1}, the system size is very small, and all possible Slater determinants are sampled up to the fifth steps. In {\bf 2}, chemical precision ($\Delta E < 1.0\ \mathrm{kcal\ mol^{-1}}$) was achieved with three steps of the Hamiltonian simulations. In {\bf 3}, we found that 10 steps of time evolution with $1 \times 10^5$ shots are not enough to achieve chemical precision. We expect that the number of important Slater determinants increases with $\mathcal{O}(M^4)$, which is the same as the scaling of Hamiltonian terms, when the target state wave function is well approximated by the single determinant as in the ground state of oligoacenes and the energy gap between the ground and the first excited states is unchanged. Considering the fact that the energy gap is smaller in {\bf 3} than in {\bf 2}, we need more than 3.81 times more shots than in the calculation of {\bf 2} to obtain the HSB-QSCI energy of {\bf 3} with the same accuracy. We have also studied the dependence of the HSB-QSCI energy on the number of shots and the number of time steps in {\bf 3}. By increasing the number of shots from $1 \times 10^5$ and simulating more time steps, the number of Slater determinants sampled from the Hamiltonian simulation increases, and the HSB-QSCI energy decreases systematically (see Figure S14 in Supporting Information). Although we need more steps and shots to achieve chemical precision for larger systems, it should be noted that up to 7-electron excited determinants were sampled in a Hamiltonian simulation with $k\Delta t = 1$ in {\bf 3}. It is known that higher-order excitations are important to compute the energy in a quantitative manner, especially in strongly correlated systems~\cite{Sharma-2017}. The ability of Hamiltonian simulations to capture higher-order excitations is remarkable and is an important feature for applications to large and strongly correlated systems. These results illustrate the ability of the Hamiltonian simulation to sample a polynomial number of important Slater determinants from exponentially large Hilbert space. Note that similar trend has also been reported by other groups~\cite{Mikkelsen-2024, Yu-2025}.
When the \texttt{kernel\_fixed\_space} subroutine in PySCF is used for subspace Hamiltonian diagonalization, chemical precision was achieved with only three steps of time evolutions with $1 \times 10^5$ shots in {\bf 3}. 
However, the HSB-QSCI wave function considers about 57\% and 46\% of the Slater determinants, hence the HSB-QSCI wave function is less compact (see Figure S11 in Supporting Information). 

\subsection{The S$_0$ and T$_1$ states of phenylene-1,4-dinitrene}

\begin{figure}[ht]
    \centering
    \includegraphics[width=\linewidth]{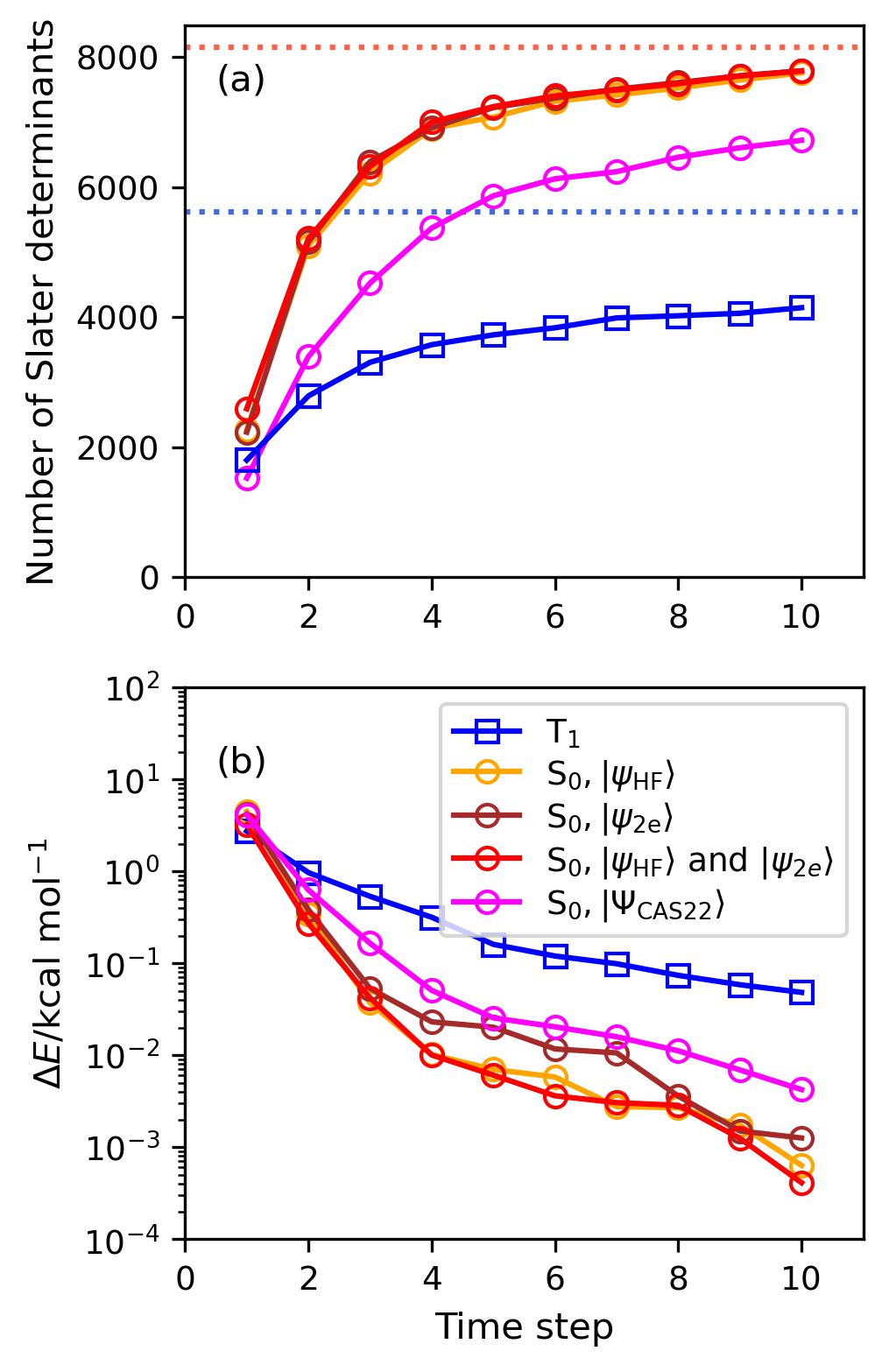}
    \caption{HSB-QSCI results of {\bf 4}. (a) The number of Slater determinants sampled from Hamiltonian simulations. The red and blue dotted lines indicate the number of Slater determinants in the CAS-CI wave function. (b) The difference of the HSB-QSCI energy from the CAS-CI values in units of kcal mol$^{-1}$, on the logarithmic scale. }
    \label{fig:fig5}
\end{figure}

{\bf 4} has two major resonance structures, the diradical with a quinonoid skeleton as shown in Figure~\ref{fig:fig2}, and the dinitrene with an aromatic ring. Its electronic ground state is an open-shell spin-singlet state with two unpaired electrons in in-plane 2p orbitals of the nitrogen atoms. The thermally excited spin-triplet state of {\bf 4} has been observed by ESR spectroscopy~\cite{Singh-1971, Nicolaides-1998, Nimura-2006}, and the singlet--triplet energy gap is determined to be $-$0.82 kcal mol$^{-1}$ from the Curie analysis of the ESR signal~\cite{Ichimura-1995}. The experimentally determined zero-field splitting parameter in the excited spin-triplet state is $|D|$ = 0.171 cm$^{-1}$~\cite{Singh-1971}, which is about 10 times larger than the value estimated from a point-dipole approximation. This deviation was explained by the lack of electron correlation effect in the point-dipole approximation~\cite{Sugisaki-2013}, insisting that sophisticated consideration of electronic correlation is essential to describe its electronic structure. 

In the CASSCF optimized orbital basis, the CAS-CI wave function of the S$_0$ state is mainly described by the HF configuration ($\ket{\psi_{\mathrm{HF}}}$) and the HOMO--LUMO two-electron excited configuration from the HF configuration ($\ket{\psi_{\mathrm{2e}}}$)
(see Table S17 in Supporting Information). In the HSB-QSCI simulations of the S$_0$ state, we performed Hamiltonian simulations with three different initial wave functions, $\ket{\psi_{\mathrm{HF}}}$, $\ket{\psi_{2e}}$, and the CAS-CI(2e,2o) wave function $\ket{\Psi_{\mathrm{CAS22}}} = 0.7138 \ket{\psi_{\mathrm{HF}}} - 0.7003 \ket{\psi_{\mathrm{2e}}}$, to investigate the dependence of the initial wave function on the energies and the convergence behavior of the HSB-QSCI method. The QSCI simulations were performed under four different conditions: (1) use the $\ket{\psi_{\mathrm{HF}}}$ results with $1 \times 10^5$ shots for each time step, (2) use the $\ket{\psi_{\mathrm{2e}}}$ with $1 \times 10^5$ shots, (3) use the samples from $\ket{\Psi_{\mathrm{CAS22}}}$ with $1 \times 10^5$ shots, and (4) merge the samples from $\ket{\psi_{\mathrm{HF}}}$ and $\ket{\psi_{\mathrm{2e}}}$ with $5 \times 10^4$ shots for each step. For the T$_1$ state, we used the ROHF-like single determinant as the initial wave function for the Hamiltonian simulation. The HSB-QSCI results are summarized in Figure~\ref{fig:fig5}. Our numerical simulations suggest that almost the same number of Slater determinants are sampled when $\ket{\psi_{\mathrm{HF}}}$ and $\ket{\psi_{\mathrm{2e}}}$ are used. Interestingly, the number of Slater determinants sampled from the Hamiltonian simulation is smaller when $\ket{\Psi_{\mathrm{CAS22}}}$ is used as the initial wave function. The accuracy of the HSB-QSCI energy with $\ket{\Psi_{\mathrm{CAS22}}}$ is about an order of magnitude worse than the HF-reference one, but still chemical precision was achieved in 2 steps of time evolution. 
Thus, using the CAS-CI wave function with a smaller active space can help generate a more compact HSB-QSCI wave function for strongly correlated systems. However, preparing the CAS-CI reference state itself may become a bottleneck when the number of quasi-degenerate orbitals is large. One possible solution to this challenge is to adopt a technique for constructing a multiconfigurational wave function without performing post-HF calculations~\cite{Sugisaki-2019}.
The singlet--triplet energy gap is calculated to be $-$1.11 kcal mol$^{-1}$, in good agreement with the CAS-CI value ($-$1.06 kcal mol$^{-1}$) and the experimental one~\cite{Ichimura-1995}. Using PySCF for the subspace Hamiltonian diagonalization slightly increases the number of Slater determinants and yields lower HSB-QSCI energies. The trends of the initial wave function dependence are similar between PyCI- and PySCF-based implementations (see Figure S12 in the Supporting Information).

\subsection{The S$_0$ and T$_1$ states of hexa-1,2,3,4,5-pentaene}
Our last example of numerical simulations is the S$_0$ and T$_1$ states of {\bf 5}. From the DFT calculations, the carbyne with =CH$_2$ termination is predicted to have a helical $\pi$ conjugation in the non-planar geometries~\cite{Liu-2013, Liu-2013-correction}. In the S$_0$ state, the planar structure is the energy minimum, but in the T$_1$ state, the planar geometry is a saddle point and the twisted geometry becomes stable. As a result, the singlet--triplet energy gap strongly depends on the dihedral angle between two =CH$_2$ terminations. In this study, we calculated the S$_0$ and T$_1$ states of the geometries with the dihedral angles of $0^\circ$ and $90^\circ$. 

Because carbyne is a one-dimensional molecule, using LMOs and adopting Hamiltonian truncation by maximum locality may be a good option to reduce the computational cost of Hamiltonian simulation. In this work, we investigated Hamiltonian term truncation by operator locality of the Pauli string in the qubit Hamiltonian. By reordering the LMOs by relative distances before constructing a qubit Hamiltonian, operator locality-based truncation is roughly equivalent to Hamiltonian term truncation based on spatial distances. To assess the effect of Hamiltonian truncation with the locality, we first calculated the number of Hamiltonian terms in the second quantized Hamiltonian and the fidelity of the ground state wave function, $|\mathrm{\braket{\Psi_0(truncated)|\Psi_0(full)}}|^2$, where $\ket{\Psi_0(\mathrm{truncated})}$ and $\ket{\Psi_0(\mathrm{full})}$ are the ground state wave functions with the truncated and untruncated Hamiltonians, respectively. The results are summarized in Tables~\ref{tab:table1} and \ref{tab:table2} for planar and twisted geometries, respectively. 

\begin{table}[ht]
\caption{\label{tab:table1} Number of terms in the second quantized Hamiltonian and fidelity of the electronic ground state wave function of {\bf 5} in planar geometry. }
\begin{tabular*}{8.5cm}{@{\extracolsep{\fill}}ccc}
\hline
Maximum locality $m$ & \#Terms & Fidelity \\
\hline
20 & 20072 & 1.0 \\
18 & 19676 & 0.999816 \\
16 & 18596 & 0.998764 \\
14 & 17592 & 0.998571 \\
12 & 16248 & 0.996394 \\
10 & 12708 & 0.992327 \\
8  & 7836  & 0.946569 \\
6  & 4840  & 0.941506 \\
4  & 3048  & 0.912734 \\
\hline
\end{tabular*}
\end{table}

\begin{table}[ht]
\caption{\label{tab:table2} Number of terms in the second quantized Hamiltonian and fidelity of the electronic ground state wave function of {\bf 5} in twisted geometry. }
\begin{tabular*}{8.5cm}{@{\extracolsep{\fill}}ccc}
\hline
Maximum locality $m$ & \#Terms & Fidelity \\
\hline
20 & 40200 & 1.0 \\
18 & 39772 & 0.999576 \\
16 & 38244 & 0.997395 \\
14 & 35232 & 0.996860 \\
12 & 30640 & 0.995021 \\
10 & 24660 & 0.987597 \\
8  & 17772 & 0.898829 \\
6  & 10744 & 0.891525 \\
4  & 4632  & 0.843284 \\
\hline
\end{tabular*}
\end{table}

\begin{figure*}[ht]
    \centering
    \includegraphics[width=\textwidth]{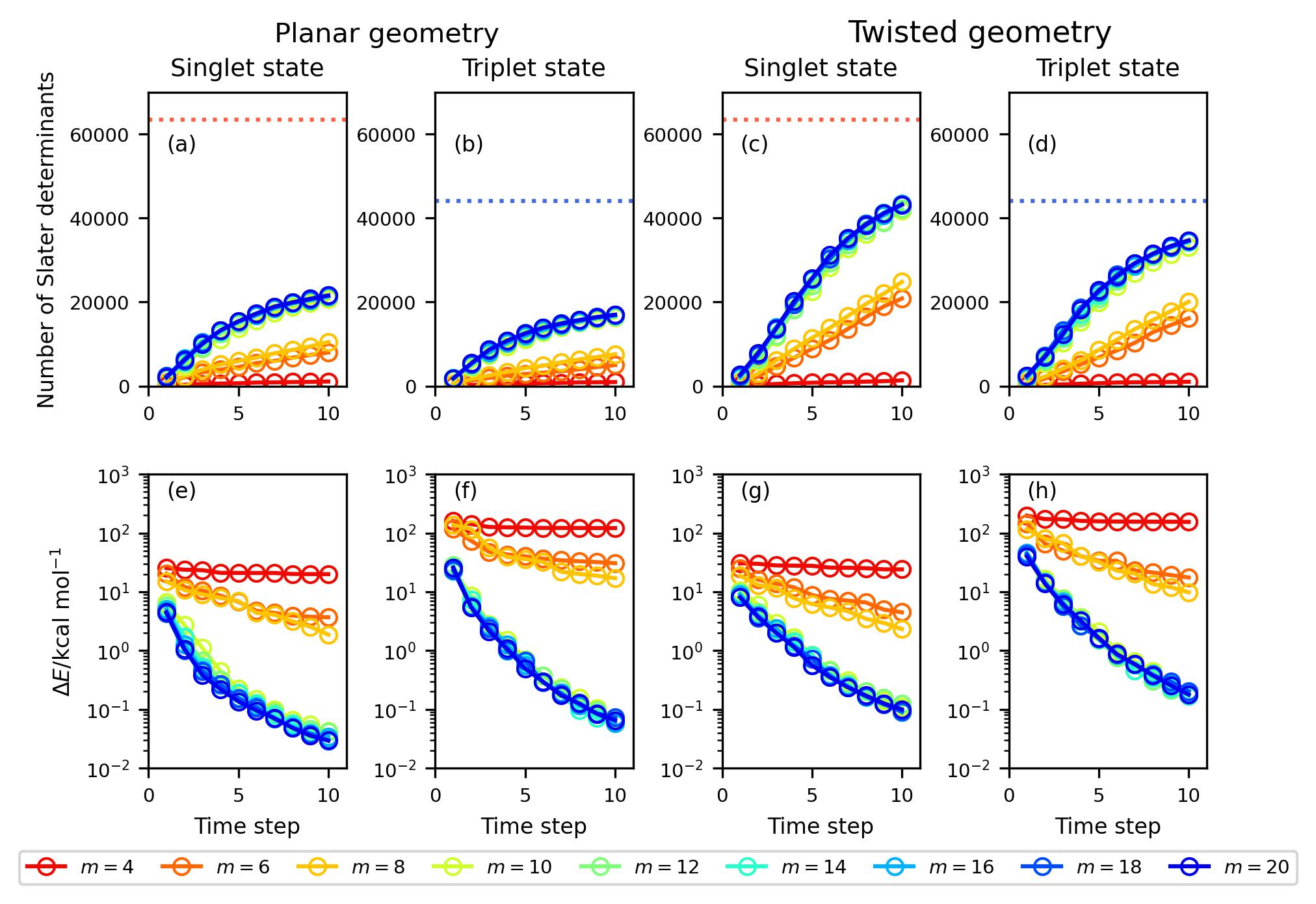}
    \caption{HSB-QSCI results of {\bf 5} in its planar and twisted geometries with different maximum locality values $m$. The number of Slater determinants included in the Hamiltonian diagonalization is given in (a), (b), (c), and (d). The red and blue dotted lines represent the number of Slater determinants in the CAS-CI wave function of the S$_0$ and T$_1$ states, respectively. The difference of the HSB-QSCI energy from the CAS-CI values in units of kcal mol$^{-1}$ are given in (e), (f), (g), and (h) on the logarithmic scale.}
    \label{fig:fig6}
\end{figure*}

Note that the number of Hamiltonian terms is larger in the twisted geometry than in the planar one, which is due to the difference in the spatial distribution of $\pi$ orbitals. In planar geometry, the one-electron MO integrals between the nearest neighbor $\pi$ orbitals (e.g., $\pi_1$ and $\pi_2$ in Figure \ref{fig:fig3}) are zero, while in twisted geometry, they are non-zero due to helical $\pi$ conjugations. As can be seen from Tables \ref{tab:table1} and \ref{tab:table2}, the number of Hamiltonian terms is reduced by about 40\% when considering up to 10-local terms, with the fidelity of the ground state wave function being about 0.99. The rapid decrease of the fidelity for the maximum locality smaller than 10 can be explained by the fact that the $\pi_x$--$\pi_x$ and $\pi_y$--$\pi_y$ interactions (here we assumed that one-dimensional carbon atom chain is parallel to the z-axis) with the next nearest neighbor C=C bond (for example, $\pi_1$--$\pi_3$ in Figure \ref{fig:fig3}) are described as 10-local operators in the qubit Hamiltonian. 

The results of the HSB-QSCI simulations are shown in Figure~\ref{fig:fig6}. The number of shots for each time step is $1 \times 10^5$. In the T$_1$ state calculations, we set the starting wave function to carry two unpaired electrons in the central $\pi$ bonds ($\pi_3$ and $\pi_8^*$ in Figure~\ref{fig:fig3}). It is clear that the number of Slater determinants sampled from the Hamiltonian simulations and the convergence behavior of the HSB-QSCI energies are almost the same for $m \geq 10$, where $m$ is the maximum locality of the qubit Hamiltonian terms. These results are consistent with the trend of the S$_0$ state fidelity value. The singlet--triplet energy gap $\Delta E_{\mathrm{S-T}} = E_{\mathrm{S}}-E_\mathrm{T}$ calculated using the real-time evolution with the untruncated Hamiltonian ($m = 20$) is $-53.96$ and $-20.33$ kcal mol$^{-1}$ for planar and twisted geometries, respectively, and the difference from the CAS-CI singlet--triplet energy gap is less than 0.1 kcal mol$^{-1}$ ($\Delta E\mathrm{_{S-T}(CAS\mathchar`-CI)}$ = $-53.93$ and $-20.26$ kcal mol$^{-1}$ for planar and twisted geometry, respectively). The number of sampled Slater determinants is larger for twisted geometry than for planar geometry, reflecting the spatial distribution of the $\pi$ orbitals, as discussed above.
When the maximum locality is set to be $m$ = 6 or 8, the error in the HSB-QSCI energy becomes larger. However, our numerical simulations revealed that chemical precision can be achieved even with the maximum locality $m = 6$, by increasing the number of shots to more than $3\times 10^6$ and setting the number of maximum time steps to $k = 20$ (see Figure S15 in Supporting Information). In this case, we observed an abrupt improvement in energy convergence. Since we did not observe such behavior when the untruncated Hamiltonian was used for the time evolution, we suspect that Hamiltonian truncation is responsible for this. When a Hamiltonian truncated to maximum locality is used for time evolution, the probability of measuring excited determinants with long-range excitations becomes smaller, because long-range interaction terms are discarded by the truncation. Consequently, the use of the truncated Hamiltonian in the time evolution operator causes an imbalance in the description of short-range and long-range interactions, which is likely responsible for the observed behavior.

Note that when using the PySCF \texttt{kernel\_fixed\_space} subroutine for subspace Hamiltonian diagonalization, chemical precision was achieved with up to five steps of time evolution with $10^5$ shots (see Figure S14 in Supporting Information), but this is a consequence of the quadratic increase of the Slater determinants in PySCF.

It should be emphasized that the ROHF-like single configurational wave function in the LMO basis used in the T$_1$ state calculation has a rather small overlap with the CAS-CI wave function. The squared overlap values $|\braket{\Phi_0|\Psi_{\mathrm{CAS\mathchar`-CI}}}|^2$ in the T$_1$ state are 0.1691 and 0.1240 for planar and twisted geometries, respectively (see Tables S16 and S18 in Supporting Information for the CAS-CI wave function). The fact that the HSB-QSCI provides accurate energy with such a small overlap is promising because the overlap between the HF and the full-CI wave functions decreases with system size, and the preparation of a sophisticated approximate wave function becomes challenging for larger molecules.  

\subsection{Hardware demonstration of the HSB-QSCI calculations of the S$_0$ and T$_1$ states of carbyne molecules}

\begin{figure*}[ht]
    \centering
    \includegraphics[width=\linewidth]{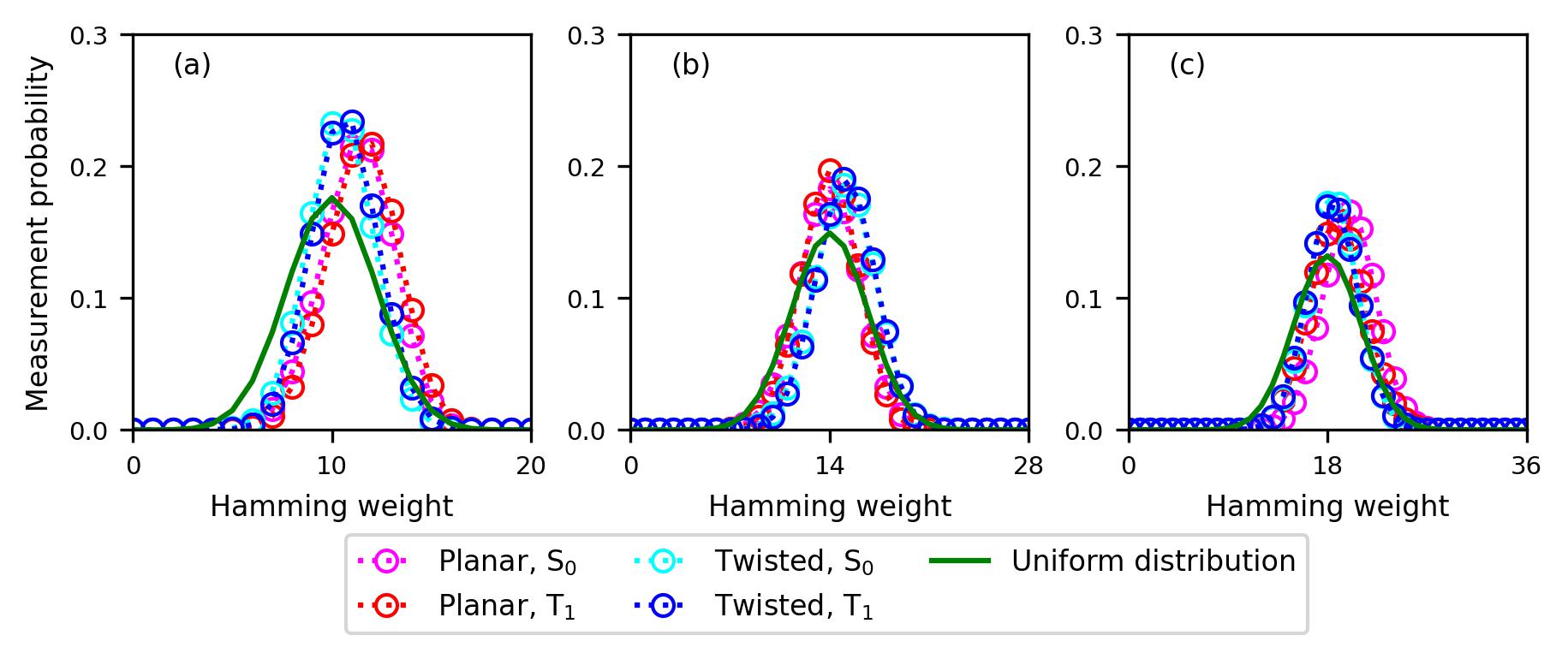}
    \caption{Hamming weight distribution of the bit strings obtained from measurements after Hamiltonian simulation of carbyne molecules with $k\Delta t = 10$, using \texttt{ibm\_kawasaki}. (a) {\bf 5}, (b) {\bf 6}, and (c) {\bf 7}}
    \label{fig:fig7}
\end{figure*}

\begin{figure*}[ht]
    \centering
    \includegraphics[width=\textwidth]{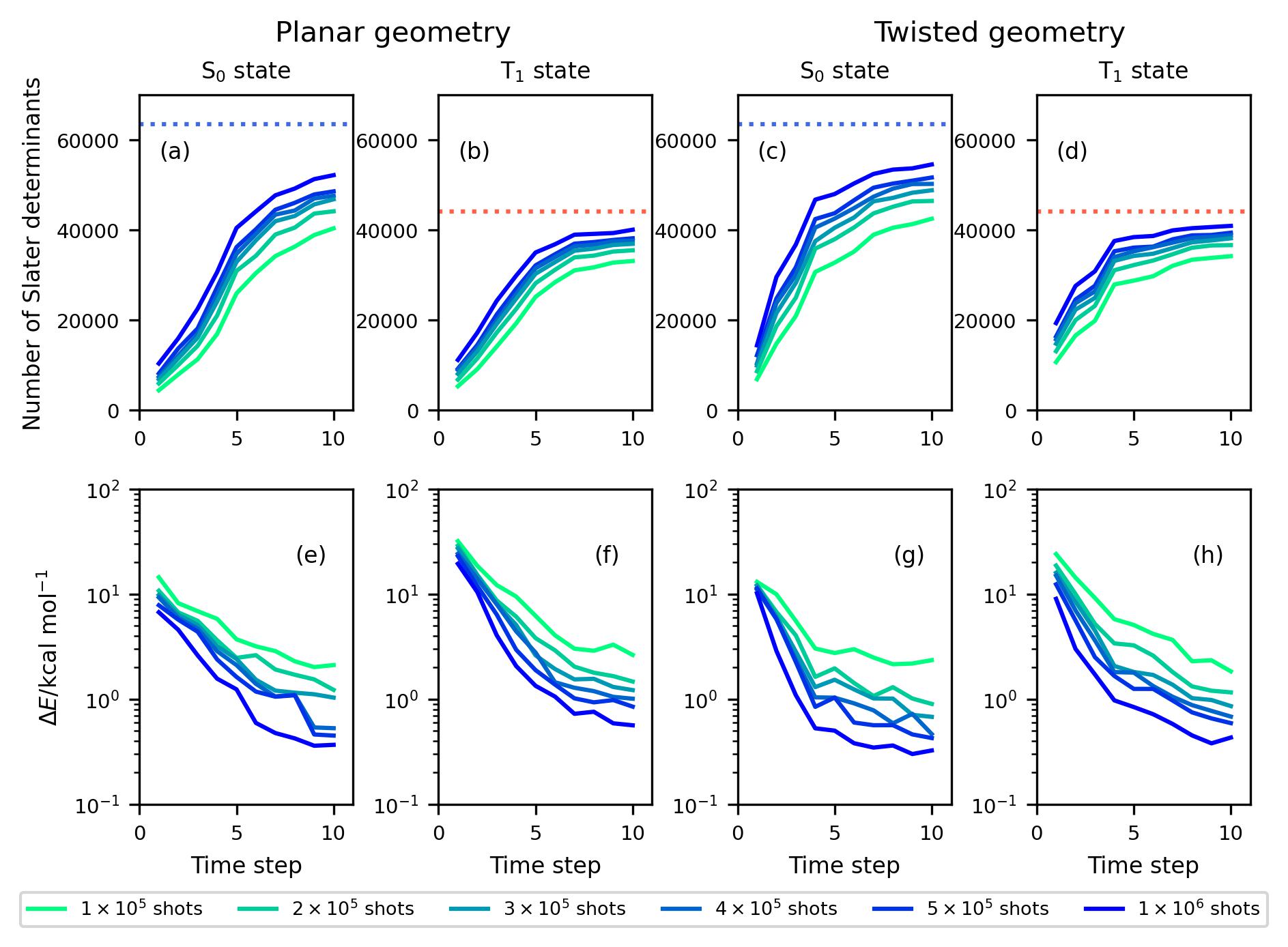}
    \caption{HSB-QSCI results of {\bf 5} obtained with a Hamiltonian truncated by the maximum locality $m$ = 10 and \texttt{ibm\_kawasaki}, with SCCR. 
    The number of Slater determinants included in the Hamiltonian diagonalization are given in (a), (b), (c), and (d), and the difference of the HSB-QSCI energy from the CAS-CI values in units of kcal mol$^{-1}$ are given in (e), (f), (g), and (h) on the logarithmic scale.}
    \label{fig:fig8}
\end{figure*}

Since the Hamiltonian term truncation based on the locality of the qubit Hamiltonian terms works excellently in {\bf 5}, we performed the HSB-QSCI calculations of {\bf 5}, {\bf 6}, and {\bf 7} on an IBM Quantum Eagle processor, namely \texttt{ibm\_kawasaki}. To reduce the quantum circuit depth of the Hamiltonian simulation, we performed the MPO-based classical compression of the quantum circuit. This approach was demonstrated in our previous study on the quantum phase difference estimation-based energy gap calculations~\cite{Kanno-2024}. The details of the MPO-based quantum circuit optimization are given in the reference~\cite{Kanno-2024}. In this work, we used a Hamiltonian truncated by the maximum locality ($m$ = 10), and the 10-layer brick wall type quantum circuit is generated for the time evolution operator of $\Delta t = 1$ in atomic unit. The number of sweeps on the optimizations was set to 10000. In constructing the MPOs, the second-order Trotter decomposition with a single Trotter slice was employed, and the singular value cutoff for the SVD was set to $10^{-4}$. The order of terms in the Trotter decomposition is set as follows. First, we generate the truncated second quantized Hamiltonian by checking the one- and two-electron integrals in the lexicographic order: $q$ after $p$ in $h_{pq}$, and $p$, $r$, $q$, $s$, in that order in $g_{pqrs}$. Then the truncated second-quantized Hamiltonian is transformed into a qubit Hamiltonian while retaining the order of the terms, and the Trotterized time evolution operator is constructed subsequently. 

We first checked the Hamming weight distribution of the bit strings obtained in the measurements and compared it to the uniform distribution. The Hamming weight distributions at $k\Delta t = 10$ are plotted in Figure~\ref{fig:fig7}. We confirmed that the Hamming weight distributions of the bit strings obtained from the Hamiltonian simulation with $k\Delta t = 10$ are significantly different from the uniform distribution. However, due to hardware noise, only 11--23\% of the measured bit strings have correct Hamming weights (10, 14, and 18 for {\bf 5}, {\bf 6}, and {\bf 7}, respectively). 

The HSB-QSCI results of {\bf 5} are summarized in Figure \ref{fig:fig8}. Here, the Hamiltonian simulations were carried out with $k = 10$, and the number of shots was set from $1 \times 10^5$ to $1 \times 10^6$. The depth of the quantum circuit after transpilation with optimization level = 3 in qiskit is 21, regardless of the evolution time steps. The SCCR~\cite{Robledo-2024} was adopted to recover configurations with a correct number of electrons, before constructing the subspace Hamiltonian matrix. 
In the first five to eight steps of time evolution, the number of Slater determinants sampled increases rapidly, but thereafter the number of additionally sampled Slater determinants becomes gradual. When the HSB-QSCI is performed up to ten steps of time evolution and $1 \times 10^5$ shots for each time step, the energy error is about 2 kcal mol$^{-1}$. Increasing the number of shots further improves the energy, and the error in the total energy with $1 \times 10^6$ shots were 0.37, 0.56, 0.33, and 0.43 kcal mol$^{-1}$ for planar (S$_0$), planar (T$_1$), twisted (S$_0$), and twisted (T$_1$), respectively, achieving chemical precision. 

\begin{table*}[ht]
\caption{\label{tab:table3} The HF energy, number of Slater determinants and total energies of the CAS-CI, and the number of selected Slater determinants and the error of the HSB-QSCI energy calculated using \texttt{ibm\_kawasaki}.}
\begin{tabular*}{18.5cm}{@{\extracolsep{\fill}}ccccccc}
\hline
\multirow{2}{*}{System} & HF          & \multicolumn{2}{c}{CAS-CI} & \multicolumn{3}{c}{HSB-QSCI} \\
                        & $E$/Hartree & \#Dets  &  $E$/Hartree     & \%\#Dets$^a$ & $\Delta E/\mathrm{kcal\ mol^{-1}}$ & \%$E_\text{Corr}$$^{b}$ \\
\hline
{\bf 5} (Planar, S$_0$)  & $-226.4832222338$ &      63504 & $-229.5024030030$ & 82.21 &  0.37 & 99.98 \\
{\bf 5} (Planar, T$_1$)  & $-226.4299826264$ &      44100 & $-229.4164659619$ & 90.91 &  0.56 & 99.97 \\
{\bf 5} (Twisted, S$_0$) & $-226.4033087159$ &      63504 & $-229.4452184415$ & 85.96 &  0.33 & 99.98 \\
{\bf 5} (Twisted, T$_1$) & $-226.4440971899$ &      44100 & $-229.4129399110$ & 92.84 &  0.43 & 99.99 \\
{\bf 6} (Planar, S$_0$)  & $-301.1974823957$ &   11778624 & $-305.2359857689$ & 52.20 &  2.00 & 99.92 \\
{\bf 6} (Planar, T$_1$)  & $-301.1536301615$ &    9018009 & $-305.1666727092$ & 63.80 &  2.93 & 99.88 \\
{\bf 6} (Twisted, S$_0$) & $-301.1365655879$ &   11778624 & $-305.1941677868$ & 49.57 &  2.71 & 99.89 \\
{\bf 6} (Twisted, T$_1$) & $-301.1634490372$ &    9018009 & $-305.1698864405$ & 60.33 &  2.84 & 99.89 \\
{\bf 7} (Planar, S$_0$)  & $-375.9128437860$ & 2363904400 & $-380.9720622554$ &  0.95 & 14.26 & 99.55 \\
{\bf 7} (Planar, T$_1$)  & $-375.8222817220$ & 1914762564 & $-380.9133659533$ &  0.77 & 23.36 & 99.27 \\
{\bf 7} (Twisted, S$_0$) & $-375.8635891161$ & 2363904400 & $-380.9394312061$ &  1.09 & 13.97 & 99.56 \\
{\bf 7} (Twisted, T$_1$) & $-375.8817559215$ & 1914762564 & $-380.9186151034$ &  0.87 & 25.79 & 99.18 \\
\hline
\end{tabular*}
$^a$ Percentage of the number of Slater determinants calculated as 100 $\times \frac{\#Dets(\text{HSB-QSCI})}{\#Dets(\text{CAS-CI})}.$ \\
$^b$ Percentage of the correlation energy in the active space considered in the HSB-QSCI, calculated as 100 $\times$ $\frac{E(\mathrm{HF}) - E(\text{HSB-QSCI})}{E(\mathrm{HF})-E(\text{CAS-CI})}$.
\end{table*}

The dimensions of the Hamiltonian matrix and the total energies of the HSB-QSCI and the CAS-CI methods are summarized in Table \ref{tab:table3}. We used $k = 10$ and performed $3\times 10^6$ shots for each time step in the HSB-QSCI for {\bf 6} and {\bf 7}. In {\bf 6}, the HSB-QSCI considered about 50--64\% of the entire Slater determinant, and it calculated the total energy with an error of about 2.0--2.9 kcal mol$^{-1}$. Although chemical precision for the total energy was not achieved, the HSB-QSCI provides the singlet--triplet energy gap with an error of 0.93 and 0.12 kcal mol$^{-1}$ for planar and twisted geometries, respectively, of {\bf 6}. The energy difference between the planar and twisted geometries is also calculated with an error of about 0.72 and 0.09 kcal mol$^{-1}$ for the S$_0$ and T$_1$ states, respectively. These results exemplify the ability of the HSB-QSCI to accurately calculate energy differences between two electronic states and two geometries. We expect that the total energy can be further improved by increasing the number of shots in the Hamiltonian simulation. 
In {\bf 7}, the error of the HSB-QSCI energy is about 14 kcal mol$^{-1}$ for the S$_0$ state, and it increases to 23.36 and 25.79 kcal mol$^{-1}$ for planar and twisted geometries, respectively, of the T$_1$ state. Again, chemical precision for the total energy was not achieved in {\bf 7}. In addition, the HSB-QSCI overestimated the singlet--triplet energy gap about 9--12 kcal mol$^{-1}$. As we pointed out above, the CAS-CI wave function of the T$_1$ state in the localized orbital basis shows an inherent multiconfigurational character, and the single Slater determinant used as the initial wave function of the time evolution has small overlap with the CAS-CI wave function. As a result, it is expected that the number of time steps or the number of shots will have to be increased to account for more Slater determinants in order to accurately describe the electronic structure of the T$_1$ state. Nevertheless, it is worth noting that the HSB-QSCI is able to compute the energy difference between the planar and twisted geometries of the S$_0$ state with chemical precision. Note that when PySCF is used for the subspace Hamiltonian diagonalization, the HSB-QSCI provided the total energy of the S$_0$ state with about 0.3 kcal mol$^{-1}$ of an error, by considering 46--48\% of the Slater determinants, but it failed to compute the total energy of the T$_1$ state with chemical precision. See Table S23 in Supporting Information for details.

Table~\ref{tab:table3} also summarizes the percentages of the number of Slater determinants and the correlation energies covered by the HSB-QSCI method in {\bf 5}--{\bf 7}. Here, we calculated the correlation energies as the energy difference between HF (RHF and ROHF for the S$_0$ and T$_1$ state, respectively, in the canonical orbital basis) and the HSB-QSCI energies. It is worth emphasizing that the HSB-QSCI covers more than 99.18\% of the correlation energies in the active space in {\bf 7}, by considering only ca. 1\% of all the Slater determinants.

\section{Conclusion}
In this study, we proposed an HSB-QSCI, which uses a quantum computer to perform Hamiltonian simulation and subsequent measurement of the quantum state in the computational basis to sample the important Slater determinants used to span the subspace Hamiltonian to be diagonalized on a classical computer. 
Compared to the reported VQE-based state preparation approaches for the QSCI~\cite{Kanno-2023, Nakagawa-2023, Nutzel-2024}, the HSB-QSCI is free from variational optimizations of the quantum states, and it works with the simple initial wave function like the HF. 
Since long-time evolution is in general difficult to simulate on a classical computer, we expect this approach to become powerful as the system size increases. The fact that higher-order excitations can be incorporated even in a short time evolution also makes HSB-QSCI suitable for simulation of large and strongly correlated systems. Proof-of-concept simulations were performed for the S$_0$ and T$_1$ states of organic molecules {\bf 1}--{\bf 5}, and we achieved chemical precision in all molecules studied. The quantum hardware demonstrations of the HSB-QSCI for the S$_0$ and T$_1$ states of three carbyne molecules {\bf 5}--{\bf 7} were also reported using \texttt{ibm\_kawasaki} with 20, 28, and 36 qubits, respectively, with the aid of MPO-based classical optimization of the quantum circuit for the time evolution operator. The HSB-QSCI achieved chemical precision for the total energy in {\bf 5}, for the singlet--triplet energy difference in {\bf 6}, and for the energy difference between planar and twisted geometries of the S$_0$ state in {\bf 7}. In {\bf 7}, by considering about 1\% of the Slater determinants, the HSB-QSCI captured more than 99.18\% of correlation energies in the active space. 
The largest system studied in this work is {\bf 7} with 36 qubits, which is the largest system capable of performing the CAS-CI calculation in a typical computational environment without a supercomputer~\cite{PySCF, Gao-2024}. We expect that the HSB-QSCI can handle the system larger than 36 qubits, by introducing batch-based implementations in the Hamiltonian matrix diagonalization part, and calculating the variance of the energy expectation value $\Delta H = \braket{\Psi|H^2|\Psi} - \braket{\Psi|H|\Psi}^2$ as explored in Reference ~\cite{Robledo-2024}. 
Note that, as discussed by Reinholdt and coworkers in their recent paper~\cite{Reinholdt-2025}, the QSCI wave function is in some cases less compact than that obtained by classical heuristics. From Eq. (\ref{eq6}), the probability that a particular Slater determinant is measured in the HSB-QSCI depends not only on the CI coefficient of the ground state, but also on the gap to the excited state and the CI coefficient of the excited state, and screening out unimportant Slater determinants from the subspace is very important in the application to larger systems.
Applying the auxiliary-field quantum Monte Carlo (AFQMC) method with the QSCI wave function as the input can be another direction towards larger-scale quantum chemical calculations with a quantum computer, as suggested in recent studies~\cite{Yoshida-2025, Danilov-2025}. 
Application of the HSB-QSCI to larger systems is currently in progress.

\section{Acknowledgments}
This work was supported by Quantum-LEAP Flagship Program (Grant No. JPMXS0120319794 and JPMXS0118067285) from Ministry of Education, Culture, Sports, Science and Technology (MEXT), Japan. K.S. acknowledges support from Center of Innovations for Sustainable Quantum AI (JPMJPF2221) from Japan Science and Technology Agency (JST), Japan and Grants-in-Aid for Scientific Research C (21K03407) and for Transformative Research Area B (23H03819) from Japan Society for the Promotion of Science (JSPS), Japan. 
A part of this work was performed for Council for Science, Technology and Innovation (CSTI), Cross-ministerial Strategic Innovation Promotion Program (SIP), “Promoting the application of advanced quantum technology platforms to social issues” (Funding agency: QST).
The authors thank Takashi Abe, Yoshiharu Mori, and Hajime Nakamura for useful discussions. 
The part of calculations was performed on the Mitsubishi Chemical Corporation (MCC) high-performance computer (HPC) system “NAYUTA”, where “NAYUTA” is a nickname for MCC HPC and is not a product or service name of MCC.
We acknowledge the use of IBM Quantum services for this work. The views expressed are those of the authors, and do not reflect the official policy or position of IBM or the IBM Quantum team.

\bibliography{ref}

\providecommand{\noopsort}[1]{}\providecommand{\singleletter}[1]{#1}%
\begin{thebibliography}{10}

\bibitem{Alan-2005}
A.~Aspuru-Guzik, A.~D. Dutoi, P.~J. Love, and M.~Head-Gordon.
\newblock Simulated quantum computation of molecular energies.
\newblock {\em Science}, 309:1704--1707, 2005.

\bibitem{Lanyon-2010}
B.~P. Lanyon, J.~D. Whitfield, G.~G. Gillett, M.~E. Goggin, M.~P. Almeida, I.~Kassal, J.~D. Biamonte, M.~Mohseni, B.~J. Powell, M.~Barbieri, A.~Aspuru-Guzik, and A.~G. White.
\newblock Towards quantum chemistry on a quantum computer.
\newblock {\em Nat. Chem.}, 2:106--111, 2010.

\bibitem{Du-2010}
J.~Du, N.~Xu, X.~Peng, P.~Wang, S.~Wu, and D.~Lu.
\newblock {NMR} implementation of a molecular hydrogen quantum simulation with adiabatic state preparation.
\newblock {\em Phys. Rev. Lett.}, 104:030502, 2010.

\bibitem{Wang-2015}
Y.~Wang, F.~Dolde, J.~Biamonte, R.~Babbush, V.~Bergholm, S.~Yang, I.~Jakobi, P.~Neumann, A.~Aspuru-Guzik, J.~D. Whitfield, and J.~Wrachtrup.
\newblock Quantum simulation of helium hydride cation in a sold-state spin register.
\newblock {\em ACS Nano}, 9:7769--7774, 2015.

\bibitem{OMalley-2016}
P.~J.~J. O'Malley, R.~Babbush, I.~D. Kivlichan, J.~Romero, J.~R. McClean, R.~Barends, J.~Kelly, P.~Roushan, A.~Tranter, N.~Ding, B.~Campbell, Y.~Chen, Z.~Chen adn B.~Chiaro, A.~Dunsworth, A.~G. Fowler, E.~Jeffrey, E.~Lucero, A.~Megrant, J.~Y. Mutus, M.~Neeley, C.~Neill, C.~Quintana, D.~Sank, A.~Vainsencher, J.~Wenner, T.~C. White, P.~V. Coveney, P.~J. Love, H.~Neven, A.~Aspuru-Guzik, and J.~M. Martinis.
\newblock Scalable quantum simulation of molecular energies.
\newblock {\em Phys. Rev. X}, 6:031007, 2016.

\bibitem{Santagati-2018}
R.~Santagati, J.~Wang, A.~A. Gentile, S.~Paesani, N.~Wiebe, J.~R. McClean, S.~Morley-Short, P.~J. Shatbolot, D.~Bonneau, J.~W. Silverstone, D.~P. Tew, X.~Zhou, J.~L. O'Brien, and M.~G. Thompson.
\newblock Witnessing eigenstates for quantum simulation of {H}amiltonian spectra.
\newblock {\em Sci. Adv.}, 4:eaap9646, 2018.

\bibitem{Blunt-2023}
N.~S. Blunt, L.~Caune, R.~Izsak, E.~T. Campbell, and N.~Holzmann.
\newblock Statistical phase estimation and error mitigation on a superconducting quantum processor.
\newblock {\em PRX Quantum}, 4:040341, 2023.

\bibitem{Yamamoto-2024}
K.~Yamamoto, S.~Duffield, Y.~Kikuchi, and D.~{Mu{\~n}oz Ramo}.
\newblock Demonstrating {B}ayesian quantum phase estimation with quantum error detection.
\newblock {\em Phys. Rev. Res.}, 6:013221, 2024.

\bibitem{Kanno-2024}
S.~Kanno, K.~Sugisaki, H.~Nakamura, H.~Yamauchi, R.~Sakuma, T.~Kobayashi, Q.~Gao, and N.~Yamamoto.
\newblock Tensor-based quantum phase difference estimation for large-scale demonstration.
\newblock {\em Proc. Natl. Acad. Sci. U.S.A.}, 122:e2425026122, 2025.

\bibitem{Peruzzo-2014}
A.~Peruzzo, J.~McClean, P.~Shadbolt, M.-H. Yung, X.-Q. Zhou, P.~J. Love, A.~Aspuru-Guzik, and J.~L. O'Brien.
\newblock A variational eigenvalue solver on a photonic quantum processor.
\newblock {\em Nat. Comm.}, 5:4213, 2014.

\bibitem{Tilly-2022}
J.~Tilly, H.~Chen, S.~Cao, D.~Picozzi, K.~Setia, Y.~Li, E.~Grant, L.~Wossnig, I.~Rungger, G.~H. Booth, and J.~Tennyson.
\newblock The variational quantum eigensolver: a review of methods and best practices.
\newblock {\em Phys. Rep.}, 986:1--128, 2022.

\bibitem{Gonthier-2022}
J.~F. Gonthier, M.~D. Radin, C.~Buda, E.~J. Doskocil, C.~M. Abuan, and J.~Romero.
\newblock Measurements as a roadblock to near-term physical quantum advantage in chemistry: Resource analysis.
\newblock {\em Phys. Rev. Res.}, 4:033154, 2022.

\bibitem{McClean-2018}
J.~R. McClean, S.~Boixo, V.~N. Smelyanskiy, R.~Babbush, and H.~Neven.
\newblock Barren plateaus in quantum neural network training landscapes.
\newblock {\em Nat. Comm.}, 9:4812, 2018.

\bibitem{Kanno-2023}
K.~Kanno, M.~Kohda, R.~Imai, S.~Koh, K.~Mitarai, W.~Mizukami, and Y.~O. Nakagawa.
\newblock Quantum-selected confiugration interaction: classical diagonalization of {H}amiltonians in subspaces selected by quantum computers.
\newblock arXiv:2302.11320, 2023.

\bibitem{Huron-1973}
B.~Huron, J.~P. Malrieu, and P.~Rancurel.
\newblock Iterative perturbation calculations of ground and excited state energies from multiconfigurational zeroth-order wavefunctions.
\newblock {\em J. Chem. Phys.}, 58:5745--5759, 1973.

\bibitem{Holmes-2016}
A.~A. Holmes, N.~M. Tubman, and C.~J. Umrigar.
\newblock Heat-bath configuration interaction: an efficient selected configuration interaction algorithm inspired by heat-bath sampling.
\newblock {\em J. Chem. Theory Comput.}, 12:3674--3680, 2016.

\bibitem{Tubman-2020}
N.~M. Tubman, C.~Daniel Freeman, D.~S. Levine, M.~Head-Gordon, and K.~Birgitta Whaley.
\newblock Modern approaches to exact diagonalization and selected configuration interaction with the adaptive sampling {CI} method.
\newblock {\em J. Chem. Theory Comput.}, 16:2139--2159, 2020.

\bibitem{Garniron-2018}
Y.~Garniron, A.~Scemama, E.~Giner, M.~Caffarel, and P.-F. Loos.
\newblock Selected configuration interaction dressed by perturbation.
\newblock {\em J. Chem. Phys.}, 149:064103, 2018.

\bibitem{Sharma-2017}
S.~Sharma, A.~A. Holmes, G.~Jeanmairet, A.~Alavi, and C.~J. Umrigar.
\newblock Semistochastic heat-bath configuration interaction method: Selected configuration interaction with semistochastic perturbation theory.
\newblock {\em J. Chem. Theory Comput.}, 13:1595--1604, 2017.

\bibitem{Yoshida-2025}
Y.~Yoshida, L.~Erhart, T.~Murokoshi, R.~Nakagawa, C.~Mori, T.~Miyanaga, T.~Mori, and W.~Mizukami.
\newblock Auxiliary-field quantum {M}onte {C}arlo method with quantum selected configuration interaction.
\newblock arXiv:2502.21081, 2025.

\bibitem{Danilov-2025}
D.~Danilov, J.~Robledo-Moreno, K.~J. Sung, M.~Motta, and J.~Shee.
\newblock Enhancing the accuracy and efficiency of sample-based quantum diagonalization with phaseless auxiliary-field quantum {M}onte {C}arlo.
\newblock arXiv:2503.05967.

\bibitem{Shirai-2025}
S.~Shirai, S.-Y. Tseng, H.~Iwakiri, T.~Horiba, H.~Hirai, and S.~Koh.
\newblock Enhancing accuracy of quantum-selected configuration interaction calculations using multireference perturbation theory: Application to aromatic molecules.
\newblock {\em ACS Omega}, 10:39736--39750, 2025.

\bibitem{Grimsley-2019}
H.~R. Grimsley, S.~E. Economou, E.~Barnes, and N.~J. Mayhall.
\newblock An adaptive variational algorithm for exact molecular simulations on a quantum computer.
\newblock {\em Nat. Comm.}, 10:3007, 2019.

\bibitem{Nakagawa-2023}
Y.~O. Nakagawa, M.~Kamoshita, W.~Mizukami, S.~Sudo, and Y.~Ohnishi.
\newblock {ADAPT-QSCI}: adaptive construction of input state for quantum-selected configuration interaction.
\newblock {\em J. Chem. Theory Comput.}, 20:10817--10825, 2024.

\bibitem{Nutzel-2024}
L.~N{\"u}tzel, A.~Gresch, L.~Hehn, L.~Marti, R.~Freund, A.~Steiner, C.~D. Marciniak, T.~Eckstein, N.~Stockinger, S.~Wolf, T.~Monz, M.~K{\"u}hn, and M.~J. Hartmann.
\newblock Solving an industrially relevant quantum chemistry problem on quantum hardware.
\newblock {\em Quantum Sci. Technol}, 10:015066, 2025.

\bibitem{Motta-2023}
M.~Motta, K.~J. Sung, K.~Birgitta Whaley, M.~Head-Gordon, and J.~Shee.
\newblock Bridging physical intuition and hardware efficiency for correlated electronic states: the local unitary cluster {J}astrow ansatz for electronic structure.
\newblock {\em Chem. Sci.}, 14:11213--11227, 2023.

\bibitem{Robledo-2024}
J.~Robledo-Moreno, M.~Motta, H.~Haas, A.~Javadi-Abhari, P.~Jurcevic, W.~Kirby, S.~Martiel, K.~Sharma, S.~Sharma, T.~Shirakawa, I.~Sitdikov, R.-Y. Sun, K.~J. Sung, M.~Takita, M.~C. Tran, S.~Yunoki, and A.~Mezzacapo.
\newblock Chemistry beyond the scale of exact diagonalization on a quantum-centric supercomputer.
\newblock {\em Sci. Adv.}, 11:eadu9991, 2025.

\bibitem{Kaliakin-2024}
D.~Kaliakin, A.~Shajan, J.~{Robledo Moreno}, Z.~Li, A.~Mitra, M.~Motta, C.~Johnson, A.~Ash Saki, S.~Das, I.~Sitdikov, A.~Mezzacapo, and K.~M. {Merz Jr.}
\newblock Accurate quantum-centric simulations of supramolecular interactions.
\newblock arXiv:2410.09209, 2024.

\bibitem{Barison-2024}
S.~Barison, J.~{Robledo Moreno}, and M.~Motta.
\newblock Quantum-centric computation of molecular excited states with extended sample-based quantum diagonalization.
\newblock {\em Quantum Sci. Technol.}, 10:025034, 2025.

\bibitem{Liepuoniute-2024}
I.~Liepuoniute, K.~D. Doney, J.~{Robledo Moreno}, J.~A. Job, W.~S. Friend, and G.~O. Jones.
\newblock Quantum-centric computational study of methylene singlet and triplet states.
\newblock {\em J. Chem. Theory Comput.}, 21:5062--5070, 2025.

\bibitem{Shajan-2024}
A.~Shajan, D.~Kaliakin, A.~Mitra, J.~{Robledo Moreno}, Z.~Li, M.~Motta, C.~Johnsonand A.~Ash Saki, S.~Das, I.~Sitdikov, A.~Mezzacapo, and K.~M. {Merz Jr.}
\newblock Towards quantum-centric simulations of extended molecules: sample-based quantum diagonalization enhanced with density matrix embedding theory.
\newblock {\em J. Chem. Theory Comput.}, 21:6801--6810, 2025.

\bibitem{footnote}
After posting this manuscript to arXiv, Mikkelsen and Nakagawa, and Yu and coworkers independently submitted their own manuscripts to arXiv, presenting the same idea explored in this study. M. Mikkelsen, Y. O. Nakagawa, arXiv:2412.13839, 2024; J. Yu, et al, arXiv:2501.09702, 2025.

\bibitem{Cortes-2022}
C.~L. Cortes and S.~K. Gray.
\newblock Quantum {K}rylov subspace algorithms for ground- and excited-state energy estimation.
\newblock {\em Phys. Rev. A}, 105:022417, 2022.

\bibitem{Zhang-2024}
Z.~Zhang, A.~Wang, X.~Xu, and Y.~Li.
\newblock Measurement-efficient quantum {K}rylov subspace diagonalization.
\newblock {\em Quantum}, 8:1438, 2023.

\bibitem{Lee-2025}
G.~Lee, S.~Choi, J.~Huh, and A.~F. Izmaylov.
\newblock Efficient strategies for reducing sampling error in quantum {K}rylov subspace diagonalization.
\newblock {\em Digital Discovery}, 4:954--969, 2025.

\bibitem{Lockwood-2024}
O.~Lockwood, P.~Weiss, F.~Aronshtein, and G.~Verdon.
\newblock Quantum dynamical {H}amiltonian {M}onte {C}arlo.
\newblock {\em Phys. Rev. Res.}, 6:033142, 2024.

\bibitem{Schriber-2016}
J.~B. Schriber and F.~A. Evangelista.
\newblock An adaptive configuration interaction approach for strongly correlated electrons with tunable accuracy.
\newblock {\em J. Chem. Phys.}, 144:161106, 2016.

\bibitem{Zimmerman-2017}
P.~M. Zimmerman.
\newblock Incremental full configuration interaction.
\newblock {\em J. Chem. Phys.}, 146:104102, 2017.

\bibitem{Jordan-1928}
P.~Jordan and E.~Wigner.
\newblock {\"U}ber das paulische {\"a}quivalenzverbot.
\newblock {\em Z. Phys}, 47:631--651, 1928.

\bibitem{Trotter-1959}
H.~F. Trotter.
\newblock On the product of semi-groups of operators.
\newblock {\em Proc. Am. Math. Soc.}, 10:545--551, 1959.

\bibitem{Suzuki-1976}
M.~Suzuki.
\newblock Relationship between d-dimensional quantal spin systems and (d + 1)-dimensional {I}sing systems: equivalence, critical exponents and systematic approximants of the partition function and spin correlations.
\newblock {\em Prog. Theor. Phys.}, 56:1454--1469, 1976.

\bibitem{Ouyand-2020}
Y.~Ouyang, D.~R. White, and E.~T. Campbell.
\newblock Compilation by stochastic {H}amiltonian sparsification.
\newblock {\em Quantum}, 4:235, 2020.

\bibitem{Kurogi-2024}
H.~Kurogi, K.~Endo, Y.~Sato, M.~Sugawara, K.~Wada, K.~Sugisaki, S.~Kanno, H.~C. Watanabe, and H.~Nakano.
\newblock Optimizing a parameterized controlled gate with {F}ree {Q}uaternion {S}election.
\newblock arXiv:2409.13547, 2024.

\bibitem{Becke-1993}
A.~D. Becke.
\newblock A new mixing of {H}artree--{F}ock and local density functional theories.
\newblock {\em J. Chem. Phys.}, 98:1372--1377, 1993.

\bibitem{Stephens-1994}
P.~J. Stephens, F.~J. Devlin, C.~F. Chabalowski, and M.~J. Frisch.
\newblock Ab initio calculation of vibrational absroption and circular dichroism spectra using density functional force fields.
\newblock {\em J. Phys. Chem.}, 98:11623--11627, 1994.

\bibitem{Ditchfield-1971}
R.~Ditchfield, W.~J. Hehre, and J.~A. Pople.
\newblock Self-consistent molecular-orbital methods. {IX.} an extended {G}aussian-type basis for molecular-orbital studies of organic molecules.
\newblock {\em J. Chem. Phys.}, 54:724--728, 1971.

\bibitem{Hehre-1972}
W.~J. Hehre, R.~Ditchfield, and J.~A. Pople.
\newblock Self-consistent molecular orbital methods. {XII.} further extensions of {G}aussian-type basis ses for use in molecular orbital studies of organic molecules.
\newblock {\em J. Chem. Phys.}, 56:2257--2261, 1972.

\bibitem{g16}
M.~J. Frisch, G.~W. Trucks, H.~B. Schlegel, G.~E. Scuseria, M.~A. Robb, J.~R. Cheeseman, G.~Scalmani, V.~Barone, G.~A. Petersson, H.~Nakatsuji, X.~Li, M.~Caricato, A.~V. Marenich, J.~Bloino, B.~G. Janesko, R.~Gomperts, B.~Mennucci, H.~P. Hratchian, J.~V. Ortiz, A.~F. Izmaylov, J.~L. Sonnenberg, D.~Williams-Young, F.~Ding, F.~Lipparini, F.~Egidi, J.~Goings, B.~Peng, A.~Petrone, T.~Henderson, D.~Ranasinghe, V.~G. Zakrzewski, J.~Gao, N.~Rega, G.~Zheng, W.~Liang, M.~Hada, M.~Ehara, K.~Toyota, R.~Fukuda, J.~Hasegawa, M.~Ishida, T.~Nakajima, Y.~Honda, O.~Kitao, H.~Nakai, T.~Vreven, K.~Throssell, J.~A. Montgomery, {Jr.}, J.~E. Peralta, F.~Ogliaro, M.~J. Bearpark, J.~J. Heyd, E.~N. Brothers, K.~N. Kudin, V.~N. Staroverov, T.~A. Keith, R.~Kobayashi, J.~Normand, K.~Raghavachari, A.~P. Rendell, J.~C. Burant, S.~S. Iyengar, J.~Tomasi, M.~Cossi, J.~M. Millam, M.~Klene, C.~Adamo, R.~Cammi, J.~W. Ochterski, R.~L. Martin, K.~Morokuma, O.~Farkas, J.~B. Foresman, and D.~J. Fox.
\newblock Gaussian 16 {R}evision {C}.01, 2016.
\newblock Gaussian Inc. Wallingford CT.

\bibitem{GAMESS}
G.~M.~J. Barca, C.~Bertoni, L.~Carrington, D.~Datta, N.~{De Silva}, J.~{Emiliano Deustua}, D.~G. Fedorov, J.~R. Gour, A.~O. Gunina, E.~Guidez, T.~Harville, S.~Irle, J.~Ivanic, K.~Kowalski, S.~S. Leang, H.~Li, W.~Li, J.~J. Lutz, I.~Magoulas, J.~Mato, V.~Mironov, H.~Nakata, B.~Q. Pham, P.~Piecuch, D.~Poole, S.~R. Pruitt, A.~P. Rendell, L.~B. Roskop, K.~Ruedenberg, T.~Sattasathuchana, M.~W. Schmidt, J.~Shen, L.~Slipchenko, M.~Sosonkina, V.~Sundriyal, A.~Tiwari, J.~L. {Galvez Vallejo}, B.~Westheimer, M.~Wloch, P.~Xu, F.~Zahariev, and M.~S. Gordon.
\newblock Recent developments in the general atomic and molecular electronic structure system.
\newblock {\em J. Chem. Phys.}, 152:154102, 2020.

\bibitem{Hehre-1969}
W.~J. Hehre, R.~F. Stewart, and J.~A. Pople.
\newblock Self-consistent molecular-orbital methods. {I.} use of {G}aussian expansions of {S}later-type atomic orbitals.
\newblock {\em J. Chem. Phys.}, 51:2657--2664, 1969.

\bibitem{Pipek-1989}
J.~Pipek and P.~G. Mezey.
\newblock A fast intrinsic localization procedure applicable for ab initio and semiempirical linear combination of atomic orbital wave functions.
\newblock {\em J. Chem. Phys.}, 90:4916--4926, 1989.

\bibitem{OpenFermion}
J.~R. McClean, N.~C. Rubin, K.~J. Sung, I.~D. Kivlichan, X.~Bonet-Monroig, Y.~Cao, C.~Dai, E.~{Schuyler Fried}, C.~Gidney, B.~Gimby, P.~Gokhale, T.~H\"{a}ner, T.~Hardikar, V.~Havl\`{i}\v{c}ek, O.~Higgott, C.~Huang, J.~Izaac, Z.~Jiang, X.~Liu, S.~McArdle, M.~Neeley, T.~O'Brien, B.~O'Gorman, I.~Ozfidan, M.~D. Radin, J.~Romero, N.~P.~D. Sawaya, B.~Senjean, K.~Setia, S.~Sim, D.~S. Steiger, M.~Steudtner, Q.~Sun, W.~Sun, D.~Wang, F.~Zhang, and R.~Babbush.
\newblock Open{F}ermion: the electronic structure package for quantum computers.
\newblock {\em Quantum Sci. Technol.}, 5:034014, 2020.

\bibitem{Tranter-2018}
A.~Tranter, P.~J. Love, F.~Mintert, and P.~V. Coveney.
\newblock A comparison of the {B}ravyi--{K}itaev and {J}ordan--{W}igner transformations for the quantum simulation of quantum chemistry.
\newblock {\em J. Chem. Theory Comput.}, 14:5617--5630, 2018.

\bibitem{qsim}
{Quantum AI team and collaborators}.
\newblock qsim, September 2020.

\bibitem{PyCI}
M.~Richer, G.~S{\'a}nchez-D{\'i}az, M.~Mart{\'i}nez-Gonz{\'a}lez, V.~Chuiko, T.~D. Kim, A.~Tehrani, S.~Wang, P.~B. Gaikwad, C.~E.~V. de~Moura, C.~Masschelein, R.~A. Miranda-Quintana, A.~Gerolin, F.~Heider-Zadeh, and P.~W. Ayers.
\newblock Py{CI}: a python-scriptable library for arbitrary determinant {CI}.
\newblock {\em J. Chem. Phys.}, 161:132502, 2024.

\bibitem{qiskit-addon-sqd}
A.~Ash Saki, S.~Barison, B.~Fuller, J.~R. Garrison, J.~R. Glick, C.~Johnson, A.~Mezzacapo, J.~Robledo-Moreno, M.~Rossmannek, P.~Schweigert, I.~Sitdikov, and K.~J. Sung.
\newblock {Qiskit addon: sample-based quantum diagonalization}.
\newblock \url{https://github.com/Qiskit/qiskit-addon-sqd}, 2024.

\bibitem{PySCF}
Q.~Sun, X.~Zhang adn S.~Banerjee, P.~Bao, M.~Barbry, N.~S. Blunt, N.~A. Bogdanov, G.~H. Booth, J.~Chen, Z.-H. Cui, J.~J.~Eriksena dn~Y.~Gao, S.~Guo, J.~Hermann, M.~R. Hermes, K.~Koh, P.~Koval, S.~Lehtola, Z.~Li, Z.~Liu, N.~Mardirossian, J.~D. McClain, M.~Motta, B.~Mussard, H.~Q. Pham, A.~Pulkin, W.~Purwanto, P.~J. Robinson, E.~Ronca, E.~R. Sayfutyarova, M.~Scheurer, H.~F. Schurkus, J.~E.~T. Smith, C.~Sun, S.-N. Sun, S.~Upadhyay, L.~K. Wagner, X.~Wang, A.~White, J.~D. Whitfield, M.~J. Williamson, S.~Wouters, J.~Yang, J.~M. Yu, T.~Zhu, T.~C .Berkelbach, S.~Sharma, A.~Y. Sokolov, and G.~K.-L. Chan.
\newblock Recent developments in the {P}y{SCF} program package.
\newblock {\em J. Chem. Phys.}, 153:024109, 2020.

\bibitem{Zeng-2021}
W.~Zeng and J.~Wu.
\newblock Open-shell graphene fragments.
\newblock {\em Chem}, 7:358--386, 2021.

\bibitem{Baldacchino-2022}
A.~J. Baldacchino, M.~I. Collins, M.~P. Nielsen, T.~W. Schmidt, D.~R. McCamey, and M.~J.~Y. Tayebjee.
\newblock Singlet fission photovoltaics: Progress and promising pathways.
\newblock {\em Chem. Phys. Rev.}, 3:021304, 2022.

\bibitem{Mikkelsen-2024}
M.~Mikkelsen and Y.~O. Nakagawa.
\newblock Quantum-selected configuration interaction with time-evolved state.
\newblock arXiv:2412.13839, 2024.

\bibitem{Yu-2025}
J.~Yu, J.~{Robledo Moreno}, J.~Iosue, L.~Bertels, D.~Claudino, B.~Fuller, P.~Groszkowski, T.~S. Humble, P.~Jurcevic, W.~Kirby, T.~A. Maier, M.~Motta, B.~Pokharel, A.~Seif, A.~Shehata, K.~J. Sung, M.~C. Tran, V.~Tripathi, A.~Mezzacapo, and K.~Sharma.
\newblock Quantum-centric algorithm for sample-based {K}rylov diagonalization.
\newblock arXiv:2501.09702.

\bibitem{Singh-1971}
B.~Singh and J.~S. Brinen.
\newblock Low-temperature photochemistry of {\it p}-diazidobenzene and 4,4'-diazidoazobenzene.
\newblock {\em J. Am. Chem. Soc.}, 93:540--542, 1971.

\bibitem{Nicolaides-1998}
A.~Nicolaides, H.~Tomioka, and S.~Murata.
\newblock Direct observation and characterization of {\it p}-phenylenebisnitrene. a labile quinoidal diradical.
\newblock {\em J. Am. Chem. Soc.}, 120:11530--11531, 1998.

\bibitem{Nimura-2006}
S.~Nimura, O.~Kikuchi, T.~Ohana, A.~Yabe, and M.~Kaise.
\newblock Singlet-triplet energy gaps of quinonoidal dinitrenes.
\newblock {\em Chem. Lett.}, 25:125--126, 2016.

\bibitem{Ichimura-1995}
A.~S. Ichimura, K.~Sato, T.~Kinoshita, T.~Takui, K.~Itoh, and P.~M. Lahti.
\newblock An electron spin resonance study of persistent dinitrenes.
\newblock {\em Mol. Cryst. Liq. Cryst.}, 271/272:279--288, 1995.

\bibitem{Sugisaki-2013}
K.~Sugisaki, K.~Toyota, K.~Sato, D.~Shiomi, M.~Kitagawa, and T.~Takui.
\newblock Quantum chemical calculations of the zero-field splitting tensors for organic spin multiplets.
\newblock In A.~Lund and M.~Shiotani, editors, {\em EPR of Free Radicals in Solids I: Trends in Methods and Applications, 2nd Edition}, pages 363--392. Springer, Dordrecht, 2013.

\bibitem{Sugisaki-2019}
K.~Sugisaki, S.~Nakazawa, K.~Toyota, K.~Sato, D.~Shiomi, and T.~Takui.
\newblock Quantum chemistry on quantum computers: a method for preparation of multiconfigurational wave functions on quantum computers without performing post-{H}artree--{F}ock calculations.
\newblock {\em ACS Cent. Sci.}, 5:167--175, 2019.

\bibitem{Liu-2013}
M.~Liu, V.~I. Artyukhov, H.~Lee, F.~Xu, and B.~I. Yakobson.
\newblock Carbyne from first principles; chain of {C} atoms, a nanorod or a nanorope.
\newblock {\em ACS Nano}, 7:10075--10082, 2013.

\bibitem{Liu-2013-correction}
M.~Liu, V.~I. Artyukhov, H.~Lee, F.~Xu, and B.~I. Yakobson.
\newblock Correction to carbyne from first principles; chain of {C} atoms, a nanorod or a nanorope.
\newblock {\em ACS Nano}, 11:5186, 2017.

\bibitem{Gao-2024}
H.~Gao, S.~Imamura, A.~Kasagi, and E.~Yoshida.
\newblock Distributed implementation of full configuration interaction for one trillion determinants.
\newblock {\em J. Chem. Theory Comput.}, 20:1185--1192, 2024.

\bibitem{Reinholdt-2025}
P.~Reinholdt, K.~M. Ziems, E.~R. Kjellgren, S.~Coriani, S.~P.~A. Sauer, and J.~Kongsted.
\newblock Critical limitations in quantum-selected configuration interaction methods.
\newblock {\em J. Chem. Theory Comput.}, 21:6811--6822, 2025.

\end{thebibliography}
\bibliographystyle{unsrt}

\newpage

\renewcommand{\thesection}{S\arabic{section}} 
\setcounter{section}{0}
\section{Supporting Information}

\renewcommand{\thepage}{S\arabic{page}}
\setcounter{page}{1}
\renewcommand{\thefigure}{S\arabic{figure}}
\setcounter{figure}{0}
\renewcommand{\thetable}{S\arabic{table}}
\setcounter{table}{0}

\subsection{Active spaces}

\begin{figure}[H]
    \centering
    \includegraphics[width=0.6\linewidth]{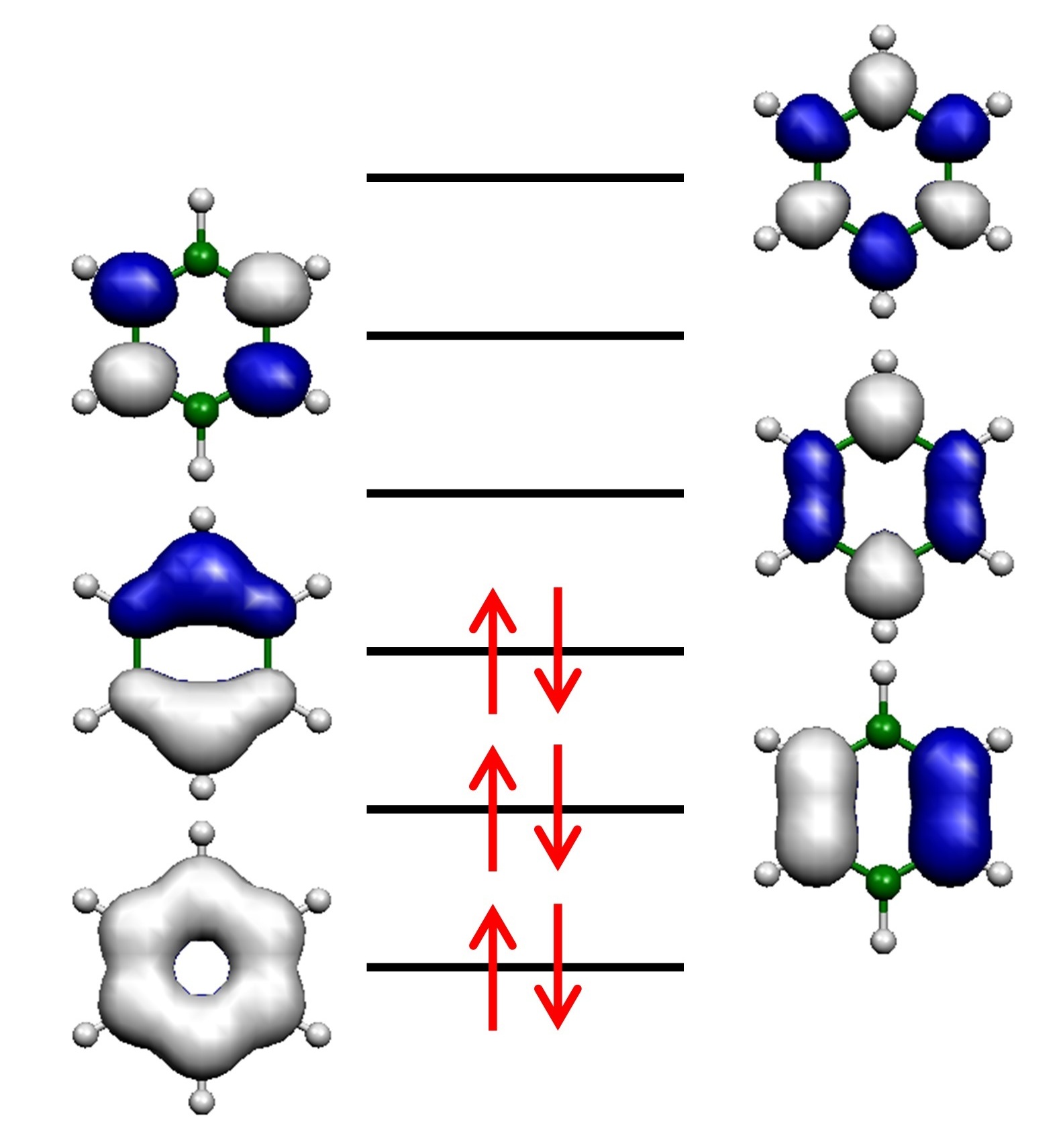}
    \caption{(6e,6o) active orbitals of {\bf 1}. Red arrows specify the electron occupancy in the RHF configuration.}
    \label{fig:figs1}
\end{figure}

\begin{figure}[H]
    \centering
    \includegraphics[width=0.7\linewidth]{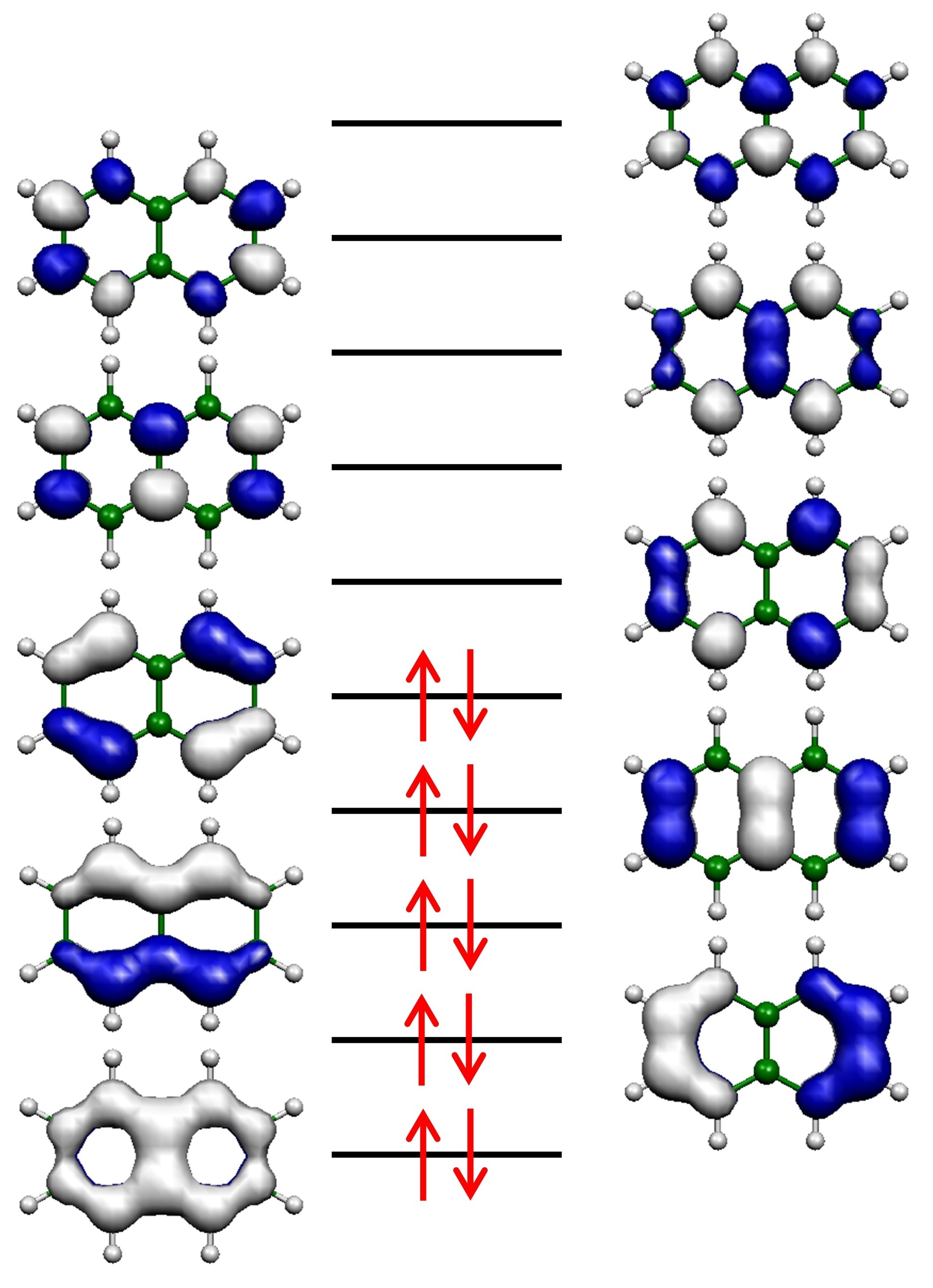}
    \caption{(10e,10o) active space of {\bf 2}. Red arrows specify the electron occupancy in the RHF configuration. }
    \label{fig:figs2}
\end{figure}

\begin{figure}[H]
    \centering
    \includegraphics[width=0.9\linewidth]{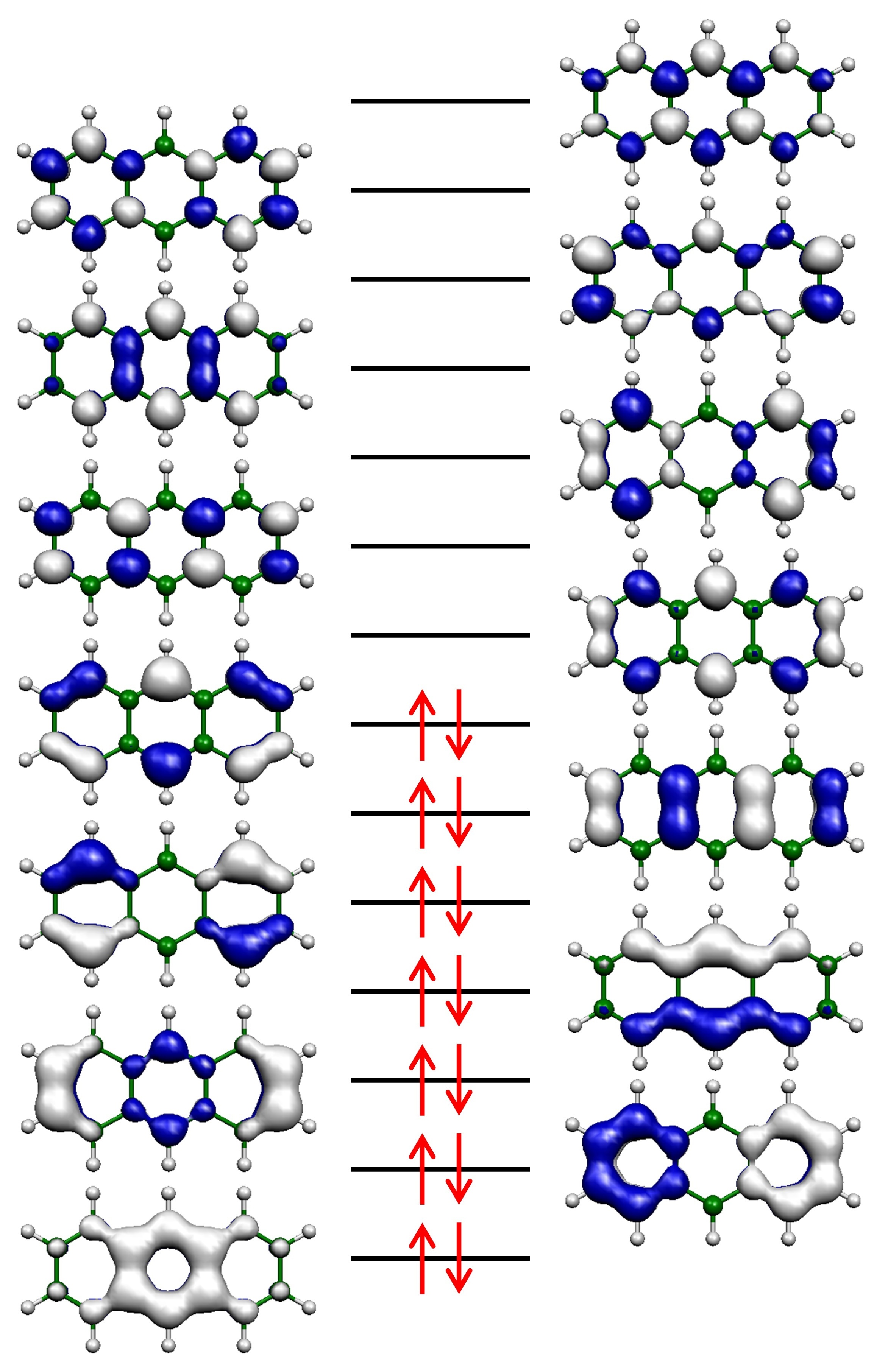}
    \caption{(14e,14o) active space of {\bf 3}. Red arrows specify the electron occupancy in the RHF configuration. }
    \label{fig:figs3}
\end{figure}

\begin{figure}[H]
    \centering
    \includegraphics[width=0.8\linewidth]{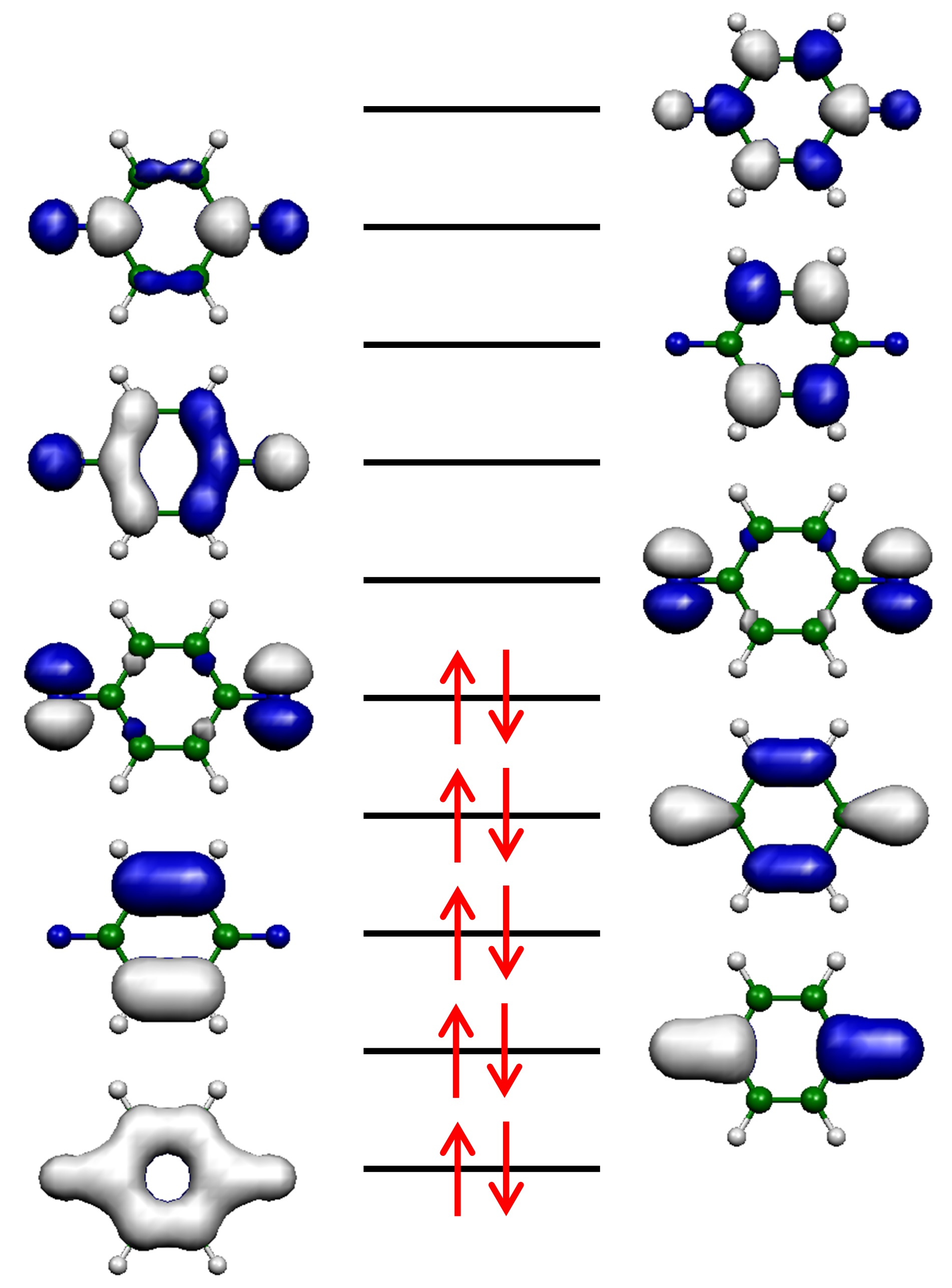}
    \caption{(10e,10o) active space of {\bf 4}. Red arrows specify the electron occupancy in the RHF configuration.}
    \label{fig:figs4}
\end{figure}

\begin{figure}[H]
    \centering
    \includegraphics[width=0.6\linewidth]{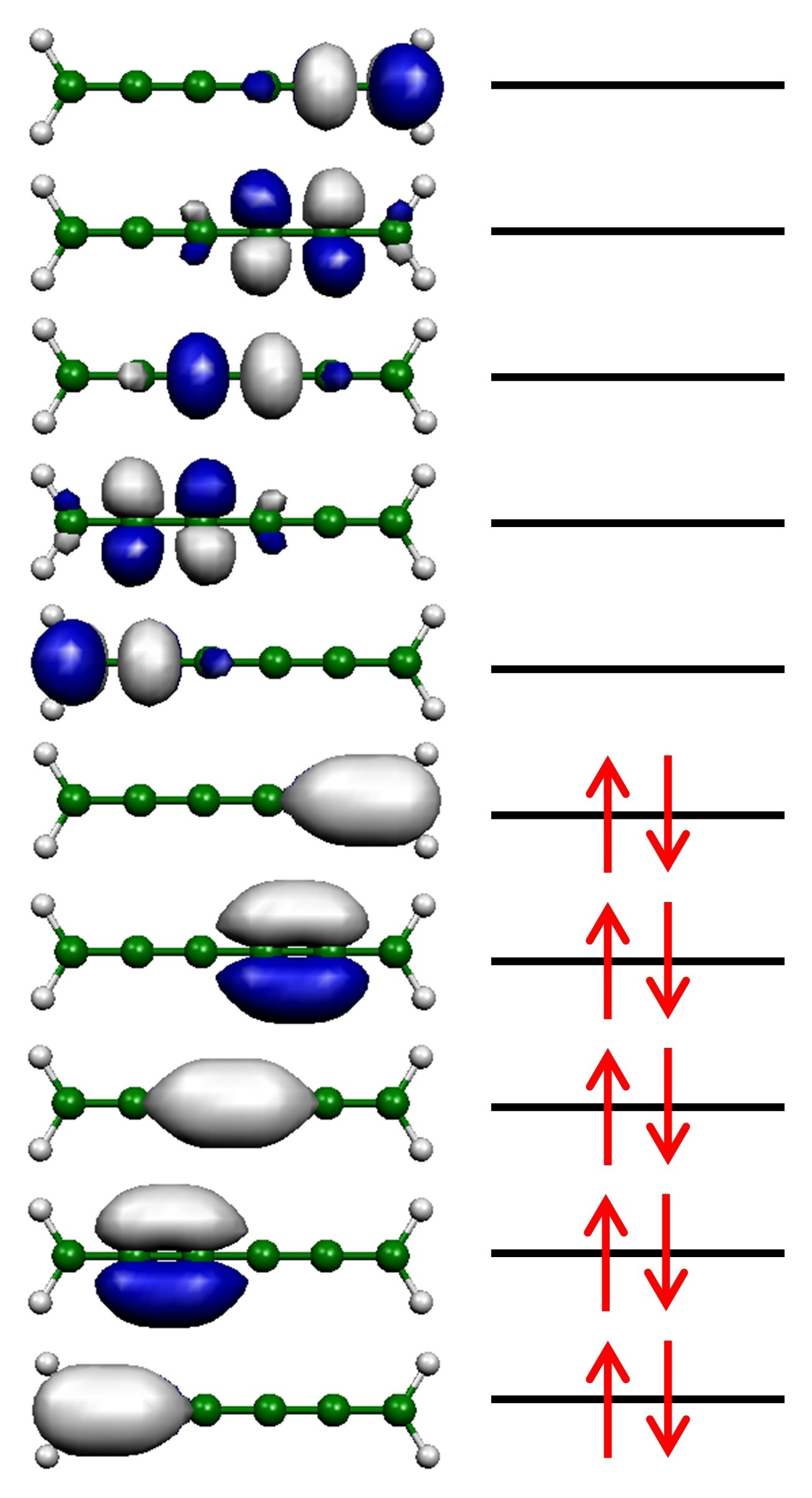}
    \caption{(10e,10o) active space of {\bf 5} in planar geometry. Red arrows specify the electron occupancy in the RHF configuration. }
    \label{fig:figs5}
\end{figure}

\begin{figure}[H]
    \centering
    \includegraphics[width=0.6\linewidth]{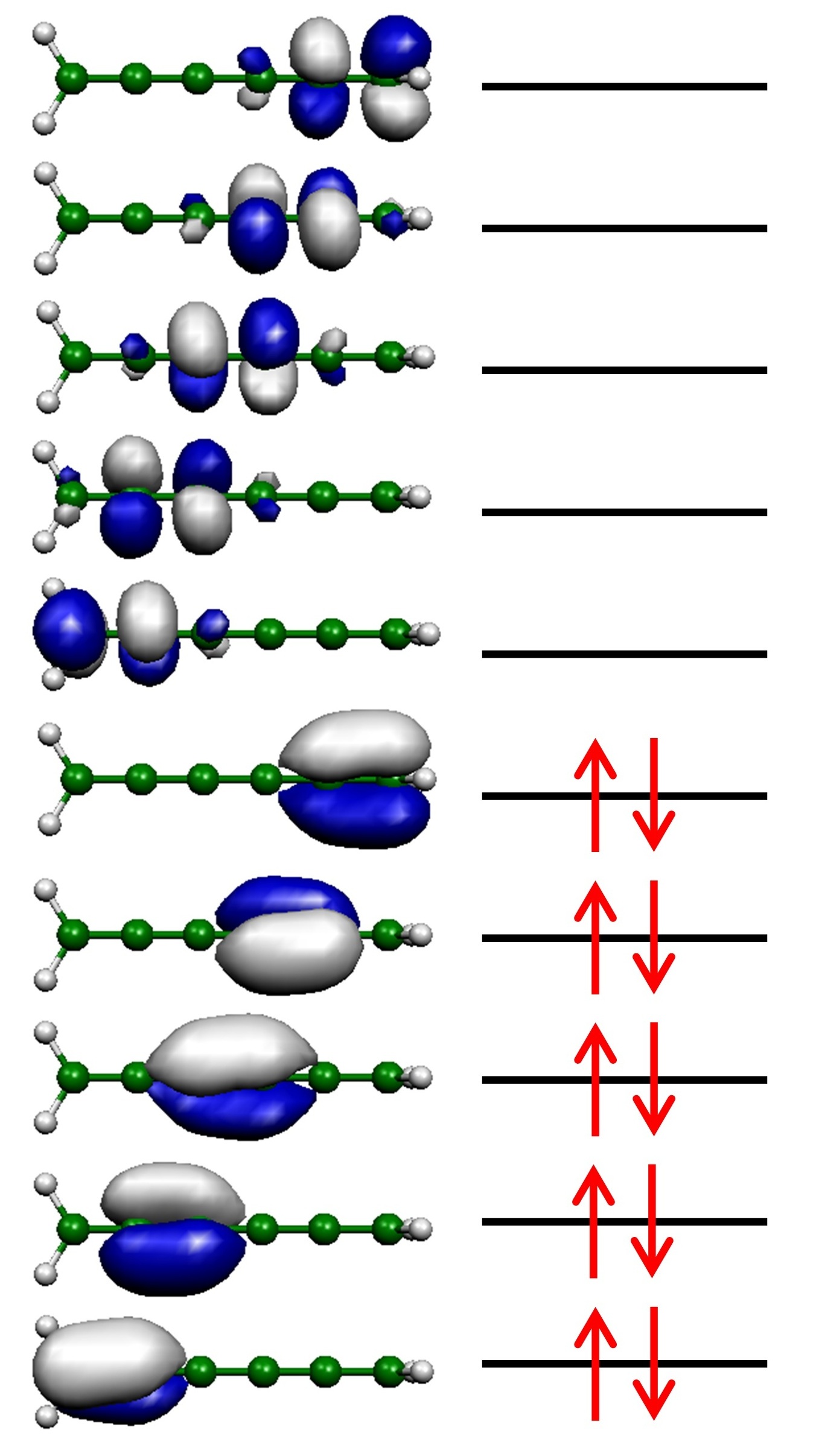}
    \caption{(10e,10o) active space of {\bf 5} in twisted geometry. Red arrows specify the electron occupancy in the RHF configuration. }
    \label{fig:figs6}
\end{figure}

\begin{figure}[H]
    \centering
    \includegraphics[width=0.6\linewidth]{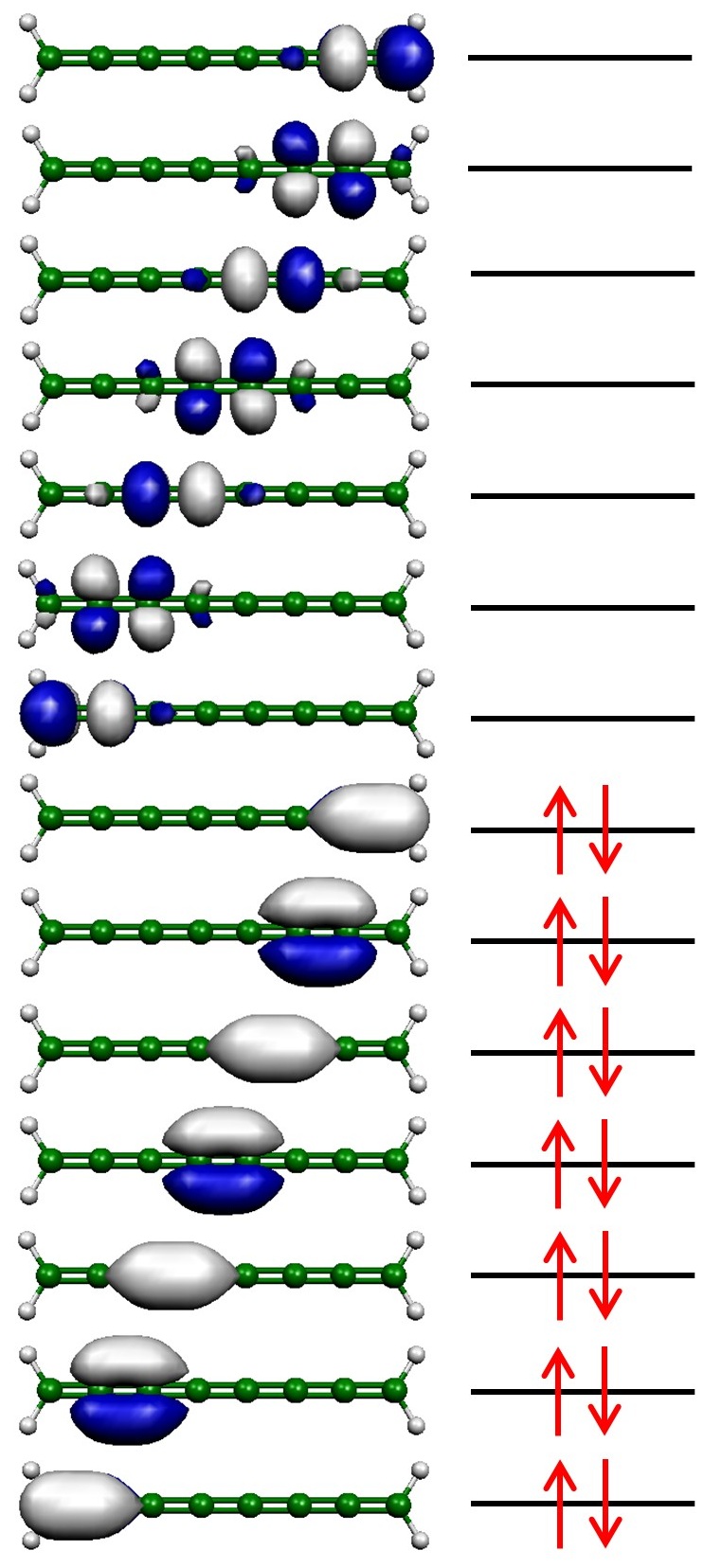}
    \caption{(14e,14o) active space of {\bf 6} in planar geometry. Red arrows specify the electron occupancy in the RHF configuration. }
    \label{fig:figs7}
\end{figure}

\begin{figure}[H]
    \centering
    \includegraphics[width=0.6\linewidth]{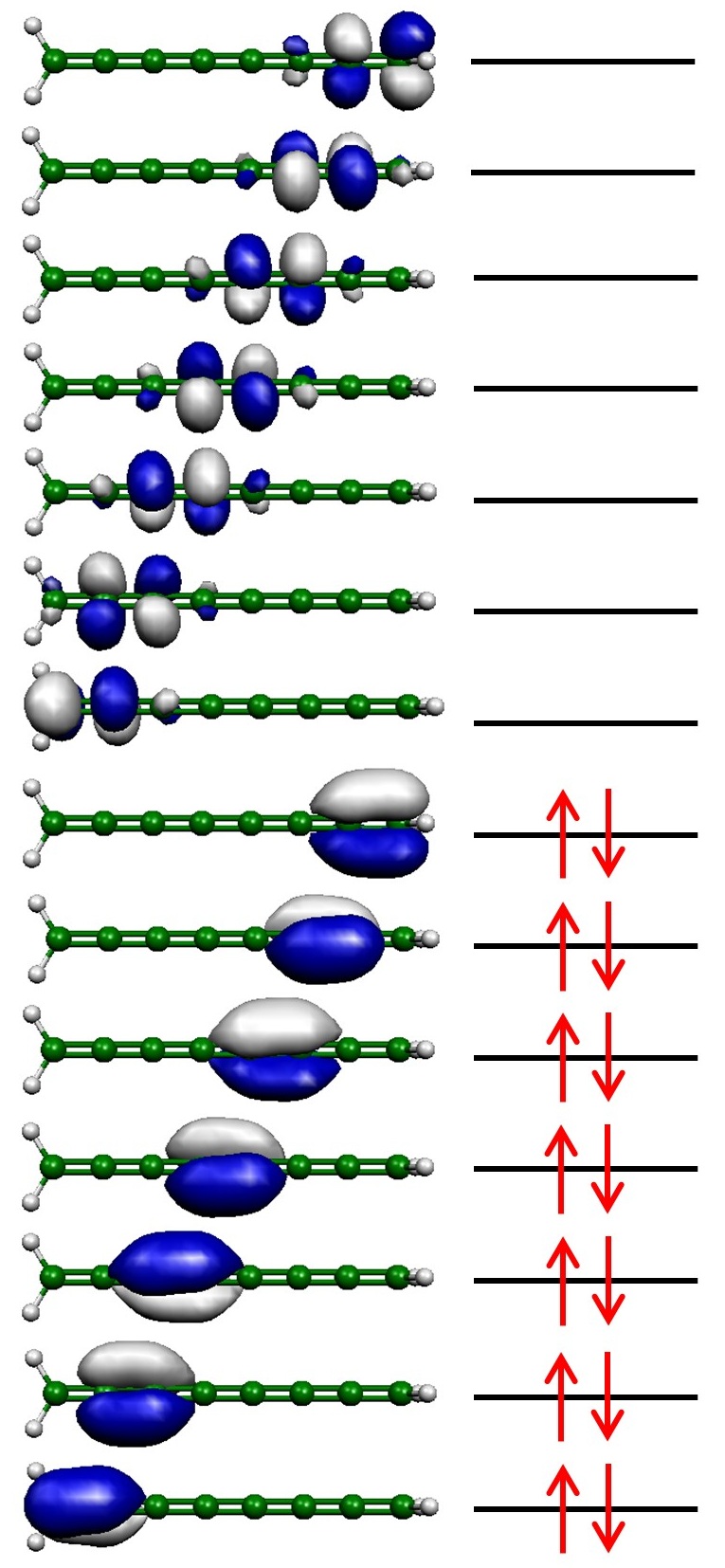}
    \caption{(14e,14o) active space of {\bf 6} in twisted geometry. Red arrows specify the electron occupancy in the RHF configuration. }
    \label{fig:figs8}
\end{figure}

\begin{figure}[H]
    \centering
    \includegraphics[width=0.6\linewidth]{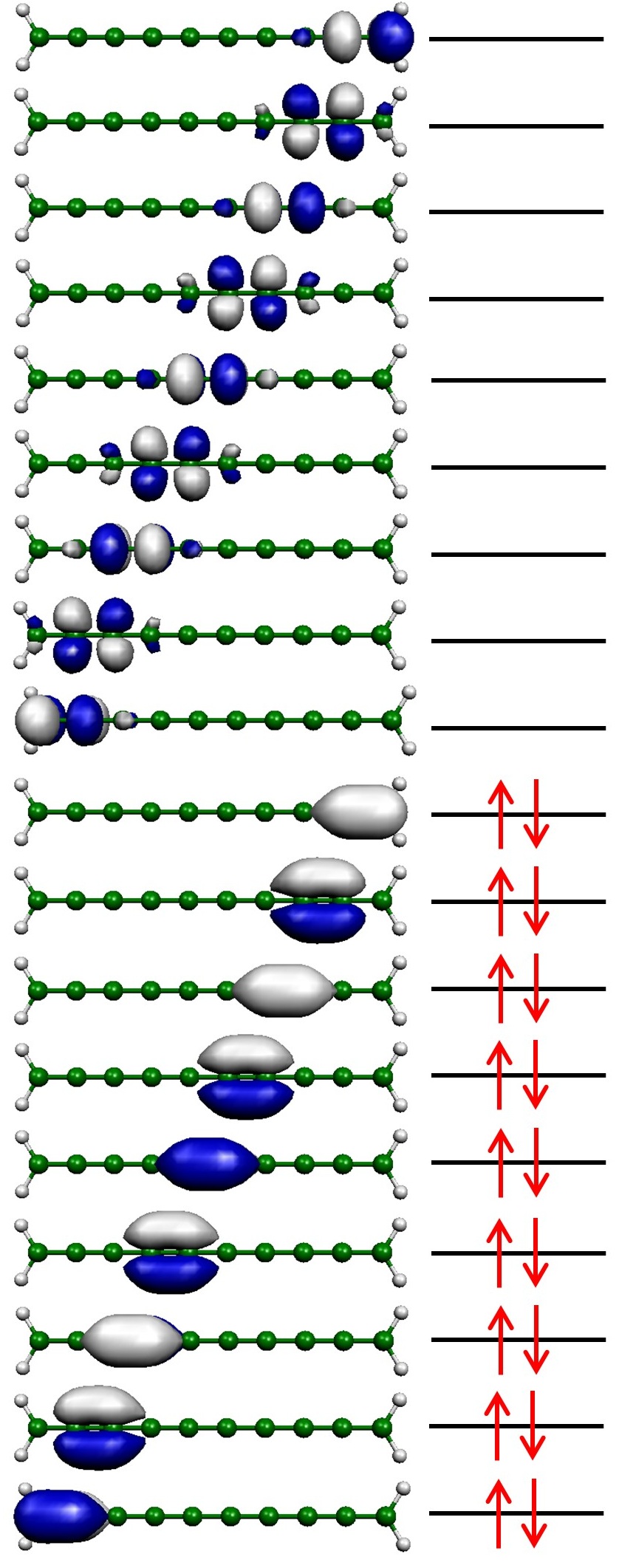}
    \caption{(18e,18o) active space of {\bf 7} in planar geometry. Red arrows specify the electron occupancy in the RHF configuration. }
    \label{fig:figs9}
\end{figure}

\begin{figure}[H]
    \centering
    \includegraphics[width=0.6\linewidth]{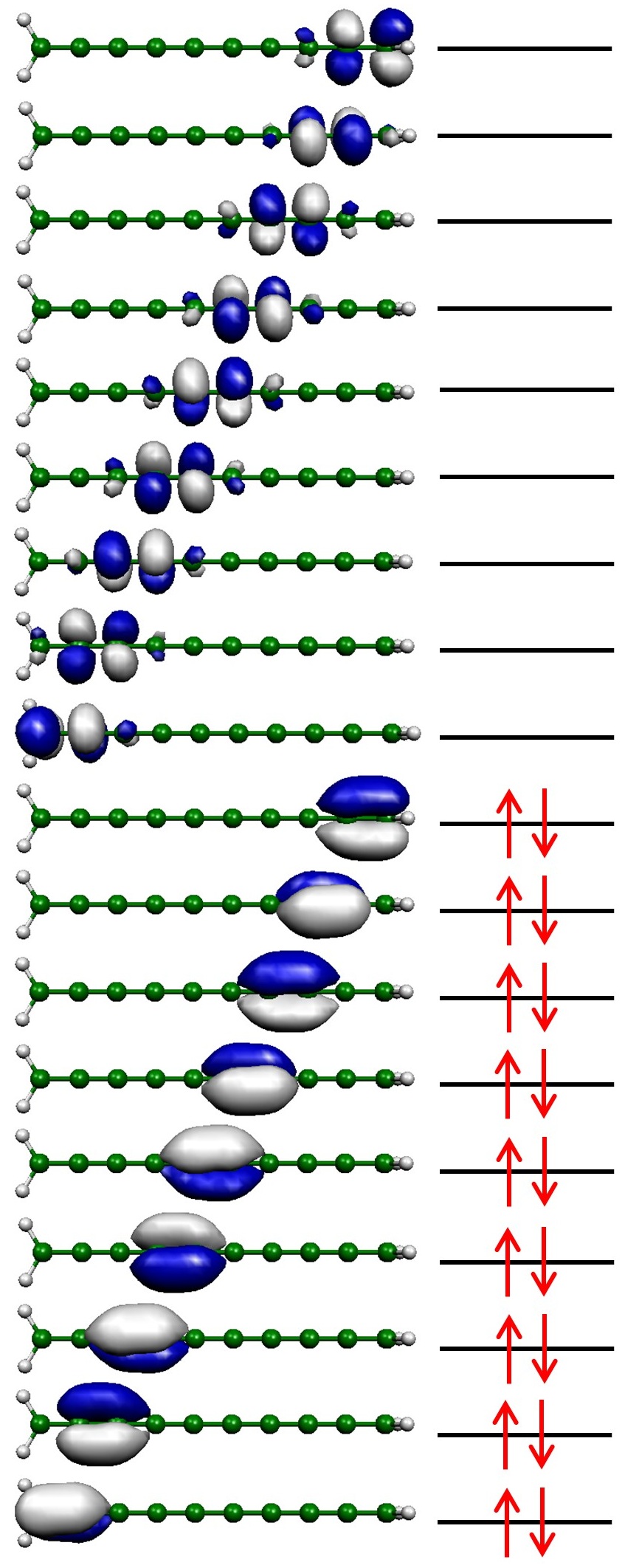}
    \caption{(18e,18o) active space of {\bf 7} in twisted geometry. Red arrows specify the electron occupancy in the RHF configuration. }
    \label{fig:figs10}
\end{figure}

\subsection{Cartesian coordinates} 

\begin{table}[H]
    \caption{Cartesian coordinates of the B3LYP/6-31G* optimized geometry of {\bf 1}, in units of {\r A}.}
    \begin{tabular*}{\linewidth}{@{\extracolsep{\fill}}crrr}
    \hline
    Atom & \multicolumn{1}{c}{X} & \multicolumn{1}{c}{Y} & \multicolumn{1}{c}{Z} \\
    \hline
    C & 0.000000    & 1.396602    & 0.000000 \\
    C & 0.000000    & $-$1.396602 & 0.000000 \\
    C & 1.209493    & 0.698301    & 0.000000 \\
    C & 1.209493    & $-$0.698301 & 0.000000 \\
    C & $-$1.209493 & 0.698301    & 0.000000 \\
    C & $-$1.209493 & $-$0.698301 & 0.000000 \\
    H & 0.000000    & 2.483720    & 0.000000 \\
    H & 0.000000    & $-$2.483720 & 0.000000 \\
    H & 2.150965    & 1.241860    & 0.000000 \\
    H & 2.150965    & $-$1.241860 & 0.000000 \\
    H & $-$2.150965 & 1.241860    & 0.000000\\
    H & $-$2.150965 & $-$1.241860 & 0.000000 \\
    \hline
    \end{tabular*}
    \label{tab:tables1}
\end{table}

\begin{table}[H]
    \caption{Cartesian coordinates of the B3LYP/6-31G* optimized geometry of {\bf 2}, in units of {\r A}.}
    \begin{tabular*}{\linewidth}{@{\extracolsep{\fill}}crrr}
    \hline
    Atom & \multicolumn{1}{c}{X} & \multicolumn{1}{c}{Y} & \multicolumn{1}{c}{Z} \\
    \hline
    C & 0.000000 &    0.000000 &    0.717063 \\
    C & 0.000000 &    0.000000 & $-$0.717063 \\
    C & 0.000000 &    1.244791 &    1.402547 \\
    C & 0.000000 &    1.244791 & $-$1.402547 \\
    C & 0.000000 & $-$1.244791 &    1.402547 \\
    C & 0.000000 & $-$1.244791 & $-$1.402547 \\
    C & 0.000000 &    2.433546 &    0.708438 \\
    C & 0.000000 &    2.433546 & $-$0.708438 \\
    C & 0.000000 & $-$2.433546 &    0.708438 \\
    C & 0.000000 & $-$2.433546 & $-$0.708438 \\
    H & 0.000000 &    1.242227 &    2.490221 \\
    H & 0.000000 &    1.242227 & $-$2.490221 \\
    H & 0.000000 & $-$1.242227 &    2.490221 \\
    H & 0.000000 & $-$1.242227 & $-$2.490221 \\
    H & 0.000000 &    3.378258 &    1.245514 \\
    H & 0.000000 &    3.378258 & $-$1.245514 \\
    H & 0.000000 & $-$3.378258 &    1.245514 \\
    H & 0.000000 & $-$3.378258 & $-$1.245514 \\
    
    \hline
    \end{tabular*}
    \label{tab:tables2}
\end{table}

\begin{table}[H]
    \caption{Cartesian coordinates of the B3LYP/6-31G* optimized geometry of {\bf 3}, in units of {\r A}.}
    \begin{tabular*}{\linewidth}{@{\extracolsep{\fill}}crrr}
    \hline
    Atom & \multicolumn{1}{c}{X} & \multicolumn{1}{c}{Y} & \multicolumn{1}{c}{Z} \\
    \hline
    C & 0.000000 &    0.000000 &    1.403608 \\
    C & 0.000000 &    0.000000 & $-$1.403608 \\
    C & 0.000000 &    1.223844 &    0.722822 \\
    C & 0.000000 &    1.223844 & $-$0.722822 \\
    C & 0.000000 & $-$1.223844 &    0.722822 \\
    C & 0.000000 & $-$1.223844 & $-$0.722822 \\
    C & 0.000000 &    2.479411 &    1.407056 \\
    C & 0.000000 &    2.479411 & $-$1.407056 \\
    C & 0.000000 & $-$2.479411 &    1.407056 \\
    C & 0.000000 & $-$2.479411 & $-$1.407056 \\
    C & 0.000000 &    3.660505 &    0.713049 \\
    C & 0.000000 &    3.660505 & $-$0.713049 \\
    C & 0.000000 & $-$3.660505 &    0.713049 \\
    C & 0.000000 & $-$3.660505 & $-$0.713049 \\
    H & 0.000000 &    0.000000 &    2.491948 \\
    H & 0.000000 &    0.000000 & $-$2.491948 \\
    H & 0.000000 &    2.477069 &    2.494589 \\
    H & 0.000000 &    2.477069 & $-$2.494589 \\
    H & 0.000000 & $-$2.477069 &    2.494589 \\
    H & 0.000000 & $-$2.477069 & $-$2.494589 \\
    H & 0.000000 &    4.607098 &    1.246673 \\
    H & 0.000000 &    4.607098 & $-$1.246673 \\
    H & 0.000000 & $-$4.607098 &    1.246673 \\
    H & 0.000000 & $-$4.607098 & $-$1.246673 \\
    \hline
    \end{tabular*}
    \label{tab:tables3}
\end{table}

\begin{table}[H]
    \caption{Cartesian coordinates of the CASSCF(10e,10o)/6-31G* optimized geometry of {\bf 4}, in units of {\r A}.}
    \begin{tabular*}{\linewidth}{@{\extracolsep{\fill}}crrr}
    \hline
    Atom & \multicolumn{1}{c}{X} & \multicolumn{1}{c}{Y} & \multicolumn{1}{c}{Z} \\
    \hline
    C & 0.000000 &    0.000000 &    1.429275 \\
    C & 0.000000 &    0.000000 & $-$1.429275 \\
    C & 0.000000 &    1.259487 &    0.674849 \\
    C & 0.000000 &    1.259487 & $-$0.674849 \\
    C & 0.000000 & $-$1.259487 &    0.674849 \\
    C & 0.000000 & $-$1.259487 & $-$0.674849 \\
    N & 0.000000 &    0.000000 &    2.710369 \\
    N & 0.000000 &    0.000000 & $-$2.710369 \\
    H & 0.000000 &    2.176735 &    1.233246 \\
    H & 0.000000 &    2.176735 & $-$1.233246 \\
    H & 0.000000 & $-$2.176735 &    1.233246 \\
    H & 0.000000 & $-$2.176735 & $-$1.233246 \\
    \hline
    \end{tabular*}
    \label{tab:tables4}
\end{table}

\begin{table}[H]
    \caption{Cartesian coordinates of the planar geometry of {\bf 5} optimized at the B3LYP/6-31G* level, in units of {\r A}.}
    \begin{tabular*}{\linewidth}{@{\extracolsep{\fill}}crrr}
    \hline
    Atom & \multicolumn{1}{c}{X} & \multicolumn{1}{c}{Y} & \multicolumn{1}{c}{Z} \\
    \hline
    C & 0.000000 &    0.000000 &    3.236615 \\
    C & 0.000000 &    0.000000 &    1.918246 \\
    C & 0.000000 &    0.000000 &    0.645266 \\
    C & 0.000000 &    0.000000 & $-$0.645266 \\
    C & 0.000000 &    0.000000 & $-$1.918246 \\
    C & 0.000000 &    0.000000 & $-$3.236615 \\
    H & 0.000000 &    0.927096 &    3.807401 \\
    H & 0.000000 &    0.927096 & $-$3.807401 \\
    H & 0.000000 & $-$0.927096 &    3.807401 \\
    H & 0.000000 & $-$0.927096 & $-$3.807401 \\
    \hline
    \end{tabular*}
    \label{tab:tables5}
\end{table}

\begin{table}[H]
    \caption{Cartesian coordinates of the twisted geometry of {\bf 5}, in units of {\r A}.}
    \begin{tabular*}{\linewidth}{@{\extracolsep{\fill}}crrr}
    \hline
    Atom & \multicolumn{1}{c}{X} & \multicolumn{1}{c}{Y} & \multicolumn{1}{c}{Z} \\
    \hline
    C &    0.000000 &    0.000000 &    3.236615 \\
    C &    0.000000 &    0.000000 &    1.918246 \\
    C &    0.000000 &    0.000000 &    0.645266 \\
    C &    0.000000 &    0.000000 & $-$0.645266 \\
    C &    0.000000 &    0.000000 & $-$1.918246 \\
    C &    0.000000 &    0.000000 & $-$3.236615 \\
    H &    0.000000 &    0.927096 &    3.807401 \\
    H &    0.927096 &    0.000000 & $-$3.807401 \\
    H &    0.000000 & $-$0.927096 &    3.807401 \\
    H & $-$0.927096 &    0.000000 & $-$3.807401 \\
    \hline
    \end{tabular*}
    \label{tab:tables6}
\end{table}

\begin{table}[H]
    \caption{Cartesian coordinates of the planar geometry of {\bf 6} optimized at the B3LYP/6-31G* level, in units of {\r A}.}
    \begin{tabular*}{\linewidth}{@{\extracolsep{\fill}}crrr}
    \hline
    Atom & \multicolumn{1}{c}{X} & \multicolumn{1}{c}{Y} & \multicolumn{1}{c}{Z} \\
    \hline
    C & 0.000000 &    0.000000 &    4.519296 \\
    C & 0.000000 &    0.000000 &    3.200579 \\
    C & 0.000000 &    0.000000 &    1.926668 \\
    C & 0.000000 &    0.000000 &    0.637785 \\
    C & 0.000000 &    0.000000 & $-$0.637785 \\
    C & 0.000000 &    0.000000 & $-$1.926668 \\
    C & 0.000000 &    0.000000 & $-$3.200579 \\
    C & 0.000000 &    0.000000 & $-$4.519296 \\
    H & 0.000000 &    0.927134 &    5.090044 \\
    H & 0.000000 &    0.927134 & $-$5.090044 \\
    H & 0.000000 & $-$0.927134 &    5.090044 \\
    H & 0.000000 & $-$0.927134 & $-$5.090044 \\
    \hline
    \end{tabular*}
    \label{tab:tables7}
\end{table}

\begin{table}[H]
    \caption{Cartesian coordinates of the twisted geometry of {\bf 6}, in units of {\r A}.}
    \begin{tabular*}{\linewidth}{@{\extracolsep{\fill}}crrr}
    \hline
    Atom & \multicolumn{1}{c}{X} & \multicolumn{1}{c}{Y} & \multicolumn{1}{c}{Z} \\
    \hline
    C &    0.000000 &    0.000000 &    4.519296 \\
    C &    0.000000 &    0.000000 &    3.200579 \\
    C &    0.000000 &    0.000000 &    1.926668 \\
    C &    0.000000 &    0.000000 &    0.637785 \\
    C &    0.000000 &    0.000000 & $-$0.637785 \\
    C &    0.000000 &    0.000000 & $-$1.926668 \\
    C &    0.000000 &    0.000000 & $-$3.200579 \\
    C &    0.000000 &    0.000000 & $-$4.519296 \\
    H &    0.000000 &    0.927134 &    5.090044 \\
    H &    0.927134 &    0.000000 & $-$5.090044 \\
    H &    0.000000 & $-$0.927134 &    5.090044 \\
    H & $-$0.927134 &    0.000000 & $-$5.090044 \\
    \hline
    \end{tabular*}
    \label{tab:tables8}
\end{table}

\begin{table}[H]
    \caption{Cartesian coordinates of the planar geometry of {\bf 7} optimized at the B3LYP/6-31G* level, in units of {\r A}.}
    \begin{tabular*}{\linewidth}{@{\extracolsep{\fill}}crrr}
    \hline
    Atom & \multicolumn{1}{c}{X} & \multicolumn{1}{c}{Y} & \multicolumn{1}{c}{Z} \\
    \hline
    C & 0.000000 &    0.000000 &    5.801617 \\
    C & 0.000000 &    0.000000 &    4.482729 \\
    C & 0.000000 &    0.000000 &    3.208380 \\
    C & 0.000000 &    0.000000 &    1.920044 \\
    C & 0.000000 &    0.000000 &    0.643194 \\
    C & 0.000000 &    0.000000 & $-$0.643194 \\
    C & 0.000000 &    0.000000 & $-$1.920044 \\
    C & 0.000000 &    0.000000 & $-$3.208380 \\
    C & 0.000000 &    0.000000 & $-$4.482729 \\
    C & 0.000000 &    0.000000 & $-$5.801617 \\
    H & 0.000000 &    0.927308 &    6.372154 \\
    H & 0.000000 &    0.927308 & $-$6.372154 \\
    H & 0.000000 & $-$0.927308 &    6.372154 \\
    H & 0.000000 & $-$0.927308 & $-$6.372154 \\
    \hline
    \end{tabular*}
    \label{tab:tables9}
\end{table}

\begin{table}[H]
    \caption{Cartesian coordinates of the twisted geometry of {\bf 7}, in units of {\r A}.}
    \begin{tabular*}{\linewidth}{@{\extracolsep{\fill}}crrr}
    \hline
    Atom & \multicolumn{1}{c}{X} & \multicolumn{1}{c}{Y} & \multicolumn{1}{c}{Z} \\
    \hline
    C &    0.000000 &    0.000000 &    5.801617 \\
    C &    0.000000 &    0.000000 &    4.482729 \\
    C &    0.000000 &    0.000000 &    3.208380 \\
    C &    0.000000 &    0.000000 &    1.920044 \\
    C &    0.000000 &    0.000000 &    0.643194 \\
    C &    0.000000 &    0.000000 & $-$0.643194 \\
    C &    0.000000 &    0.000000 & $-$1.920044 \\
    C &    0.000000 &    0.000000 & $-$3.208380 \\
    C &    0.000000 &    0.000000 & $-$4.482729 \\
    C &    0.000000 &    0.000000 & $-$5.801617 \\
    H &    0.000000 &    0.927308 &    6.372154 \\
    H &    0.927308 &    0.000000 & $-$6.372154 \\
    H &    0.000000 & $-$0.927308 &    6.372154 \\
    H & $-$0.927308 &    0.000000 & $-$6.372154 \\
    \hline
    \end{tabular*}
    \label{tab:tables10}
\end{table}

\subsection{The CAS-CI wave function}

\begin{table}[H]
    \caption{CAS-CI(6e,6o)/STO-3G wave function of the S$_0$ state of {\bf 1}. Only the Slater determinants having the coefficients larger than 0.05 in absolute value are listed.}
    \begin{tabular*}{\linewidth}{@{\extracolsep{\fill}}ccr}
    \hline
    Alpha & Beta & \multicolumn{1}{c}{Coefficients}\\
    \hline
    111000 & 111000 & 0.9166\\
    110100 & 110100 & $-$0.1733\\
    101010 & 101010 & $-$0.1733\\
    110010 & 101100 & $-$0.1153 \\
    101100 & 110010 & $-$0.1153 \\
    111000 & 100110 & 0.0829 \\
    100110 & 111000 & 0.0829 \\
    110001 & 011100 & 0.0776 \\ 
    011100 & 110001 & 0.0776 \\
    101001 & 011010 & 0.0776 \\
    011010 & 101001 & 0.0776 \\
    101010 & 011001 & 0.0558 \\
    011001 & 101010 & 0.0558 \\
    110100 & 011001 & 0.0558 \\
    011001 & 110100 & 0.0558 \\
    \hline
    \end{tabular*}
    \label{tab:tables11}
\end{table}

\begin{table}[H]
    \caption{CAS-CI(6e,6o)/STO-3G wave function of the T$_1$ state of {\bf 1}. Only the Slater determinants having the coefficients larger than 0.05 in absolute value are listed.}
    \begin{tabular*}{\linewidth}{@{\extracolsep{\fill}}ccr}
    \hline
    Alpha & Beta & \multicolumn{1}{c}{Coefficients}\\
    \hline
    111100 & 110000 & 0.6456 \\
    111010 & 101000 & 0.6456 \\
    111001 & 011000 & $-$0.2225 \\
    101110 & 100010 & $-$0.1341 \\
    110110 & 100100 & $-$0.1341 \\
    011110 & 100001 & 0.0872 \\
    110101 & 010100 & 0.0811 \\
    101011 & 001010 & 0.0811 \\
    101101 & 010010 & 0.0795 \\
    110011 & 001100 & 0.0795 \\
    111100 & 010100 & 0.0577 \\
    111010 & 010010 & $-$0.0577 \\
    111100 & 001010 & $-$0.0577 \\
    111010 & 001100 & $-$0.0577 \\
    110101 & 110000 & 0.0573 \\
    101101 & 101000 & $-$0.0573 \\
    101011 & 110000 & $-$0.0573 \\
    110011 & 101000 & $-$0.0573 \\
    \hline
    \end{tabular*}
    \label{tab:tables12}
\end{table}

\begin{table}[H]
    \caption{CAS-CI(10e,10o)/STO-3G wave function of the S$_0$ state of {\bf 2}. Only the Slater determinants having the coefficients larger than 0.05 in absolute value are listed.}
    \begin{tabular*}{\linewidth}{@{\extracolsep{\fill}}ccr}
    \hline
    Alpha & Beta & \multicolumn{1}{c}{Coefficients}\\
    \hline
    1111100000 & 1111100000 & 0.8615 \\
    1111010000 & 1111010000 & $-$0.1498 \\
    1110101000 & 1110101000 & $-$0.1338 \\
    1111001000 & 1110110000 & 0.0973 \\
    1110110000 & 1111001000 & 0.0973 \\
    1111000100 & 1101110000 & $-$0.0856 \\
    1101110000 & 1111000100 & $-$0.0856 \\
    1111010000 & 1101100100 & $-$0.0780 \\
    1101100100 & 1111010000 & $-$0.0780 \\
    1111100000 & 1110011000 & $-$0.0735 \\
    1110011000 & 1111100000 & $-$0.0735 \\
    1110100100 & 1101101000 & 0.0706 \\
    1101101000 & 1110100100 & 0.0706 \\
    1111000010 & 1011110000 & $-$0.0692 \\
    1011110000 & 1111000010 & $-$0.0692 \\
    1101100100 & 1101100100 & $-$0.0671 \\
    1101110000 & 1101110000 & $-$0.0652 \\
    1111000100 & 1111000100 & $-$0.0625 \\
    1110100010 & 1011101000 & 0.0587 \\
    1011101000 & 1110100010 & 0.0587 \\
    1110100001 & 0111101000 & $-$0.0501 \\
    0111101000 & 1110100001 & $-$0.0501 \\
    \hline
    \end{tabular*}
    \label{tab:tables13}
\end{table}

\begin{table}[H]
    \caption{CAS-CI(10e,10o)/STO-3G wave function of the T$_1$ state of {\bf 2}. Only the Slater determinants having the coefficients larger than 0.05 in absolute value are listed.}
    \begin{tabular*}{\linewidth}{@{\extracolsep{\fill}}ccr}
    \hline
    Alpha & Beta & \multicolumn{1}{c}{Coefficients}\\
    \hline
    1111110000 & 1111000000 & 0.7111 \\
    1111101000 & 1110100000 & $-$0.3703 \\
    1111100100 & 1101100000 & 0.2860 \\
    1111100010 & 1011100000 & 0.1859 \\
    1111100001 & 0111100000 & $-$0.1176 \\
    1110111000 & 1110001000 & $-$0.1101 \\
    1111010100 & 1101010000 & $-$0.1016 \\
    1111011000 & 1110010000 & 0.0952 \\
    1110110100 & 1101001000 & 0.0840 \\
    1101110100 & 1101000100 & $-$0.0770 \\
    1111010010 & 1011010000 & $-$0.0715 \\
    1101111000 & 1110000100 & 0.0690 \\
    1110110010 & 1011001000 & 0.0668 \\
    1011111000 & 1110000010 & 0.0637 \\
    1111110000 & 1101001000 & 0.0599 \\
    1110110100 & 1111000000 & 0.0583 \\
    1110101100 & 1100101000 & $-$0.0556 \\
    1111110000 & 1011010000 & $-$0.0555 \\
    1111001100 & 1100110000 & 0.0551 \\
    1111010010 & 1111000000 & 0.0542 \\
    1110101010 & 1010101000 & $-$0.0535 \\
    1111101000 & 1011001000 & 0.0511 \\
    1110110010 & 1110100000 & $-$0.0505 \\
    1011110010 & 1011000010 & $-$0.0503 \\
    \hline
    \end{tabular*}
    \label{tab:tables14}
\end{table}

\begin{table}[H]
    \caption{CAS-CI(14e,14o)/STO-3G wave function of the S$_0$ state of {\bf 3}. Only the Slater determinants having the coefficients larger than 0.05 in absolute value are listed.}
    \begin{tabular*}{\linewidth}{@{\extracolsep{\fill}}ccr}
    \hline
    Alpha & Beta & \multicolumn{1}{c}{Coefficients}\\
    \hline
    11111110000000 & 11111110000000 & 0.8021 \\
    11111101000000 & 11111101000000 & $-$0.1588 \\
    11111010100000 & 11111010100000 & $-$0.1066 \\
    11111100100000 & 11111011000000 & 0.0891 \\
    11111011000000 & 11111100100000 & 0.0891 \\
    11110110010000 & 11110110010000 & $-$0.0741 \\
    11111100010000 & 11110111000000 & $-$0.0719 \\
    11110111000000 & 11111100010000 & $-$0.0719 \\
    11111110000000 & 11111001100000 & $-$0.0681 \\
    11111001100000 & 11111110000000 & $-$0.0681 \\
    11111100001000 & 11101111000000 & 0.0673 \\
    11101111000000 & 11111100001000 & 0.0673 \\
    11111100000100 & 11011111000000 & $-$0.0627 \\
    11011111000000 & 11111100000100 & $-$0.0627 \\
    11111101000000 & 11110110010000 & $-$0.0617 \\
    11110110010000 & 11111101000000 & $-$0.0617 \\
    11111101000000 & 11101110001000 & 0.0579 \\
    11101110001000 & 11111101000000 & 0.0579 \\
    11111010010000 & 11110110100000 & 0.0552 \\
    11110110100000 & 11111010010000 & 0.0552 \\
    11111010001000 & 11101110100000 & $-$0.0514 \\
    11101110100000 & 11111010001000 & $-$0.0514 \\
    \hline
    \end{tabular*}
    \label{tab:tables15}
\end{table}

\begin{table}[H]
    \caption{CAS-CI(14e,14o)/STO-3G wave function of the T$_1$ state of {\bf 3}. Only the Slater determinants having the coefficients larger than 0.05 in absolute value are listed.}
    \begin{tabular*}{\linewidth}{@{\extracolsep{\fill}}ccr}
    \hline
    Alpha & Beta & \multicolumn{1}{c}{Coefficients}\\
    \hline
    11111111000000 & 11111100000000 & 0.7043 \\
    11111110010000 & 11110110000000 & 0.2348 \\
    11111110100000 & 11111010000000 & $-$0.2284 \\
    11111110001000 & 11101110000000 & $-$0.1778 \\
    11111110000100 & 11011110000000 & 0.1439 \\
    11111011100000 & 11111000100000 & $-$0.0886 \\
    11111110000010 & 10111110000000 & 0.0886 \\
    11111101010000 & 11110101000000 & $-$0.0865 \\
    11111110000001 & 01111110000000 & 0.0771 \\
    11110111010000 & 11110100010000 & $-$0.0767 \\
    11111101100000 & 11111001000000 & 0.0749 \\
    11111011010000 & 11110100100000 & 0.0722 \\
    11111101001000 & 11101101000000 & 0.0688 \\
    11111101000100 & 11011101000000 & $-$0.0634 \\
    11111111000000 & 11011101000000 & $-$0.0598 \\
    11111101000100 & 11111100000000 & 0.0584 \\
    11111111000000 & 11110100100000 & $-$0.0581 \\
    11111110001000 & 11111100000000 & 0.0581 \\
    11111111000000 & 11101110000000 & $-$0.0581 \\
    11111011000100 & 11011100100000 & 0.0571 \\
    11111011010000 & 11111100000000 & $-$0.0562 \\
    11110111100000 & 11111000010000 & 0.0542 \\
    11111011001000 & 11101100100000 & $-$0.0533 \\
    11011111100000 & 11111000000100 & 0.0503 \\
    \hline
    \end{tabular*}
    \label{tab:tables16}
\end{table}

\begin{table}[H]
    \caption{CAS-CI(10e,10o)/6-31G* wave function of the S$_0$ state of {\bf 4}. Only the Slater determinants having the coefficients larger than 0.05 in absolute value are listed.}
    \begin{tabular*}{\linewidth}{@{\extracolsep{\fill}}ccr}
    \hline
    Alpha & Beta & \multicolumn{1}{c}{Coefficients}\\
    \hline
    1111100000 & 1111100000 & 0.6291 \\
    1111010000 & 1111010000 & $-$0.6134 \\
    1110101000 & 1110101000 & $-$0.1358 \\
    1110011000 & 1110011000 & 0.1346 \\
    1111001000 & 1110110000 & 0.0926 \\
    1110110000 & 1111001000 & 0.0926 \\
    1101100100 & 1101100100 & $-$0.0788 \\
    1101010100 & 1101010100 & 0.0769 \\
    1110100010 & 1011101000 & $-$0.0666 \\
    1011101000 & 1110100010 & $-$0.0666 \\
    1110010010 & 1011011000 & 0.0653 \\
    1011011000 & 1110010010 & 0.0653 \\
    1011100010 & 1011100010 & $-$0.0542 \\
    1011010010 & 1011010010 & 0.0531 \\
    1101100100 & 1110101000 & $-$0.0513 \\
    1110101000 & 1101100100 & $-$0.0513 \\
    1110100100 & 1101101000 & $-$0.0510 \\
    1101101000 & 1110100100 & $-$0.0510 \\
    1110011000 & 1101010100 & 0.0504 \\
    1101010100 & 1110011000 & 0.0504 \\
    \hline
    \end{tabular*}
    \label{tab:tables17}
\end{table}

\begin{table}[H]
    \caption{CAS-CI(10e,10o)/6-31G* wave function of the T$_1$ state of {\bf 4}. Only the Slater determinants having the coefficients larger than 0.05 in absolute value are listed.}
    \begin{tabular*}{\linewidth}{@{\extracolsep{\fill}}ccr}
    \hline
    Alpha & Beta & \multicolumn{1}{c}{Coefficients}\\
    \hline
    1111110000 & 1111000000 & 0.8848 \\
    1110111000 & 1110001000 & $-$0.1835 \\
    1101110100 & 1101000100 & $-$0.1105 \\
    1011111000 & 1110000010 & $-$0.0967 \\
    1110110010 & 1011001000 & 0.0891 \\
    1111101000 & 1110100000 & $-$0.0840 \\
    1111011000 & 1110010000 & 0.0824 \\
    1101111000 & 1110000100 & $-$0.0758 \\
    1011110010 & 1011000010 & $-$0.0742 \\
    1101110100 & 1110001000 & $-$0.0718 \\
    1110111000 & 1101000100 & $-$0.0718 \\
    1110110100 & 1101001000 & $-$0.0671 \\
    1011111000 & 1011001000 & $-$0.0670 \\
    1011110010 & 1110001000 & $-$0.0644 \\
    1110111000 & 1011000010 & $-$0.0633 \\
    0111111000 & 1110000001 & $-$0.0554 \\
    1101111000 & 1101001000 & $-$0.0525 \\
    1110110001 & 0111001000 & $-$0.0500 \\
    \hline
    \end{tabular*}
    \label{tab:tables18}
\end{table}

\begin{table}[H]
    \caption{CAS-CI(10e,10o)/6-31G* wave function of the S$_0$ state of {\bf 5} in planar geometry. Only the Slater determinants having the coefficients larger than 0.05 in absolute value are listed.}
    \begin{tabular*}{\linewidth}{@{\extracolsep{\fill}}ccr}
    \hline
    Alpha & Beta & \multicolumn{1}{c}{Coefficients}\\
    \hline
    1111100000 & 1111100000 & 0.8943 \\
    1111000001 & 1111000001 & $-$0.1638 \\
    0111110000 & 0111110000 & $-$0.1638 \\
    1101100100 & 1101100100 & $-$0.1349 \\
    1011101000 & 1011101000 & $-$0.1266 \\
    1110100010 & 1110100010 & $-$0.1266 \\
    1101100001 & 1111000100 & 0.0503 \\
    1111000100 & 1101100001 & 0.0503 \\
    0111100100 & 1101110000 & $-$0.0503 \\
    1101110000 & 0111100100 & $-$0.0503 \\
    \hline
    \end{tabular*}
    \label{tab:tables19}
\end{table}

\begin{table}[H]
    \caption{CAS-CI(10e,10o)/6-31G* wave function of the T$_1$ state of {\bf 5} in planar geometry. Only the Slater determinants having the coefficients larger than 0.05 in absolute value are listed.}
    \begin{tabular*}{\linewidth}{@{\extracolsep{\fill}}ccr}
    \hline
    Alpha & Beta & \multicolumn{1}{c}{Coefficients}\\
    \hline
    1111100100 & 1101100000 & 0.4112 \\
    1111110000 & 0111100000 & 0.3648 \\
    1111100001 & 1111000000 & $-$0.3648 \\
    1111100100 & 1111000000 & $-$0.2693 \\
    1111100100 & 0111100000 & $-$0.2693 \\
    1111110000 & 1101100000 & $-$0.2677 \\
    1111100001 & 1101100000 & 0.2677 \\
    1111110000 & 1111000000 & 0.1912 \\
    1111100001 & 0111100000 & $-$0.1912 \\
    1111100100 & 1101000001 & 0.0651 \\
    1111100100 & 0101110000 & $-$0.0651 \\
    1111000101 & 1101000001 & 0.0643 \\
    0111110100 & 0101110000 & 0.0643 \\
    1111100001 & 1101000100 & 0.0635 \\
    1111110000 & 0101100100 & $-$0.0635 \\
    1111010001 & 0111000001 & 0.0621 \\
    0111110001 & 0111010000 & $-$0.0621 \\
    1011101100 & 1001101000 & 0.0585 \\
    1110100110 & 1100100010 & 0.0585 \\
    1101100101 & 1111000000 & 0.0569 \\
    1101110100 & 0111100000 & $-$0.0569 \\
    1111000101 & 1101100000 & 0.0554 \\
    0111110100 & 1101100000 & $-$0.0554 \\
    1111110000 & 0111000100 & 0.0528 \\
    1111100001 & 0111000100 & $-$0.0528 \\
    1111100001 & 0111010000 & 0.0518 \\
    1111110000 & 0111000001 & 0.0518 \\
    1110110010 & 0110100010 & 0.0512 \\
    1011101001 & 1011001000 & $-$0.0512 \\
    1111100001 & 1101000001 & 0.0508 \\
    1111110000 & 0101110000 & 0.0508 \\
    1011111000 & 0011101000 & 0.0503 \\
    1110100011 & 1110000010 & $-$0.0503 \\
    \hline
    \end{tabular*}
    \label{tab:tables20}
\end{table}

\begin{table}[H]
    \caption{CAS-CI(10e,10o)/6-31G* wave function of the S$_0$ state of {\bf 5} in twisted geometry. Only the Slater determinants having the coefficients larger than 0.05 in absolute value are listed.}
    \begin{tabular*}{\linewidth}{@{\extracolsep{\fill}}ccr}
    \hline
    Alpha & Beta & \multicolumn{1}{c}{Coefficients}\\
    \hline
    1111100000 & 1111100000 & 0.8519 \\
    1111000001 & 1111000001 & $-$0.1630 \\
    0111110000 & 0111110000 & $-$0.1630 \\
    1101100100 & 1101100100 & $-$0.1431 \\
    1011101000 & 1011101000 & $-$0.1295 \\
    1110100010 & 1110100010 & $-$0.1295 \\
    1101110000 & 0111100100 & 0.0618 \\
    0111100100 & 1101110000 & 0.0618 \\
    1111000100 & 1101100001 & $-$0.0618 \\
    1101100001 & 1111000100 & $-$0.0618 \\
    \hline
    \end{tabular*}
    \label{tab:tables21}
\end{table}

\begin{table}[H]
    \caption{CAS-CI(10e,10o)/6-31G* wave function of the T$_1$ state of {\bf 5} in twisted geometry. Only the Slater determinants having the coefficients larger than 0.05 in absolute value are listed.}
    \begin{tabular*}{\linewidth}{@{\extracolsep{\fill}}ccr}
    \hline
    Alpha & Beta & \multicolumn{1}{c}{Coefficients}\\
    \hline
    1111100100 & 1101100000 & 0.3522 \\
    1111100001 & 1111000000 & 0.2650 \\
    1111110000 & 0111100000 & $-$0.2650 \\
    1111100100 & 1111000000 & $-$0.2209 \\
    1111100100 & 0111100000 & $-$0.2209 \\
    1111100001 & 1101100000 & $-$0.2074 \\
    1111110000 & 1101100000 & 0.2074 \\
    1111100100 & 1011100000 & 0.1938 \\
    1111100100 & 1110100000 & 0.1938 \\
    1111100010 & 1110100000 & $-$0.1625 \\
    1111101000 & 1011100000 & 0.1625 \\
    1111110000 & 1011100000 & 0.1498 \\
    1111100001 & 1110100000 & $-$0.1498 \\
    1111100010 & 1101100000 & $-$0.1471 \\
    1111101000 & 1101100000 & 0.1471 \\
    1111100001 & 0111100000 & 0.1437 \\
    1111110000 & 1111000000 & $-$0.1437 \\
    1111100001 & 1011100000 & $-$0.1228 \\
    1111110000 & 1110100000 & 0.1228 \\
    1111100010 & 1111000000 & 0.1221 \\
    1111101000 & 0111100000 & $-$0.1221 \\
    1111100010 & 1011100000 & $-$0.1088 \\
    1111101000 & 1110100000 & 0.1088 \\
    1111100010 & 0111100000 & 0.0962 \\
    1111101000 & 1111000000 & $-$0.0962 \\
    1111000101 & 1101000001 & 0.0562 \\
    0111110100 & 0101110000 & 0.0562 \\
    1011101100 & 1001101000 & 0.0513 \\
    1110100110 & 1100100010 & 0.0513 \\
    1111100100 & 0101110000 & $-$0.0511 \\
    1111100100 & 1101000001 & 0.0511 \\
    \hline
    \end{tabular*}
    \label{tab:tables22}
\end{table}

\subsection{HSB-QSCI results calculated using \texttt{kernel\_fixed\_space} subroutine in PySCF}
\begin{figure}[H]
    \centering
    \includegraphics[width=\linewidth]{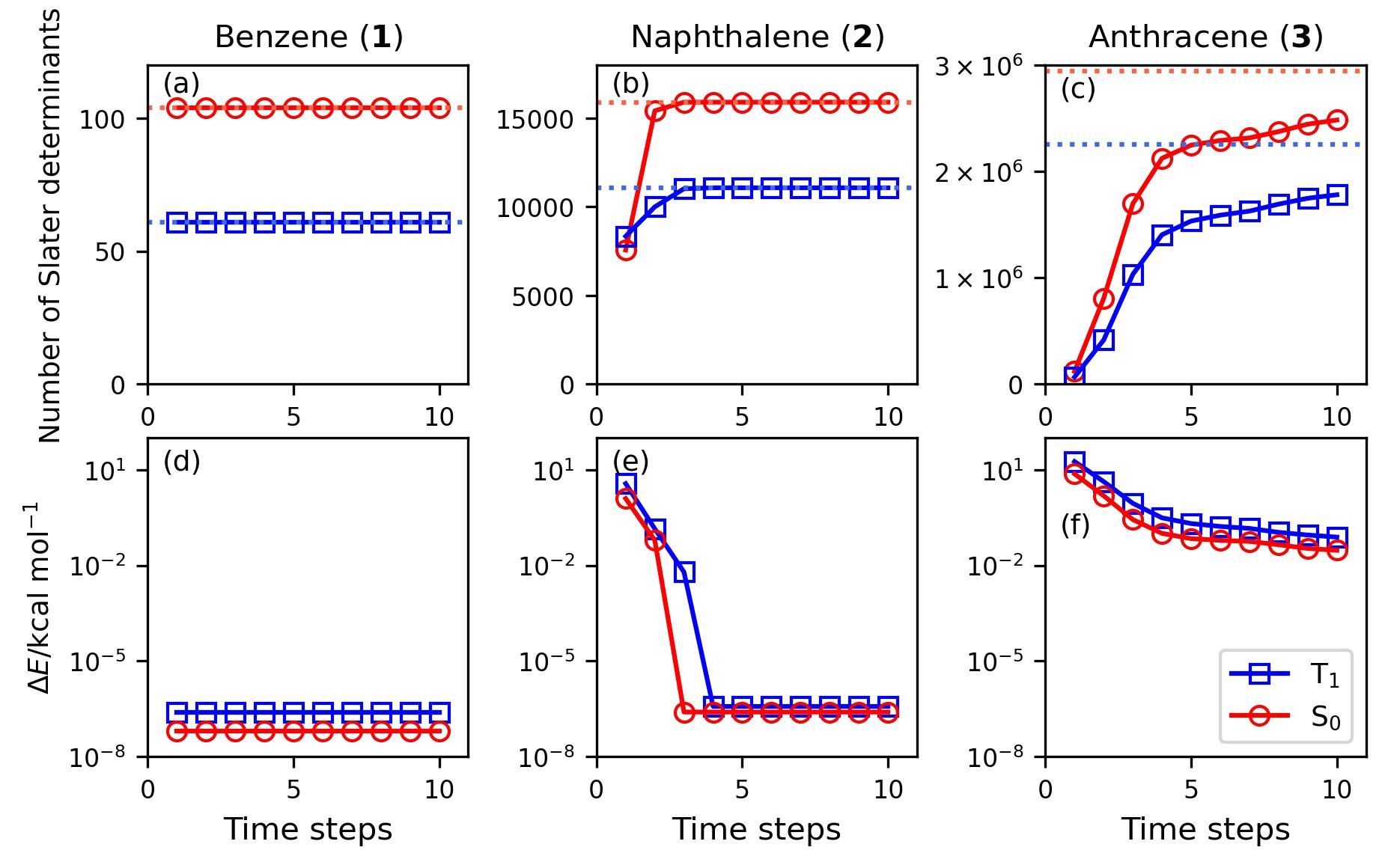}
    \caption{HSB-QSCI results of oligoacenes, calculated using \texttt{kernel\_fixed\_space} subroutine in PySCF. The number of Slater determinants included in the Hamiltonian diagonalization is given in (a), (b), and (c) for {\bf 1}, {\bf 2}, and {\bf 3}, respectively. Red and blue indicate the spin-singlet ground and spin-triplet excited states, respectively. Dotted lines represent the number of Slater determinants in the CAS-CI wave function. The difference of the HSB-QSCI energy from the CAS-CI values in units of kcal mol$^{-1}$ are given in (d), (e), and (f) on the logarithmic scale.}
    \label{fig:s_1}
\end{figure}

\begin{figure}[H]
    \centering
    \includegraphics[width=\linewidth]{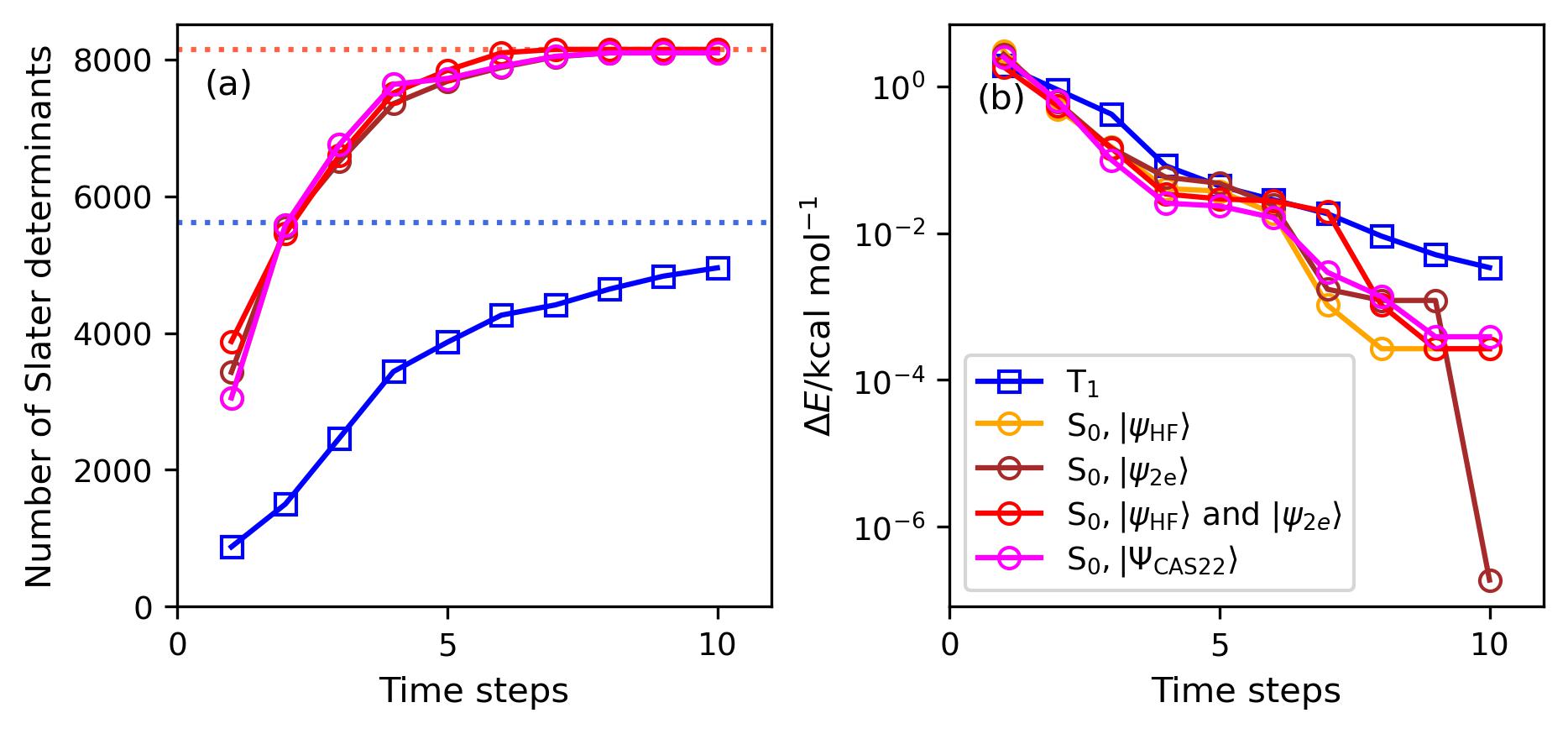}
    \caption{HSB-QSCI results of {\bf 4}, calculated using \texttt{kernel\_fixed\_space} subroutine in PySCF. (a) The number of Slater determinants sampled from Hamiltonian simulations. The red and blue dotted lines indicate the number of Slater determinants in the CAS-CI wave function. (b) The difference of the HSB-QSCI energy from the CAS-CI values in units of kcal mol$^{-1}$, on the logarithmic scale.}
    \label{fig:s_2}
\end{figure}

\begin{figure}[H]
    \centering
    \includegraphics[width=\linewidth]{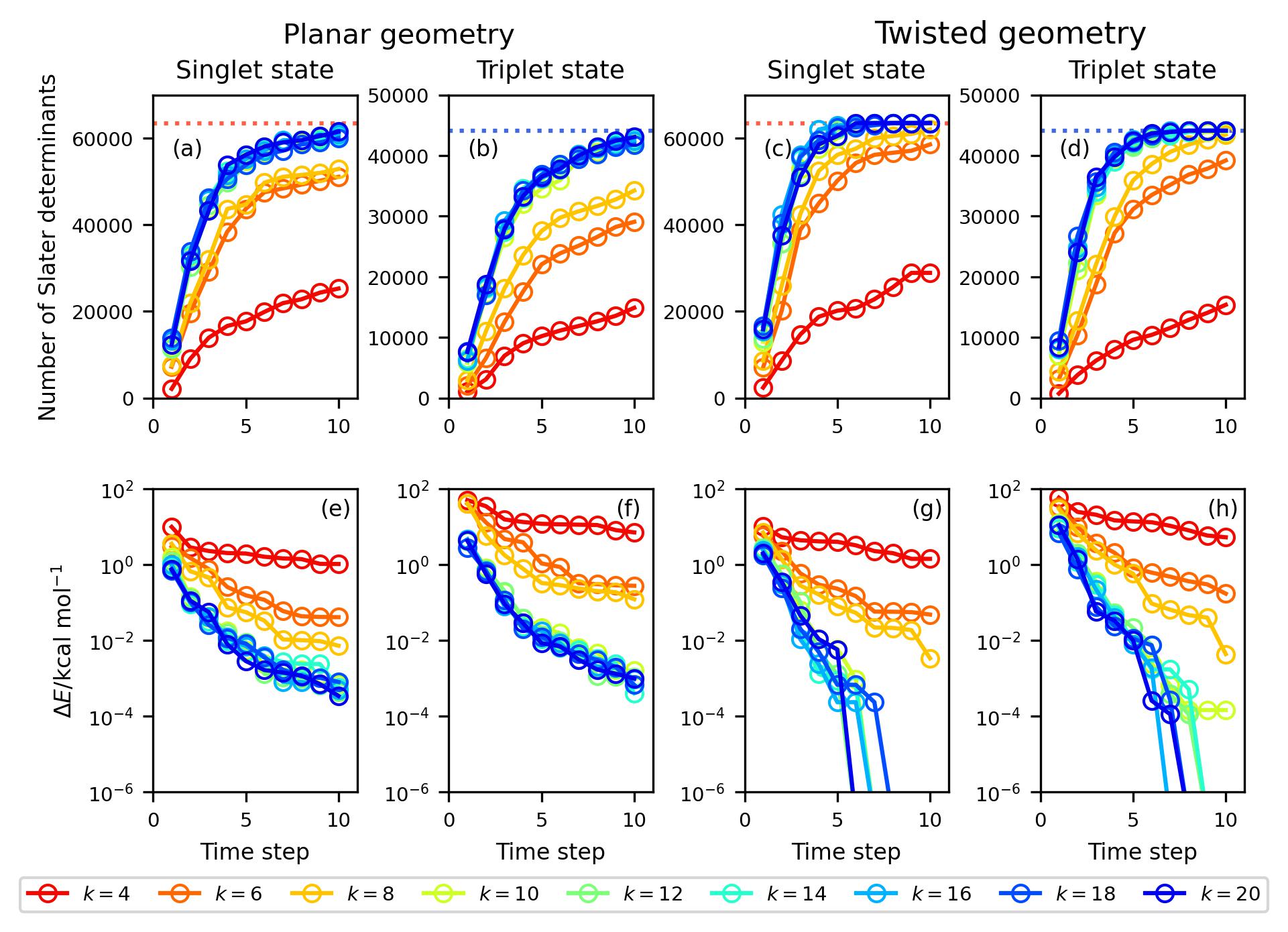}
    \caption{HSB-QSCI results of {\bf 5} in its planar and twisted geometries with different maximum locality values, calculated using \texttt{kernel\_fixed\_space} subroutine in PySCF. The number of Slater determinants included in the Hamiltonian diagonalization is given in (a), (b), (c), and (d). The red and blue dotted lines represent the number of Slater determinants in the CAS-CI wave function of the S$_0$ and T$_1$ states, respectively. The difference of the HSB-QSCI energy from the CAS-CI values in units of kcal mol$^{-1}$ are given in (e), (f), (g), and (h) on the logarithmic scale.}
    \label{fig:s_3}
\end{figure}

\subsection{Shot number dependence on the HSB-QSCI simulations of {\bf 3}}

\begin{figure}[H]
    \centering
    \includegraphics[width=\linewidth]{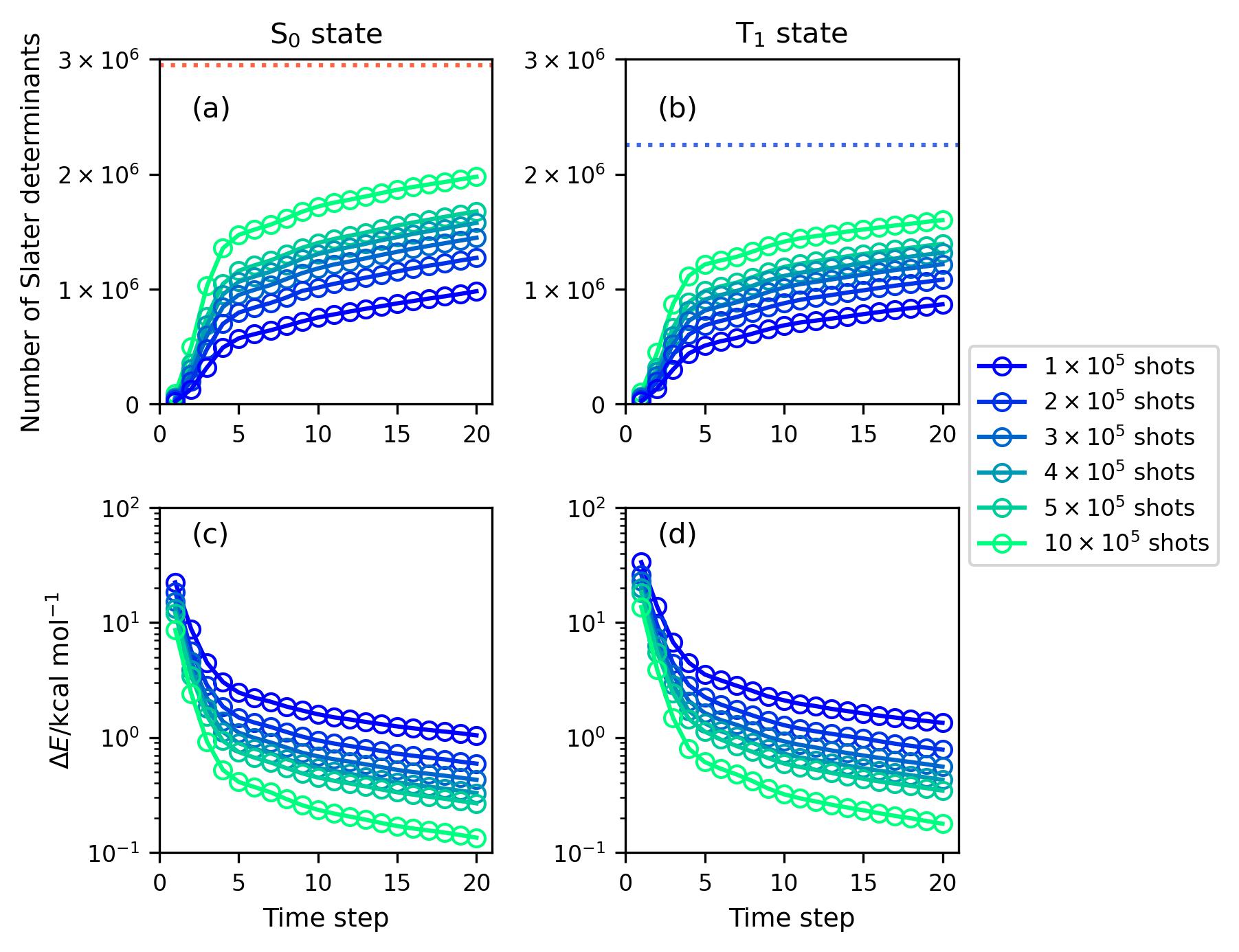}
    \caption{Shot number and time step dependence on the HSB-QSCI simulations of {\bf 3}. The number of Slater determinants sampled from Hamiltonian simulations is plotted in (a) and (b) for the S$_0$ and T$_1$ states, respectively. The red and blue dotted lines indicate the number of Slater determinants in the CAS-CI wave function. (b) The difference of the HSB-QSCI energy from the CAS-CI values in units of kcal mol$^{-1}$, in log scale is given in (c) and (d) for the S$_0$ and T$_1$ states, respectively.}
    \label{fig:s11}
\end{figure}

\begin{table*}[ht]
\caption{\label{tab:table23} The HF energy, number of Slater determinants and total energies of the CAS-CI, and the number of selected Slater determinants and the error of the HSB-QSCI energy calculated using \texttt{ibm\_kawasaki} and \texttt{kernel\_fixed\_space} in PySCF}.
\begin{tabular*}{\textwidth}{@{\extracolsep{\fill}}ccccccc}
\hline
\multirow{2}{*}{System} & HF          & \multicolumn{2}{c}{CAS-CI} & \multicolumn{3}{c}{HSB-QSCI} \\
                        & $E$/Hartree & \#Dets  &  $E$/Hartree     & \%\#Dets$^a$ & $\Delta E/\mathrm{kcal\ mol^{-1}}$ & \%$E_\text{Corr}$$^{b}$ \\
\hline
{\bf 5} (Planar, S$_0$)  & $-226.4832222338$ &      63504 & $-229.5024030030$ & 100.00 & 0.00 & 100.00 \\
{\bf 5} (Planar, T$_1$)  & $-226.4299826264$ &      44100 & $-229.4164659619$ & 100.00 & 0.00 & 100.00 \\
{\bf 5} (Twisted, S$_0$) & $-226.4033087159$ &      63504 & $-229.4452184415$ & 100.00 & 0.00 & 100.00 \\
{\bf 5} (Twisted, T$_1$) & $-226.4440971899$ &      44100 & $-229.4129399110$ & 100.00 & 0.01 & 100.00 \\
{\bf 6} (Planar, S$_0$)  & $-301.1974823957$ &   11778624 & $-305.2359857689$ & 100.00 & 0.00 & 100.00 \\
{\bf 6} (Planar, T$_1$)  & $-301.1536301615$ &    9018009 & $-305.1666727092$ &  95.50 & 0.01 & 100.00 \\
{\bf 6} (Twisted, S$_0$) & $-301.1365655879$ &   11778624 & $-305.1941677868$ &  99.83 & 0.00 & 100.00 \\
{\bf 6} (Twisted, T$_1$) & $-301.1634490372$ &    9018009 & $-305.1698864405$ &  98.57 & 0.00 & 100.00 \\
{\bf 7} (Planar, S$_0$)  & $-375.9128437860$ & 2363904400 & $-380.9720622554$ &  48.29 & 0.31 &  99.99 \\
{\bf 7} (Planar, T$_1$)  & $-375.8222817220$ & 1914762564 & $-380.9133659533$ &  27.48 & 3.70 &  99.88 \\
{\bf 7} (Twisted, S$_0$) & $-375.8635891161$ & 2363904400 & $-380.9394312061$ &  46.39 & 0.30 &  99.99 \\
{\bf 7} (Twisted, T$_1$) & $-375.8817559215$ & 1914762564 & $-380.9186151034$ &  28.96 & 2.40 &  99.92 \\
\hline
\end{tabular*}
$^a$ Percentage of the number of Slater determinants calculated as 100 $\times \frac{\#Dets(\mathrm{HSB\mathchar`-QSCI})}{\#Dets(\mathrm{CAS\mathchar`-CI})}.$ \\
$^b$ Percentage of the correlation energy in the active space considered in the HSB-QSCI, calculated as 100 $\times$ $\frac{E(\mathrm{HF}) - E(\mathrm{HSB\mathchar`-QSCI})}{E(\mathrm{HF})-E(\mathrm{CAS\mathchar`-CI})}$.
\end{table*}

\subsection{Shot number dependence on the HSB-QSCI simulations of {\bf 5} with the maximum locality $m = 6$}

\begin{figure}[H]
    \centering
    \includegraphics[width=\linewidth]{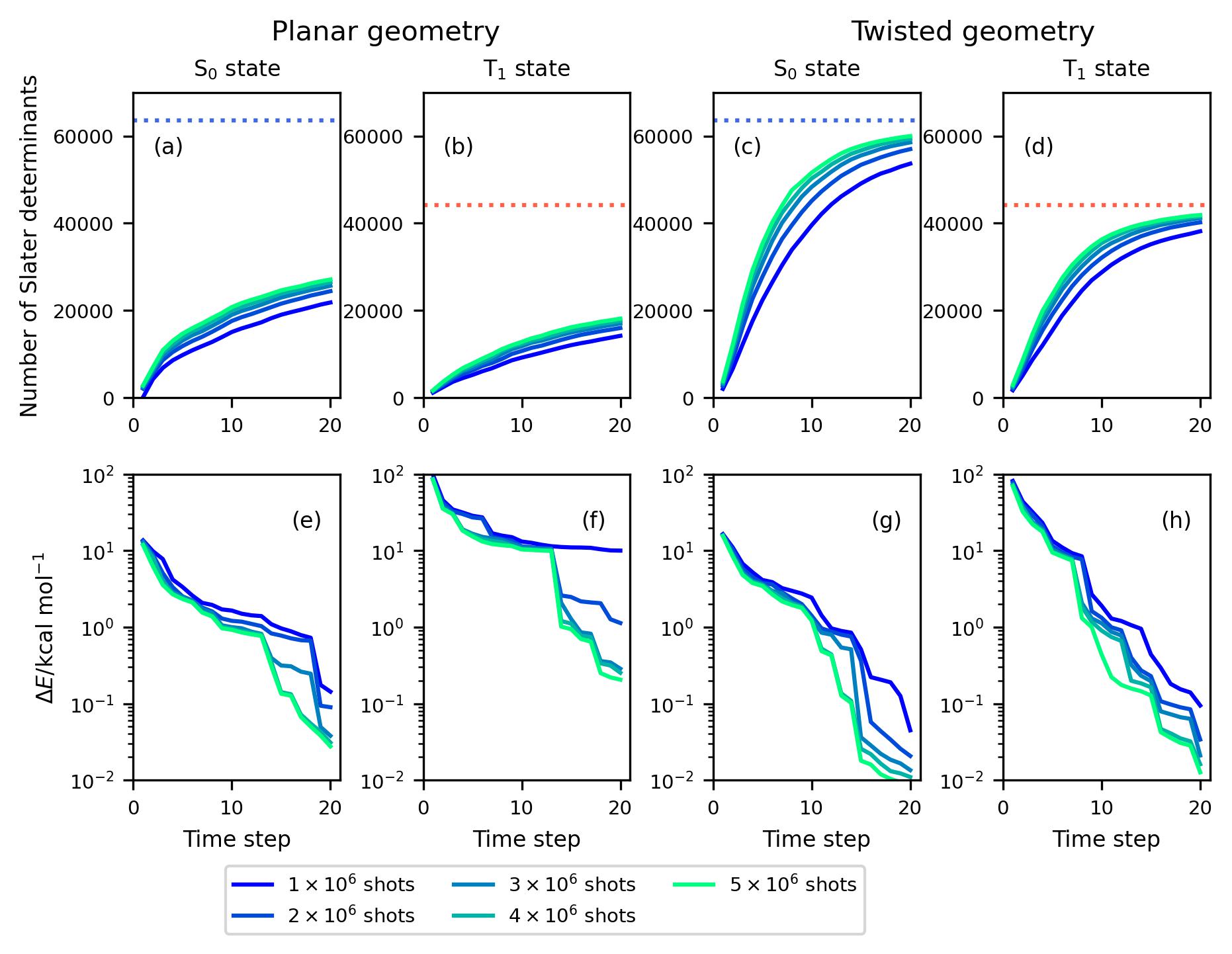}
    \caption{Shot number and time step dependence on the HSB-QSCI simulations of {\bf 5} in its planar and twisted geometries with the maximum locality of Hamiltonian terms $m = 6$. The number of Slater determinants included in the Hamiltonian diagonalization is given in (a), (b), (c), and (d). The red and blue dotted lines represent the number of Slater determinants in the CAS-CI wave function of the S$_0$ and T$_1$ states, respectively. The difference of the HSB-QSCI energy from the CAS-CI values in units of kcal mol$^{-1}$ are given in (e), (f), (g), and (h) on the logarithmic scale.}
    \label{fig:s12}
\end{figure}

\end{document}